\documentclass[a4paper,11pt]{article}
\pdfoutput=1 

\usepackage{jheppub} 

\usepackage[T1]{fontenc} 

\usepackage{graphicx}	
\usepackage[english]{babel}
\usepackage{amsmath,amsfonts,amsthm,amssymb, psfrag} 

\usepackage[fleqn,tbtags]{mathtools}
\usepackage{blkarray}
\usepackage{tikz,tikz-cd}
\usepackage{hyperref}
\usepackage{pgfplots}
\usepackage{colortbl}

\newcommand{\arrowIn}{
\tikz \draw[-latex] (-2pt,0) -- (2pt,0);
}

\def\p{\partial}

\newcommand{\dn}{\mathrm{dn}}
\newcommand{\sn}{\mathrm{sn}}
\newcommand{\cn}{\mathrm{cn}}

\def\be{\begin{equation}}
\def\ee{\end{equation}}
\def\bea{\begin{eqnarray}}
\def\eea{\end{eqnarray}}

\newcommand{\NN}{{\cal N}}

\newcommand{\tr}{\mbox{tr}}

\newcommand{\tQ}{\tilde{Q}}
\newcommand{\Qt}{\widetilde{Q}}

\newcommand{\SU}{\mathrm{SU}}
\newcommand{\SO}{\mathrm{SO}}
\newcommand{\Acal}{\mathcal{A}}
\newcommand{\Hcal}{\mathcal{H}}
\newcommand{\Wcal}{\mathcal{W}}
\newcommand{\Ncal}{\mathcal{N}}
\newcommand{\Scal}{\mathcal{S}}
\newcommand{\Rcal}{\mathcal{R}}
\newcommand{\Lcal}{\mathcal{L}}
\newcommand{\Zset}{\mathbb{Z}}
\newcommand{\Rhat}{\hat{R}}
\newcommand{\ttheta}{\tilde{\theta}}
\newcommand{\Cset}{{\,\,{{{^{_{\pmb{\mid}}}}\kern-.47em{\mathrm C}}}}}

\newcommand{\doublet}[2]{\left(\begin{array}{c}#1\\#2\end{array}\right)}

\newcommand{\ket}[1]{\left|#1\right.\rangle}
\newcommand{\sket}[1]{\left|\scriptstyle{#1}\right.\rangle}
\newcommand{\diff}{\mathrm{d}}
\newcommand{\ra}{\rightarrow}
\newcommand{\half}{\frac12}

\newcommand{\Tr}{\mathrm{Tr}}

\newcommand{\lp}{{\lambda'}}

\newcommand{\mparagraph}[1]{\paragraph{#1}\mbox{}}
\newcommand{\BW}[5]{W\left(\begin{array}{cc} #4&#3\\#1&#2\end{array}\Big|#5\right)}

\title{\boldmath Dynamical Spin Chains in 4D $\mathcal{N}=2$ SCFTs }

\preprint{DESY 21-082}

\author[a]{Elli Pomoni}
\author[b,c]{Randle Rabe}
\author[b,c]{Konstantinos Zoubos}

\affiliation[a]{DESY Theory Group\\ Notkestrasse 85  22607 Hamburg, Germany}
\affiliation[b]{Department of Physics, University of Pretoria\\
Private Bag X20, Hatfield 0028, South Africa}
\affiliation[c]{National Institute for Theoretical and Computational Sciences (NITheCS) \\
Gauteng, South Africa }

\emailAdd{elli.pomoni@desy.de}
\emailAdd{randlerabe@gmail.com}
\emailAdd{kzoubos@up.ac.za}

\abstract{ 

\vspace*{0.5cm}

This is the first in a series of papers devoted to the
study of spin chains capturing the  spectral problem of 4d $\Ncal=2$ SCFTs in the planar limit.
At one loop and in  
 the quantum plane limit, we discover a quasi-Hopf symmetry algebra, defined by the $R$-matrix read off from the superpotential.
This implies that when orbifolding the $\mathcal{N}=4$ symmetry algebra down to the  $\mathcal{N}=2$ one and then marginaly deforming, the broken generators are not lost, but get upgraded to quantum generators.
Importantly, we demonstrate that  these chains  are dynamical, in the sense that their Hamiltonian depends on a parameter which is dynamically determined along the chain. 
At one loop we map the holomorphic  $\SU(3)$ scalar sector to a dynamical 15-vertex model, which corresponds to an RSOS model, whose adjacency graph can be read off from the gauge theory quiver/brane tiling. One scalar $\SU(2)$ sub-sector is described by an alternating nearest-neighbour Hamiltonian, while another choice of  $\SU(2)$ sub-sector leads to a dynamical dilute Temperley-Lieb model. These sectors have a common vacuum state, around which the magnon dispersion relations are naturally uniformised by elliptic functions. 
Concretely, for the $\mathbb{Z}_2$ quiver theory 
we study these dynamical chains  by solving the one- and two-magnon problems with the coordinate Bethe ansatz approach. We confirm our analytic results by numerical comparison with the explicit diagonalisation of the Hamiltonian for short closed chains.

}
\begin{document} 
\maketitle
\flushbottom

\section{Introduction}
\label{sec:intro}

The maximally supersymmetric gauge theory in four dimensions (4d), $\mathcal{N}=4$ super Yang-Mills (SYM),  is often, rightfully so, referred to as the harmonic oscillator of our century. It is integrable in the planar limit  and the prototype of an exactly solvable gauge theory  \cite{Beisert:2010jr}. Integrability allows the complete determination of the exact operator spectrum of $\mathcal{N} = 4$ SYM  in the planar limit while it has also been used for the study of a number of observables like correlation functions,  Wilson loops  and scattering amplitudes \cite{ArkaniHamed:2010kv,CaronHuot:2011kk,DelDuca:2019tur}. Impressively, planar integrability can also be used to obtain non-planar results \cite{Bargheer:2017nne,Bargheer:2018jvq}.

This paper is devoted to the study of the next simplest (=most symmetric) gauge theories in 4d after $\mathcal{N}=4$ SYM, namely $\mathcal{N}=2$ superconformal field theories (SCFTs). Moving away from the harmonic oscillator of our century, $\mathcal{N}=2$ SCFTs are,
for the time being, believed to not be integrable  \cite{Gadde:2010zi,Gadde:2010ku,Liendo:2011xb,Pomoni:2011jj,Gadde:2012rv,Pomoni:2013poa,Pomoni:2019oib}. This is because their Scattering ($S$-) matrix does not obey the usual Yang-Baxter equation\footnote{Even for $\mathcal{N}=4$ SYM, away from the leading order in the YM coupling, the fact that  the $S$-matrix obeys the Yang-Baxter equation is guaranteed only after  a non-local transformation of the basis states which was constructed in \cite{Arutyunov:2006yd}.}.
In this paper we will argue that this conclusion deserves to be revisited.

Most known integrable models fall into one of three classes: rational, trigonometric and elliptic\footnote{Integrable models exist even
beyond these three classes. The hyperelliptic Chiral Potts model is a famous example \cite{BaxterBook,McCoyBook,AuYang:1987zc,McCoy:1987pt,Baxter_1998}. We will come back to it in the conclusions.}. These are based on Yangian symmetry, quantum affine algebra and elliptic quantum groups, respectively   \cite{Torrielli:2011gg,Loebbert:2016cdm,Drinfeld88,Felder:1994be}. For the first two classes the $S$-matrix obeys the Yang-Baxter equation (YBE). The $R$-matrix of  elliptic models,  depending on the choice of basis, may obey a modified, dynamical Yang-Baxter equation  (dYBE) \cite{Felder:1994be,Felder:1996xym}. The dYBE itself arises as a special case of an even more general quasi-Hopf YBE  \cite{Drinfeld90,Drinfeld91,Babelon:1995rz,Enriquez_1998,Jimbo:1999zz}.

The spin chain emerging in the planar limit of $\mathcal{N}=4$ SYM is a rational integrable model based on the Yangian symmetry built on  $PSU(2,2|4)$.  We believe, for reasons we start unraveling with this paper, that the spectral problem of $\mathcal{N}=2$ SCFTs can be mapped to dynamical\footnote{In this paper, the term dynamical spin chain is not used in the same sense as in \cite{Beisert:2003ys}, but as in \cite{Felder:1994be,Felder:1996xym}. 
Beisert in \cite{Beisert:2003ys} uses the term dynamic for the spin chains of $\mathcal{N}=4$ SYM which arise at higher loops (higher orders in $g_{YM}$) and include length changing operations.
The spin chains we will study in this paper are at one-loop where length changing operations cannot take place.
What changes is the dynamical parameter, according to the gauge group indices which are being contracted between different fields occupying neighbouring sites on the chain.
} elliptic models. What is more, we exhibit that $\mathcal{N}=2$ SCFTs' spin chains
  enjoy a quantum group deformation of $PSU(2,2|4)_\kappa \supset  SU(2,2|2)$ leading to a  quasi-Hopf YBE, providing an additional handle to their study.

 Lagrangian $\mathcal{N}=2$  SCFTs are classified \cite{Bhardwaj:2013qia} and a
 good way to obtain most of them  is via orbifolding  $\mathcal{N}=4$ SYM. 
The quivers of $\mathcal{N}=2$  SCFTs with $SU(N)$ color factors follow an ADE classification
 using finite/affine Dynkin diagrams\footnote{For $SO/Sp$ gauge groups we need to further use an orientifold plane. Exceptional gauge groups are more complicated but are also possible to get.}. 
  The orbifolding procedure \cite{Douglas:1996sw} gives  theories with all the coupling constants equal to each other and to the Yang-Mills (YM) coupling constant of the  $\mathcal{N}=4$ mother theory. This is called the orbifold point. 
Orbifold daughters of   $\mathcal{N}=4$ SYM at the orbifold point are known to be  integrable in the planar limit \cite{Beisert:2005he}, their Bethe equations were studied in \cite{Wang:2003cu,Ideguchi:2004wm,Beisert:2005he,Solovyov:2007pw}, their Y-system in \cite{Beccaria:2011qd}  and TBA in  \cite{Arutyunov:2010gu,deLeeuw:2012hp}.
To go away from the orbifold point we marginally deform the theory.
 The question of planar integrability of $\Ncal=2$ theories marginally deformed  away from the orbifold point has been addressed in a number of papers \cite{Gadde:2010zi,Gadde:2010ku,Liendo:2011xb,Pomoni:2011jj,Gadde:2012rv,Pomoni:2013poa}. See \cite{Pomoni:2019oib} for a recent review.

$\Ncal=1$ marginal deformations of $\mathcal{N}=4$ SYM \cite{Leigh:1995ep} and their spin chains in the planar limit have been well studied and some of them have been shown to be integrable \cite{Roiban:2003dw,Berenstein:2004ys}. For the most general marginal deformation, the one-loop $R$-matrix for the holomorphic scalar sector in the quantum plane limit was constructed in \cite{Bundzik:2005zg}. In  \cite{Mansson:2008xv}, the RTT relations arising from this $R$-matrix were shown to lead to quantum generators that deform the $\SU(3)$ algebra to a quantum group. More recent work \cite{Dlamini:2019zuk,Dlamini:2016aaa} took a quasi-Hopf perspective by focusing on the Drinfeld twist underlying the $R$-matrix, which allows to express the deformation in terms of a star product. The work of \cite{Bundzik:2005zg,Mansson:2008xv,Dlamini:2019zuk,Dlamini:2016aaa}  will be an inspiration and starting point for the present work.
 
 Revisiting the study of marginally deformed (away from the orbifold point) orbifold daughters of   $\mathcal{N}=4$ SYM, we will mostly concentrate on
the simplest example, the $\mathbb{Z}_2$ orbifold of $\mathcal{N}=4$ SYM, with gauge group  $SU(N)  \times SU(N)$, the quiver of which is depicted in Figure \ref{fig:z2}.
This is a theory with a well understood $AdS_5\times S^5/\mathbb{Z}_2$ gravity dual \cite{Kachru:1998ys,Lawrence:1998ja,Gadde:2009dj}.  
 For the $\mathbb{Z}_2$ orbifold there is only one marginal deformation which allows us to explore the real dimension-two conformal manifold of the $g_1 \neq g_2$ exactly marginal YM couplings.\footnote{
 In supersymmetric gauge theories the  YM couplings $g_{YM}$ together with the theta angle $\theta_{YM}$ combine to form holomorphic gauge coupling $\tau = { \theta_{YM} \over 2 \pi} + {4\pi i \over g_{YM}^2}$. 
 In this paper we are not interested in the $\theta_{YM}$ parameters, and set them to zero. The  YM couplings $g_1,g_2$ are simply real.}
 All the novel features of $\mathcal{N}=2$ SCFT spin chains which we wish to bring to light with this paper are nicely captured by this simple example.\footnote{The only big difference between $\mathbb{Z}_2$ and $\mathbb{Z}_{k>2}$ is that the  $\mathbb{Z}_2$ quiver theory has an extra $SU(2)_L$ global symmetry, as we will introduce in Section \ref{subsec:QuiverTheory}.
}

This will be the first in a series of papers with which we will revisit the one-loop spectral problem of 4d $\Ncal=2$ SCFTs in the planar limit. 
We approach the study of the spin chains which describe it from several independent fronts.
 We begin this paper with a review of the  $\mathbb{Z}_2$ quiver SCFT, its field content and its symmetries as well as by recalling the scalar Hamiltonian originally derived in  \cite{Gadde:2010zi}, in Section \ref{sec:SU(3)}.
In Section \ref{sec:QuasiHopf}, simply by looking at  the superpotential of the $\mathbb{Z}_2$ quiver theory and its $F$-term equations, and following \cite{Bundzik:2005zg,Mansson:2008xv,Dlamini:2019zuk} we will obtain  the $R$-matrix of the one-loop spin chain in
 the quantum plane limit as well as the Drinfeld twist with which it can be factored. 
 The  Drinfeld twist does not obey the usual cocycle condition and the $R$-matrix which does not satisfy the usual YBE, but rather a generalisation known as the quasi-Hopf YBE \cite{Drinfeld90}.
 Through the RTT relations \cite{FRT90}, the  $R$-matrix  defines a quantum algebra which captures the symmetry of our spin chain.
 Most importantly, as we break the $\mathcal{N}=4$ superconformal algebra (SCA) down to the  $\mathcal{N}=2$ SCA the broken generators are not lost, but they get upgraded to quantum generators. The discussion in Section \ref{sec:QuasiHopf} uses a basis which is natural from the point of view of the $\mathcal{N}=2$ daughter theories, however quite cumbersome as not all possible spin chains (corresponding to single trace local operators) made out of $\mathcal{N}=2$ letters are allowed color contractions.

In Section \ref{sec:Dynamical}, looking for a better basis which is more natural from the point of view of the $\mathcal{N}=4$ mother theory, we  discover that the spin chain capturing the spectral problem of the  $\mathbb{Z}_2$ quiver SCFT, is a dynamical spin chain. Its Hamiltonian depends on a parameter which is dynamically determined along the chain, precisely keeping track of which gauge indices are contracted between two neighbouring cites of the spin chain.
Following this logic, we discover that  the holomorphic  $\SU(3)$ scalar sector can be mapped to a dynamical 15-vertex model.
This is despite the fact that naively the $\SU(3)$ scalar sector with single-site vector space spanned by $X,Y,Z$ gives a 19 vertex model: the $U(1)_r$ R-symmetry of $\mathcal{N}=2$ SCA sets to zero four of the vertices.
What is more, this  dynamical 15-vertex model can be rewritten as an RSOS (Restricted Solid-on-Solid) model 
\cite{Pasquier:1986jc,DiFrancesco:1989ha,Roche90,Fendley:1989vt}, the adjacency graph of which we are able to determine and is depicted in Figure \ref{fig:z2adjacency}.
Amazingly, the adjacency graph of the RSOS model is dual to the brane-tiling diagram of the corresponding quiver theory \cite{Hanany:2005ve,Franco:2005rj,Franco:2005sm} (see also \cite{Yamazaki:2008bt} for a review).

In this holomorphic $\SU(3)$ scalar sector we identify one closed $\SU(2)$ sub-sector which is naturally described by an {\it alternating, nearest-neighbour} spin chain.\footnote{
The readers of this paper are most probably familiar with the alternating ABJ(M) spin chains \cite{Minahan:2008hf,Gaiotto:2008cg,Minahan:2009te}.
We wish to stress that our spin chains are different from the  ABJ(M)  ones in two ways: the spin chains of the  3d ABJ(M) theories are of alternating representation while our  spin chains of 4d  $\mathcal{N}=2$  SCFTs have alternating bonds.
What is more, we have at one-loop nearest neighbour spin chains  while for  ABJ(M) they start necessarily at  two-loops and they are next to nearest neighbour.
}  The eigenvalue problem for spin chains of this type has been studied in the past in \cite{Bell_1989,Medvedetal91}. We refer to this sector as the $XY$ sector and as we will see it enjoys an unbroken $\SU(2)$ symmetry, precisely as its image in the  $\mathcal{N}=4$ SYM mother theory. 
The study of this sector through the coordinate Bethe Ansatz is presented in Section \ref{sec:XYsector}.

What is more, within the holomorphic $\SU(3)$ sector  a second closed sub-sector is identified as a {\it dynamical dilute Temperley-Lieb model}.
This second sub-sector we will call the $XZ$ sector and in this case the $\SU(2)$ is naively broken (but in reality upgraded to a quantum group defined by the RTT relations). This sector was studied around the $Z$ vacuum in \cite{Gadde:2010zi} with the conclusion that it is not integrable as it does not satisfy the usual YBE. The study of this sector though the coordinate Bethe Ansatz is presented in Section \ref{sec:XZsector}.
We will clearly see that the spin chain of the  $XZ$ sector is elliptic and this implies that the conclusion of \cite{Gadde:2010zi} might be too naive, and one needs to understand whether a dynamical version of the YBE might be applicable instead.

Having identified these two distinct $\SU(2)$ sub-sectors we study them using the coordinate Bethe ansatz.
This part of our work is tedious but cannot be avoided as the spin chains we are dealing with are not well known models as was the the case of the Heisenberg XXX spin chain  for $\mathcal{N}=4$ SYM. We need to discover their best possible description by studying them in detail and the only straightforward method we have at our disposal is the coordinate Bethe ansatz.
The state space of both of them contains {\it the $X$ vacuum} (made out of bifundamental hypermultiplet fields). Already for the one magnon problem around the $X$ vacuum we discover, in Section \ref{sec:Elliptic}, that 
the single  magnon dispersion relations as well as the wave functions  are naturally uniformised by elliptic functions.
 Moving to  two magnons, in  Section \ref{sec:XYsector} and  \ref{sec:XZsector} respectively, we discover that in the center-of-mass frame the eigenvalue problem can be solved only after including contact terms. Away from the center-of-mass frame the eigenvector has to contain a second set of momenta (not related by permutations to the original set), similar to the method of \cite{Bibikov_2016}. 
This is due to the elliptic  nature of the problem\footnote{The appearance of the second set of momenta is natural in staggered models \cite{BaxterBook}.}. For a fixed total energy and total momentum, more than one set of momenta (beyond those obtained by permutations) are allowed, and indeed they are needed in order to parametrise the eigenvectors for the two magnon problem.
We further  unveil the rich set of symmetries of the one and most impressively of the two magnon eigensystems, which summarise in Table  \ref{Symmetries}.
The $\mathbb{Z}_2$ symmetry of the quiver (which exchanges the coupling constants as well as the fields) commutes with the Hamiltonian and  we can organise the eigenvectors in a basis where they are also eigenvectors of the  $\mathbb{Z}_2$ with very specific and interesting eigenvalues.

Obtaining from the most general 2-magnon solution with the second set of momenta, the centre-of-mass solution involving contact terms (but only one set of momenta) is a technical computation which we detail in  Appendices \ref{XYCoMlimit} and  \ref{XZCoMlimit} for the interested reader. In Appendix \ref{appendix:Alternating} we demonstrate with an example how one reproduces an alternating-type Hamiltonian from a dynamical $R$-matrix using the Algebraic Bethe ansatz framework. Although this example (an alternating $XX$ model) is of free-fermion type and thus too simple to describe our Hamiltonians, we believe that it does provide a useful illustration of how the features that we consider important for our models come together in a concrete setting.

\section{The SU(3) scalar sector}
\label{sec:SU(3)}

The spin chain for the $\mathbb{Z}_{2}$ quiver was first introduced and studied in \cite{Gadde:2010zi}.  Before we delve into the description of the spin chain, a short review of the gauge theory and its symmetries is in order.

\subsection{The $\mathbb{Z}_{2}$ quiver theory}
\label{subsec:QuiverTheory}

We will focus on the $\Ncal=2$ SCFT which interpolates between the $\Zset_2$ orbifold of $\Ncal=4$ SYM and $\Ncal=2$ superconformal QCD (SCQCD). Starting from the $\Zset_2$ orbifold where the couplings of the two gauge groups are equal, $g_1=g_2$, also known as the orbifold point,
in this paper we will be interested in the marginally deformed theory with  $g_1 \neq g_2$. The marginal deformation introduces
an one parameter family of theories parametrised by the ratio
 $\kappa=g_2/g_1$. Without loss of generality, we will take $\kappa\leq 1$. For all allowed values of $\kappa \in \left[0,1\right]$ this one parameter family of theories enjoys the full $\mathcal{N}=2$ superconformal algebra (SCA) including the $SU(2)_R \times U(1)_r$ R-symmetries but also an extra $SU(2)_L$ global symmetry which is special for the $\Zset_2$ quiver.
 
  The edge of the conformal manifold where $g_2 \to 0$ or equivalently $\kappa\ra0$ is special. In this limit, the second gauge group gets ungauged and becomes a global symmetry, which combines with the extra $SU(2)_L$ to give the $U(N_f)= U(2N)$ flavor symmetry of
SCQCD \cite{Gadde:2010zi}.\footnote{In the limit $\kappa\ra0$  the spin chain states (single trace local operators) which are not in a
$SU(2)_L$ singlet representation break.
What is more, the bifundamental hypermultiplet fields stick together forming dimers which are much more difficult to treat.
 In this work we will not consider this limit, but rather the spin chain at generic values of $\kappa$, which can be thought of as a regularisation of the SCQCD spin chain.}

\begin{figure}[t]
\begin{center}
\begin{tikzpicture}[scale=0.6]
  \draw[fill=blue] (0,0) circle (2ex);
\draw[fill=red] (6,0) circle (2ex);

\draw[->,blue,thick] (-0.5,-0.8) arc (-60:-310:1);
\draw[->,red,thick] (6.4,0.8) arc (120:-130:1);

\draw[->,purple,thick] (0.2,0.8) arc (110:70:8);
\draw[->,purple,thick] (5.7,0.5) arc (70:110:8);
\draw[->,purple,thick] (5.7,-0.8) arc (-70:-110:8);
\draw[->,purple,thick] (0.2,-0.5) arc (-110:-70:8);

  \node at (-0.7,-0) {$1$};\node at (6.8,-0) {$2$};

  \node at (3,2) {$Q_{12}$};\node at (3,-1.7) {$Q_{21}$};
 \node at (3,0.5) {$\Qt_{21}$};\node at (3,-0.4) {$\Qt_{12}$};

\node at (-2.4,-0.5) {$\phi_1$};\node at (8.2,-0.5) {$\phi_2$};

\end{tikzpicture}
\end{center}
\caption{\it The $\mathbb{Z}_{2}$ quiver. The color groups are denoted by blobs representing the field content of the $\NN=1$ vector multiplets inside the $\NN=2$ vector multiplets. We use $\NN=1$ language for the hypermultiplets. The arrow to the right is $Q^{\hat{I}}$ while the arrow to the left is $\tilde{Q}^{\hat{I}}$. \label{fig:z2}}
\end{figure}
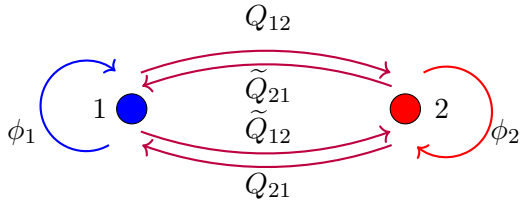

To describe the field content of the theory we consider the $\mathbb{Z}_{2}$ orbifold of $\mathcal{N}=4$ SYM, see \cite{Gadde:2010zi} or the recent review
\cite{Pomoni:2019oib}
for more details. We start with the  $2N \times 2N$ color matrices in the mother $\mathcal{N}=4$ theory, which, in $\Ncal=1$ language, contains
three chiral superfields $X,Y,Z$ in the adjoint of the $\SU(2N)$ gauge group. The action of the $\Zset_2$ is chosen such as to preserve $\Ncal=2$ supersymmetry, which means that it only acts on a $\Cset^2$ of the $\Cset^3$ transverse space spanned by the three complex scalar fields. The orbifold procedure projects out some $N\times N$ blocks of these fields \cite{Douglas:1996sw}, so that after the projection they look like
\begin{eqnarray}
\label{eq:Z2orbifold} 
 X  =
     \begin{pmatrix}
 & Q_{12}   \\
Q_{21} &   
\end{pmatrix}  
\, ,  \quad
Y =  \begin{pmatrix}
  & \tilde{Q}_{12}  \\
 \tilde{Q}_{21}&   \\
\end{pmatrix}  
\,    ,    \quad
 Z  =
\begin{pmatrix}
\phi_{1} &    \\
&   \phi_{2} \\  
\end{pmatrix}  
\, . 
\end{eqnarray}
where we indicate the surviving $N\times N$ blocks. After the orbifold the gauge group is now $\SU(N)\times \SU(N)$. We see that while on $Z$ it  acts diagonally and thus keeps us on the same node of the quiver,  on $X$ takes us clockwise around the quiver while on $Y$ anticlockwise.\footnote{This distinction is not very important here as there are only two nodes, but it becomes more relevant for $\Zset_k$ quivers. The quiver for the $\mathcal{N}=2$ preserving $\mathbb{Z}_3$ orbifold is displayed in Figure \ref{Z3}.} 
From the surviving  $N \times N$ blocks,
$Q_{12} $ and $\tilde{Q}_{12}$ have the same bi-fundamental color structure $\Box_1 \times \overline{\Box}_2$, and the same is the case for $Q_{21} $ and $\tilde{Q}_{21}$ but with the opposite orientation $\overline{\Box}_1 \times {\Box}_2$. Thus, we can put them together in a doublet of an extra $SU(2)_L$ with index $\hat{I}$, as follows
\be
\label{eq:hatdoublet}
 Q_{\hat{I}}  = \left( Q_{12} \, , \, \tilde{Q}_{12} \right)^T \qquad \mbox{and} 
 \qquad  
 \tilde{Q}_{\hat{I}}  =  \left( \tilde{Q}_{21}    \, , \,    Q_{21}\right)^T
 \,.
 \ee
The superpotential is explicitly invariant under the  extra $SU(2)_L$ rotating the  doublets of $SU(2)_L$ in \eqref{eq:hatdoublet} and can be written as
\begin{equation}
\label{eq:quiver}
\mathcal{W}_{\mathbb{Z}_{2}}
=
  ig_{1}  \, \tr_2  \left(\,  \tilde{Q}^{\hat{I}} \phi_{1} Q_{\hat{I}}  \right) -  ig_{2} \, \tr_1  \left(\,  Q_{\hat{I}} \phi_{2}   \tilde{Q}^{\hat{I}}\right) 
\end{equation} 
or more explicitly as
\be \label{Winterpolating}
\mathcal{W}_{\mathbb{Z}_{2}} =  ig_1 \,  \tr_2(\Qt_{21}\phi_1Q_{12}  -  Q_{21}\phi_1\Qt_{12})  -   ig_2  \, \tr_1(Q_{12}\phi_2\Qt_{21} -  \Qt_{12}\phi_2 Q_{21}  )
\, .
\ee
This second more explicit form may be more intuitive for some readers as it is easy to remember that the bifundamental fields $Q_{ij}$ and $\tQ_{ij}$ are labelled according to the direction of the arrows. For instance, $Q_{12}$ transforms in the fundamental of gauge group 1 and the antifundamental of gauge group 2   ($\Box_1 \times \overline{\Box}_2$).

We already wish to indicate  that we have the choice of working in the daughter $\mathcal{N}=2$ SCFT picture (with our single letter basis composed of the six fields $\phi_i$ and $Q_{ij},\tQ_{ij}$) or  in the mother $\Ncal=4$ SYM picture where the single site  basis is made out of $X,Y,Z$ . We will mostly use the latter as it allows us to simplify the discussion. As we will see, the information of whether we are working with the upper or lower component of a given field in the $\Ncal=4$ picture will be provided by a dynamical parameter $\lambda$.

\subsection{The Hamiltonian}

In this paper we will focus on the one-loop holomorphic $\SU(3)$ sector of the $\mathbb{Z}_{2}$ quiver. In the mother $\Ncal=4$ SYM this sector is made up of three complex scalar fields $X,Y,Z$ in the adjoint of the $\SU(2N)$ gauge group. 
The planar Hamiltonian of this theory has been derived, for the full scalar sector, in \cite{Gadde:2010zi}. We begin by visually rederiving the Hamiltonian in two $\SU(2)$-like sectors, the one formed by the fields $X$ and $Y$ and the one formed by $X$ and $Z$, so that we can highlight the difference between these sectors.

\mparagraph{The XY sector}

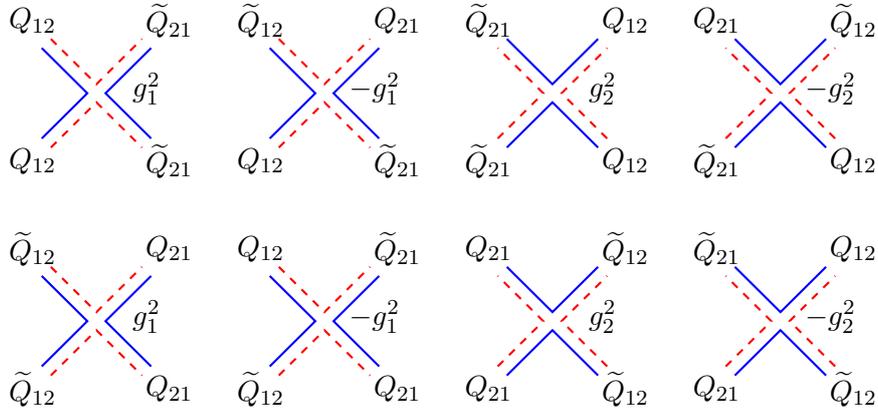
\begin{figure}[ht]
\begin{center}
\begin{tikzpicture}[scale=0.6]

  \draw[-,thick,blue] (-0.2,0)--(0.8,1)--(-0.2,2);\draw[-,thick,red,dashed] (0,-0.2) -- (1,0.8)--(2,-0.2);
  \draw[-,thick,blue] (2.2,0)--(1.2,1)--(2.2,2);\draw[-,thick,red,dashed] (0,2.2) -- (1,1.2)--(2,2.2);
  \node at (-0.4,-0.5) {$Q_{12}$};\node at (2.6,-0.5) {$\Qt_{21}$}; \node at (-0.4,2.6) {$Q_{12}$}; \node at (2.6,2.6) {$\Qt_{21}$};
  \node at (2.1,1.1) {$g_1^2$};

  \draw[-,thick,blue] (5-0.2,0)--(5+0.8,1)--(5-0.2,2);\draw[-,thick,red,dashed] (5+0,-0.2) -- (5+1,0.8)--(5+2,-0.2);
  \draw[-,thick,blue] (5+2.2,0)--(5+1.2,1)--(5+2.2,2);\draw[-,thick,red,dashed] (5+0,2.2) -- (5+1,1.2)--(5+2,2.2);
  \node at (5-0.4,-0.5) {$Q_{12}$};\node at (5+2.6,-0.5) {$\Qt_{21}$}; \node at (5-0.4,2.6) {$\Qt_{12}$}; \node at (5+2.6,2.6) {$Q_{21}$};
  \node at (5+2.1,1.1) {$-g_1^2$};

    \draw[-,thick,red,dashed] (10-0.2,0)--(10+0.8,1)--(10-0.2,2);\draw[-,thick,blue] (10+0,-0.2) -- (10+1,0.8)--(10+2,-0.2);
  \draw[-,thick,red,dashed] (10+2.2,0)--(10+1.2,1)--(10+2.2,2);\draw[-,thick,blue] (10+0,2.2) -- (10+1,1.2)--(10+2,2.2);
  \node at (10-0.4,-0.5) {$\Qt_{21}$};\node at (10+2.6,-0.5) {$Q_{12}$}; \node at (10-0.4,2.6) {$\Qt_{21}$}; \node at (10+2.6,2.6) {$Q_{12}$};
  \node at (10+2.1,1.1) {$g_2^2$};

      \draw[-,thick,red,dashed] (15-0.2,0)--(15+0.8,1)--(15-0.2,2);\draw[-,thick,blue] (15+0,-0.2) -- (15+1,0.8)--(15+2,-0.2);
  \draw[-,thick,red,dashed] (15+2.2,0)--(15+1.2,1)--(15+2.2,2);\draw[-,thick,blue] (15+0,2.2) -- (15+1,1.2)--(15+2,2.2);
  \node at (15-0.4,-0.5) {$\Qt_{21}$};\node at (15+2.6,-0.5) {$Q_{12}$}; \node at (15-0.4,2.6) {$Q_{21}$}; \node at (15+2.6,2.6) {$\Qt_{12}$};
  \node at (15+2.1,1.1) {$-g_2^2$};

\end{tikzpicture}
\end{center}

\begin{center}
\begin{tikzpicture}[scale=0.6]

  \draw[-,thick,blue] (-0.2,0)--(0.8,1)--(-0.2,2);\draw[-,thick,red,dashed] (0,-0.2) -- (1,0.8)--(2,-0.2);
  \draw[-,thick,blue] (2.2,0)--(1.2,1)--(2.2,2);\draw[-,thick,red,dashed] (0,2.2) -- (1,1.2)--(2,2.2);
  \node at (-0.4,-0.5) {$\Qt_{12}$};\node at (2.6,-0.5) {$Q_{21}$}; \node at (-0.4,2.6) {$\Qt_{12}$}; \node at (2.6,2.6) {$Q_{21}$};
  \node at (2.1,1.1) {$g_1^2$};

  \draw[-,thick,blue] (5-0.2,0)--(5+0.8,1)--(5-0.2,2);\draw[-,thick,red,dashed] (5+0,-0.2) -- (5+1,0.8)--(5+2,-0.2);
  \draw[-,thick,blue] (5+2.2,0)--(5+1.2,1)--(5+2.2,2);\draw[-,thick,red,dashed] (5+0,2.2) -- (5+1,1.2)--(5+2,2.2);
  \node at (5-0.4,-0.5) {$\Qt_{12}$};\node at (5+2.6,-0.5) {$Q_{21}$}; \node at (5-0.4,2.6) {$Q_{12}$}; \node at (5+2.6,2.6) {$\Qt_{21}$};
  \node at (5+2.1,1.1) {$-g_1^2$};

    \draw[-,thick,red,dashed] (10-0.2,0)--(10+0.8,1)--(10-0.2,2);\draw[-,thick,blue] (10+0,-0.2) -- (10+1,0.8)--(10+2,-0.2);
  \draw[-,thick,red,dashed] (10+2.2,0)--(10+1.2,1)--(10+2.2,2);\draw[-,thick,blue] (10+0,2.2) -- (10+1,1.2)--(10+2,2.2);
  \node at (10-0.4,-0.5) {$Q_{21}$};\node at (10+2.6,-0.5) {$\Qt_{12}$}; \node at (10-0.4,2.6) {$Q_{21}$}; \node at (10+2.6,2.6) {$\Qt_{12}$};
  \node at (10+2.1,1.1) {$g_2^2$};

      \draw[-,thick,red,dashed] (15-0.2,0)--(15+0.8,1)--(15-0.2,2);\draw[-,thick,blue] (15+0,-0.2) -- (15+1,0.8)--(15+2,-0.2);
  \draw[-,thick,red,dashed] (15+2.2,0)--(15+1.2,1)--(15+2.2,2);\draw[-,thick,blue] (15+0,2.2) -- (15+1,1.2)--(15+2,2.2);
  \node at (15-0.4,-0.5) {$Q_{21}$};\node at (15+2.6,-0.5) {$\Qt_{12}$}; \node at (15-0.4,2.6) {$\Qt_{21}$}; \node at (15+2.6,2.6) {$Q_{12}$};
  \node at (15+2.1,1.1) {$-g_2^2$};

\end{tikzpicture}
\end{center}
\caption{\it The vertices contributing to the Hamiltonian in the $XY$ sector. A solid blue line denotes the first and a dashed red line denotes the second gauge group. Time moves upwards. Here we have already performed the Wick contractions of the conjugate fields with the second gauge invariant operator to write the vertices directly as spin chain interactions.} \label{fig:XYvertices}
\end{figure}

This is the sector which includes all the (holomorphic) bifundamental fields. To derive the Hamiltonian, let us start by considering the $\phi_i$ F-terms:
\be
F_{\phi_1}=g_1 (Q_{12}\Qt_{21}-\Qt_{12}Q_{21})\;\;,\quad F_{\phi_2}=g_2 (Q_{21}\Qt_{12}-\Qt_{21} Q_{12})
\ee
From the potential $F\bar{F}$, following the treatment in e.g. \cite{Roiban:2003dw}, we can immediately draw the vertices  contributing to the one-loop Hamiltonian. These are shown in Figure \ref{fig:XYvertices}.

After taking out an overall factor of $g_1g_2$ and defining $\kappa=g_2/g_1$, we find the Hamiltonian:

\be \label{XYHamiltonian}
H_{\ell , \ell+1}=\begin{pmatrix}0&0&0&0&0&0&0&0\cr 0&\kappa^{-1}&-\kappa^{-1}&0&0&0&0&0\cr 0&
 -\kappa^{-1}&\kappa^{-1}&0&0&0&0&0\cr 0&0&0&0&0&0&0&0\cr 0&0&0&0&0&0&0&0\cr 0&0&0&0
 &0&\kappa&-\kappa&0\cr 0&0&0&0&0&-\kappa&\kappa&0\cr 0&0&0&0&0&0&0
 &0\cr \end{pmatrix}\;,\qquad \text{in the basis} \quad \left(\begin{array}{c}Q_{12}Q_{21}\\Q_{12}\tQ_{21}\\\tQ_{12}Q_{21}\\\tQ_{12}\tQ_{21}\\Q_{21}Q_{12}\\Q_{21}\tQ_{12}\\\tQ_{21}Q_{12}\\\tQ_{21}\tQ_{12}\end{array}\right)
\ee
where the indices $\ell , \ell+1$ denote the nearest neighbour sites of the spin chain.
Note that the basis is 8-dimensional instead of 16-dimensional as one would expect given our four fields. The remaining combinations of fields cannot occur, as they are not allowed by the gauge structure (for e.g. a $Q_{12}$ cannot be followed by a $Q_{12}$ or $\tQ_{12}$). We have chosen this truncated basis such that the upper left block of the Hamiltonian corresponds to the first gauge group to the left of the first site where the Hamiltonian acts. In other words, the upper left block acts on two bifundamental squarks which are contracted or in the singlet representation of the second gauge group and have their indices open which means that they are in the bifundamental representation of the first color group ($\Box_1 \times \overline{\Box}_1$).
On the other hand, the lower right block of the Hamiltonian acts on two squarks which have open color indices from the second gauge group ($\Box_2 \times \overline{\Box}_2$) and are color contracted with respect to the first color group.
We emphasise that, although the Hamiltonian looks block-diagonal, this is an artifact of the notation. The same fields appear in both blocks, and thus the upper and lower blocks of the Hamiltonian will mix when acting on a spin chain configuration. 

For this and other reasons to become clear later, we will prefer to work in the mother $\mathcal{N}=4$ picture, where we only deal with the $2N\times 2N$ fields $X,Y$ instead of their component fields. A spin chain state such as $\ket{XYXYYX\cdots}$ in the $\mathcal{N}=4$  picture can be decomposed into two states, in this case $\ket{Q_{12}\tQ_{21}Q_{12}\tQ_{21}\tQ_{12}Q_{21}\cdots }$ and $\ket{Q_{21}\tQ_{12}Q_{21}\tQ_{12}\tQ_{21}Q_{12}\cdots}$ in the $\mathcal{N}=2$  picture. These states can of course be mapped to each other by exchanging the gauge groups. In the $XY$ picture, which of the two chains we are considering is uniquely defined by specifying the gauge group to the left of a given site of the chain (meaning the first index of the bifundamental field at that site). Without loss of generality we can take this reference site to be the first site of the chain. 

 Similarly, the above action of the Hamiltonian is decomposed into an action of \emph{two} Hamiltonians in the $XY$ basis. Whether we are on the upper or lower block again depends on which gauge group is to the left of the first site we are acting on. We call these Hamiltonians $\Hcal_{1}$ and $\Hcal_2$, with
\be
\label{eq:HXY}
\Hcal_1
=
\left(\begin{array}{cccc}
 0 & 0 & 0 & 0 \\
 0 & \kappa^{-1} & -\kappa^{-1} & 0 \\
 0 & -\kappa^{-1} & \kappa^{-1} & 0 \\
 0 & 0 & 0 & 0 \\
\end{array}
\right)
\quad
\mbox{and}
\quad
\Hcal_2
=
\left(\begin{array}{cccc}
 0 & 0 & 0 & 0 \\
 0 & \kappa & -\kappa & 0 \\
 0 & -\kappa & \kappa & 0 \\
 0 & 0 & 0 & 0 \\
\end{array}
\right)
\;, \quad \text{in the basis} \left(\begin{array}{c}XX\\XY\\YX\\YY\\ \end{array}\right)_{i=1,2}
\ee
where the notation is that $\mathcal{H}_1$ acts on the basis labelled by $i=1$ in the representation $\Box_1 \times \overline{\Box}_1$ of the color group,
while   $\mathcal{H}_2$ acts on the basis with $i=2$ in the representation $\Box_2 \times \overline{\Box}_2$.
In other words, by $\Hcal_1$ we denote the Hamiltonian which is applicable when the gauge group to the left of a site $\ell$ along the chain is the first one, while $\Hcal_2$ is the corresponding Hamiltonian when the gauge group to the left of a site $\ell$ is the second one. Both Hamiltonians are of XXX type but with a different (ferromagnetic) coupling given by $\kappa^{-1}$ and $\kappa$, respectively.

Given that the $XY$ sector is only made up of bifundamentals, which means that the gauge group alternates at consecutive sites (regardless of whether the field at that site is an $X$ or a $Y$), we conclude that the Hamiltonian of this sector is alternately $\Hcal_1$ and $\Hcal_2$. If, for instance, we fix the gauge group to the left of the first site to be the first one, we will have $\Hcal_1$ acting on odd-even sites and $\Hcal_2$ acting on even-odd sites.

We conclude that the $XY$ sector of the interpolating theory is governed by an alternating XXX-model Hamiltonian. In section \ref{sec:XYsector} we will study this alternating spin chain in more detail using the coordinate Bethe ansatz.

For the sake of the interested reader, in order to obtain  the  Hamiltonian in \eqref{eq:HXY}  from
 \cite{Gadde:2010zi}, we can start with the form of the Hamiltonian given at the top of page 16 of \cite{Gadde:2010zi}. Firstly, we note that $\mathbb{K} = \mathbb{K}_{SU(2)_R}$ and is zero on our sector as we look only at the upper components ($IJ = ++$) of the $SU(2)_R$ triplet $Q\tilde{Q}$ or $\tilde{Q}Q$. 
Then the only contributions that are left in our sector are
\begin{equation}
\Hcal_1  |  Q\tilde{Q}  \rangle=  2 \hat{\mathbb{K}}  |  Q\tilde{Q}  \rangle
 \, , \quad
\Hcal_2
 |  \tilde{Q}  Q \rangle= 2\kappa^2  \hat{\mathbb{K}}  |  \tilde{Q} Q \rangle
\end{equation}
where
$\hat{\mathbb{K}} = \mathbb{K}_{SU(2)_L}$.
Rescaling the Hamiltonian by an overall  $2 \kappa$ and choosing the basis \eqref{XYHamiltonian} we get  \eqref{eq:HXY}.

\mparagraph{XZ sector}

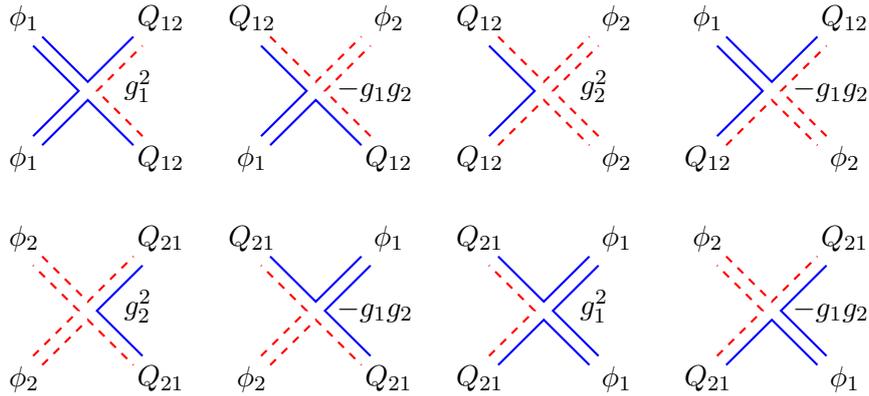
\begin{figure}[ht]
\begin{center}
\begin{tikzpicture}[scale=0.6]

  \draw[-,thick,blue] (-0.2,0)--(0.8,1)--(-0.2,2);\draw[-,thick,blue] (0,-0.2) -- (1,0.8)--(2,-0.2);
  \draw[-,thick,red,dashed] (2.2,0)--(1.2,1)--(2.2,2);\draw[-,thick,blue] (0,2.2) -- (1,1.2)--(2,2.2);
  \node at (-0.4,-0.5) {$\phi_{1}$};\node at (2.6,-0.5) {$Q_{12}$}; \node at (-0.4,2.6) {$\phi_{1}$}; \node at (2.6,2.6) {$Q_{12}$};
  \node at (2.1,1.1) {$g_1^2$};

  \draw[-,thick,blue] (5-0.2,0)--(5+0.8,1)--(5-0.2,2);\draw[-,thick,blue] (5+0,-0.2) -- (5+1,0.8)--(5+2,-0.2);
  \draw[-,thick,red,dashed] (5+2.2,0)--(5+1.2,1)--(5+2.2,2);\draw[-,thick,red,dashed] (5+0,2.2) -- (5+1,1.2)--(5+2,2.2);
  \node at (5-0.4,-0.5) {$\phi_{1}$};\node at (5+2.6,-0.5) {$Q_{12}$}; \node at (5-0.4,2.6) {$Q_{12}$}; \node at (5+2.6,2.6) {$\phi_{2}$};
  \node at (5+2.3,1.0) {$-g_1g_2$};

    \draw[-,thick,blue] (10-0.2,0)--(10+0.8,1)--(10-0.2,2);\draw[-,thick,red,dashed] (10+0,-0.2) -- (10+1,0.8)--(10+2,-0.2);
  \draw[-,thick,red,dashed] (10+2.2,0)--(10+1.2,1)--(10+2.2,2);\draw[-,thick,red,dashed] (10+0,2.2) -- (10+1,1.2)--(10+2,2.2);
  \node at (10-0.4,-0.5) {$Q_{12}$};\node at (10+2.6,-0.5) {$\phi_{2}$}; \node at (10-0.4,2.6) {$Q_{12}$}; \node at (10+2.6,2.6) {$\phi_{2}$};
  \node at (10+2.1,1.1) {$g_2^2$};

      \draw[-,thick,blue] (15-0.2,0)--(15+0.8,1)--(15-0.2,2);\draw[-,thick,red,dashed] (15+0,-0.2) -- (15+1,0.8)--(15+2,-0.2);
  \draw[-,thick,red,dashed] (15+2.2,0)--(15+1.2,1)--(15+2.2,2);\draw[-,thick,blue] (15+0,2.2) -- (15+1,1.2)--(15+2,2.2);
  \node at (15-0.4,-0.5) {$Q_{12}$};\node at (15+2.6,-0.5) {$\phi_2$}; \node at (15-0.4,2.6) {$\phi_1$}; \node at (15+2.6,2.6) {$Q_{12}$};
  \node at (15+2.3,1.0) {$-g_1g_2$};

\end{tikzpicture}
\end{center}

\begin{center}
\begin{tikzpicture}[scale=0.6]

  \draw[-,thick,red,dashed] (-0.2,0)--(0.8,1)--(-0.2,2);\draw[-,thick,red,dashed] (0,-0.2) -- (1,0.8)--(2,-0.2);
  \draw[-,thick,blue] (2.2,0)--(1.2,1)--(2.2,2);\draw[-,thick,red,dashed] (0,2.2) -- (1,1.2)--(2,2.2);
  \node at (-0.4,-0.5) {$\phi_{2}$};\node at (2.6,-0.5) {$Q_{21}$}; \node at (-0.4,2.6) {$\phi_{2}$}; \node at (2.6,2.6) {$Q_{21}$};
  \node at (2.1,1.1) {$g_2^2$};

  \draw[-,thick,red,dashed] (5-0.2,0)--(5+0.8,1)--(5-0.2,2);\draw[-,thick,red,dashed] (5+0,-0.2) -- (5+1,0.8)--(5+2,-0.2);
  \draw[-,thick,blue] (5+2.2,0)--(5+1.2,1)--(5+2.2,2);\draw[-,thick,blue] (5+0,2.2) -- (5+1,1.2)--(5+2,2.2);
  \node at (5-0.4,-0.5) {$\phi_{2}$};\node at (5+2.6,-0.5) {$Q_{21}$}; \node at (5-0.4,2.6) {$Q_{21}$}; \node at (5+2.6,2.6) {$\phi_{1}$};
  \node at (5+2.3,1.0) {$-g_1g_2$};

    \draw[-,thick,red,dashed] (10-0.2,0)--(10+0.8,1)--(10-0.2,2);\draw[-,thick,blue] (10+0,-0.2) -- (10+1,0.8)--(10+2,-0.2);
  \draw[-,thick,blue] (10+2.2,0)--(10+1.2,1)--(10+2.2,2);\draw[-,thick,blue] (10+0,2.2) -- (10+1,1.2)--(10+2,2.2);
  \node at (10-0.4,-0.5) {$Q_{21}$};\node at (10+2.6,-0.5) {$\phi_{1}$}; \node at (10-0.4,2.6) {$Q_{21}$}; \node at (10+2.6,2.6) {$\phi_{1}$};
  \node at (10+2.1,1.1) {$g_1^2$};

      \draw[-,thick,red,dashed] (15-0.2,0)--(15+0.8,1)--(15-0.2,2);\draw[-,thick,blue] (15+0,-0.2) -- (15+1,0.8)--(15+2,-0.2);
  \draw[-,thick,blue] (15+2.2,0)--(15+1.2,1)--(15+2.2,2);\draw[-,thick,red,dashed] (15+0,2.2) -- (15+1,1.2)--(15+2,2.2);
  \node at (15-0.4,-0.5) {$Q_{21}$};\node at (15+2.6,-0.5) {$\phi_1$}; \node at (15-0.4,2.6) {$\phi_2$}; \node at (15+2.6,2.6) {$Q_{21}$};
  \node at (15+2.3,1.0) {$-g_1g_2$};

\end{tikzpicture}
\end{center}
\caption{\it The vertices contributing to the Hamiltonian in the $XZ$ sector. A solid blue line denotes the first and a dashed red line denotes the second gauge group. Time moves upwards. Here we have already performed the Wick contractions of the conjugate fields with the second gauge invariant operator to write the vertices directly as spin chain interactions.} \label{fig:XZvertices}
\end{figure}

In this sector we will consider operators composed of the bifundamental field $X$ and the adjoint field $Z$. To find the Hamiltonian we will need the $\tQ_{ij}$ F-terms:
\be
F_{\tQ_{12}}=i(g_2\phi_2Q_{21}-g_1Q_{21}\phi_1) \;,\;\;F_{\tQ_{21}}=i (g_1\phi_1Q_{12}-g_2 Q_{12}\phi_2)
\ee
These lead to the interactions shown in Figure \ref{fig:XZvertices}.

We will again divide by an overall factor of $g_1g_2$, resulting in the Hamiltonian:
\be \label{XZHamiltonian}
H_{i,i+1}=\begin{pmatrix}0&0&0&0&0&0&0&0\cr 0&\kappa&-1&0&0&0&0&0\cr 0&
 -1&\kappa^{-1}&0&0&0&0&0\cr 0&0&0&0&0&0&0&0\cr 0&0&0&0&0&0&0&0\cr 0&0&0&0
 &0&\kappa^{-1}&-1&0\cr 0&0&0&0&0&-1&\kappa&0\cr 0&0&0&0&0&0&0
 &0\cr \end{pmatrix}\;,\qquad \text{in the basis} \quad \left(\begin{array}{c}Q_{12}Q_{21}\\Q_{12}\phi_2\\\phi_1 Q_{12}\\\phi_1\phi_1\\Q_{21} Q_{12}\\Q_{21}\phi_1\\\phi_2 Q_{21}\\\phi_2\phi_2\end{array}\right)\;.
\ee
As before, the upper left block contains the interactions with the first gauge group on the left, while the lower right block the opposite. The state space is again truncated as some combinations of fields (e.g. $\phi_1\phi_2$) cannot occur due to the gauge index structure. The Hamiltonian (\ref{XZHamiltonian}) can of course also be reproduced from the more general scalar Hamiltonian in \cite{Gadde:2010zi}.

We will again prefer to look at the Hamiltonian for the $\Ncal=4$ $X$ and $Z$ fields rather than their $\Ncal=2$ component fields. So for instance a state like $\ket{XXZXZZ\cdots}$ will correspond to two states $\ket{Q_{12}Q_{21}\phi_1Q_{12}\phi_2\phi_2\cdots}$ and its $\Zset_2$ conjugate $\ket{Q_{21}Q_{12}\phi_2Q_{21}\phi_1\phi_1\cdots}$ depending on the gauge group at a given reference site. To act with the Hamiltonian on the $X,Z$ basis, we again need to specify whether the gauge group is $1$ or $2$ to the left of the first site where the Hamiltonian acts. The major difference from the $XY$ sector is that the gauge group does not change on crossing a $Z$ field. We write:
\begin{equation}
\label{eq:HXZ}
\Hcal_1
=
\begin{pmatrix}
0  & 0  & 0 & 0 \\
0  &  \kappa &  -1&  0 \\
 0 & -1  &  \kappa^{-1}  & 0  \\
0  & 0  & 0  & 0
\end{pmatrix}
\, , \quad
\Hcal_2
=
\begin{pmatrix}
0  & 0  & 0 & 0 \\
0  &  \kappa^{-1} &  -1 &  0 \\
 0 & -1   &  \kappa  & 0  \\
0  & 0  & 0  & 0
\end{pmatrix}
\;, \quad \text{in the basis} \left(\begin{array}{c}XX\\XZ\\ZX\\ZZ\\ \end{array}\right)_{i=1,2} \, ,
\end{equation}
where the notation is that $\mathcal{H}_i$ acts on the basis in the gauge group representation $\Box_i \times \overline{\Box}_j$ where $j$ is not correlated to $i$ as above and can take both 1,2 values.
More explicitly, $\Hcal_1$ is the Hamiltonian acting on two sites where the gauge group to the left of the first site
is the first one, while $\Hcal_2$ acts when the gauge group to the left is the second one. Unlike the $XY$ sector, where each Hamiltonian was of Heisenberg type, in the $XZ$ sector the Hamiltonians are of Temperley-Lieb type. This immediately brings to mind the XXZ model whose quantum-group invariant Hamiltonian (obtained by adding an appropriate boundary term to the open chain) is of Temperley-Lieb type. However, unlike the XXZ case, this Hamiltonian changes dynamically along the chain, since $\Hcal_1$ is exchanged with $\Hcal_2$ (and vice versa) every time one crosses an $X$ field.  We will see in section \ref{sec:XZsector} that $Z$ excitations around the vacuum formed by the $X$ fields behave very similarly to those of the alternating $XY$ sector. On the other hand, as the Hamiltonian does not change when crossing a $Z$ field, $X$ excitations around the $Z$ vacuum can be expected to behave similarly to those of the XXZ model. This is precisely what was found in the study of this case in \cite{Gadde:2010zi}, where the $S$-matrix for two-$X$ magnon scattering in the $Z$ vacuum was found to be of $XXZ$ type. More appropriately, using the language of the current paper, the  $S$-matrix for two-$X$ magnon scattering in the $Z$ vacuum
was found to be a dynamical XXZ $S$-matrix. 

There is of course a third $\SU(2)$-like sector formed by the $Y$ and $Z$ fields but it is equivalent to the $XZ$ sector by exchanging the $Q_{ij}$ fields with the $\tQ_{ij}$ ones. So we will not consider this sector separately.

\section{Quasi-Hopf algebra and the quantum plane limit} \label{sec:QuasiHopf}

In the previous section we described how marginally deforming the $\Zset_2$ orbifold theory by allowing $\kappa=g_2/g_1$ to be different from 1 led to a deformation of the integrable model that arises at the orbifold point. The $\Ncal=1$ marginal deformations of $\Ncal=4$ SYM are known to be related to quantum group deformations of the global symmetry group of the $\Ncal=4$ theory. The best known example is the $\beta$-deformation, which is related to a $q$-deformation of the $\SU(4)\simeq \SO(6)$ symmetry group, where $q=e^{i\beta}$  \cite{Roiban:2003dw,Berenstein:2004ys}. For real $\beta$ the deformation has a quasitriangular Hopf algebra structure and in this case one can go beyond just the symmetries and show that all the integrable structures of the $\Ncal=4$ theory are preserved (see e.g. \cite{Zoubos:2010kh} for a review and further references). The complex-$\beta$ case, as well as the more general Leigh-Strassler deformations \cite{Leigh:1995ep}, go beyond the class of quasitriangular Hopf algebras, and it was recently shown that they are instead described by a quasitriangular \emph{quasi-Hopf} algebra \cite{Dlamini:2019zuk} which deforms the $\SU(3)$ part of the global symmetry group. 

Quasi-Hopf algebras were introduced by Drinfeld \cite{Drinfeld90} as the most general class of algebras which are closed under arbitrary invertible twists. In the quasitriangular Hopf algebra setting (meaning that there exists an $R$-matrix which relates the coproduct to the permuted coproduct of the theory, and which satisfies the Yang-Baxter equation), a Drinfeld twist is an element of $\Acal\otimes \Acal$ which takes an original $R$-matrix to a new one according to
\be
R_{12}^F=F_{21}R_{12}F_{12}^{-1}
\ee
If the twist satisfies a certain cocycle condition, then the twisted $R$-matrix $R^F$ also satisfies the YBE and the resulting algebra (arising from the RTT relations \cite{FRT90}) is also a quasitriangular Hopf algebra. Dropping the cocycle condition on the twist, one obtains an $R$-matrix which does not satisfy the YBE, but rather a generalisation known as the quasi-Hopf YBE \cite{Drinfeld90}. This involves an object known as a \emph{coassociator} which lives on three copies of the algebra and is defined by the twist through the relation
\be \label{coassociator}
\Phi=F_{23} (\mathrm{id}\otimes \Delta) (F)(\Delta\otimes \mathrm{id}) (F^{-1}) F^{-1}_{12} \;,
\ee
and the quasi-Hopf version of the YBE reads \cite{Drinfeld90}
\be \label{qhYBE}
R_{12}\Phi_{312}R_{13}\Phi^{-1}_{132} R_{23}\Phi_{123}=\Phi_{321}R_{23}\Phi^{-1}_{231}R_{13}\Phi_{213}R_{12}\;.
\ee 
In the Hopf setting $\Phi=I\otimes I \otimes I$ and the quasi-Hopf YBE reduces to the standard YBE.

\subsection{$R$-matrix and Drinfeld twist in the quantum plane limit} \label{sec:QPlanes}

A practical way to search for a quantum-group deformation of the symmetries of our theory is through considering the corresponding quantum plane limit.
In the gauge theory the quantum plane arises through the $F$-term relations \cite{Berenstein:2000ux,Mansson:2008xv,Dlamini:2019zuk}. As discussed in \cite{Mansson:2008xv} this is an $\SU(3)$-type quantum plane, of the type studied in \cite{EwenOgievetsky94}, and we will follow a similar approach below. For $\Ncal=4$ SYM, setting the $F_Z$ term to zero one simply obtains the relation $XY=YX$, i.e.  the classical, commutative plane, while for the $\beta$-deformation this is modified to $XY=qYX$ and can be attributed to the transverse coordinates $x^i$ seen by open strings ending on the D3-branes becoming noncommutative.\footnote{This is a statement about the transverse coordinates themselves and is independent of the fact that they also become $N\times N$ matrices when one has multiple branes.} 
The quadratic relations on the coordinates of the quantum plane can be derived via an appropriate $R$-matrix as
\be \label{qplane}
R^{i\;j}_{\;k\;l} X^k X^l=X^j X^i
\ee
In \cite{Dlamini:2019zuk} the $R$-matrix leading to the full quantum plane geometry of the Leigh-Strassler theories was studied and shown to arise from a Drinfeld twist which does not satisfy the cocycle equation. Therefore the quasi-Hopf setting is the appropriate one to understand the symmetries of the Leigh-Strassler marginal deformation of the $\Ncal=4$ theory. 

It is thus natural to ask whether one can find a similar quantum-group structure for the $\Ncal=2$ theories we are considering in the present work. To answer this, let us start by writing the superpotential (\ref{Winterpolating}) in a way that will make it easier to read off the quantum plane structure. As in \cite{Dlamini:2019zuk}, to see this structure one first needs to write out the cyclically related terms in the gauge theory traces (i.e. open up the gauge indices temporarily). Then we will collect the terms which have the first gauge group to the left:
\be
g_1\left(\phi_1Q_{12}\Qt_{21}+Q_{12}\Qt_{21}\phi_1-\phi_1\Qt_{12}Q_{21}-\Qt_{12}Q_{21}\phi_1\right)+g_2\left(\Qt_{12}\phi_2Q_{21}  -    Q_{12}\phi_{2}\Qt_{21}  \right)
\,.
\ee
Working in an $\SU(3)$ basis $X^1=X,X^2=Y,X^3=Z$, this allows us to define the deformation of the $\SU(3)$-invariant symbol $\epsilon_{ijk}$ as 
\be
E^{(1)}_{123}=g_1\;,E^{(1)}_{231}=g_2\;,E^{(1)}_{312}=g_1\;,E^{(1)}_{132}=-g_2\;,E^{(1)}_{321}=-g_1\;,E^{(1)}_{213}=-g_1\;,
\ee
Here the subscript $(1)$ indicates that the first gauge index in the corresponding terms is that of the first gauge group. Similarly one can define for the second gauge group:
\be
E^{(2)}_{123}=g_2\;,E^{(2)}_{231}=g_1\;,E^{(2)}_{312}=g_2\;,E^{(2)}_{132}=-g_1\;,E^{(2)}_{321}=-g_2\;,E^{(2)}_{213}=-g_2\;,
\ee
Using these symbols, we can write the superpotential as
\be
\Wcal=E^{(1)}_{ijk} \Tr_1\left(X^{i} X^{j} X^{k}\right)+E^{(2)}_{ijk} \Tr_2\left(X^{i} X^{j} X^{k}\right)
\ee
In this expression, the gauge index on the left is denoted by the subscript $n=1,2$\footnote{The index  $n=1,2$ can also be understood as counting the images of quantum plane due to the orbifold.} and this uniquely fixes the indices of the remaining terms (i.e. whether $X^1$ is $Q_{12}$ or $Q_{21}$). We see that as far as the quantum plane is concerned the superpotential has split into two terms, which is of course related to the block structure of our Hamiltonians in each sector. To complete the picture, one needs to also define the conjugates $F^{ijk}$ which are related to the conjugate superpotential $\overline{\Wcal}=\overline{X}_i\overline{X}_j\overline{X}_kF^{ijk}$. Since all couplings are taken to be real, we simply take $F^{ijk}_{(n)}=E_{ijk}^{(n)}$.

From these tensors we can now reconstruct the Hamiltonian of each sector \eqref{eq:HXY}  and \eqref{eq:HXZ} using the relation
\be \label{HamEF}
\left(\Hcal_n\right)^{ij}_{kl}=E^{(n)}_{klr}F^{ijr}_{(n)}  \, .
\ee
The reason that the Hamiltonian can be directly read off from the superpotential is that at one loop there is a non-renormalisation theorem at work and only the F-terms contribute. For instance, for the $XY$ sector (free indices $1,2$), we obtain
\be
\begin{split}
\left(\Hcal_1\right)^{12}_{\;12}&=E^{(1)}_{123}F^{123}_{(1)}=g_1^2\;,\left(\Hcal_1\right)^{21}_{\;12}=E^{(1)}_{123}F^{213}_{(1)}=-g_1^2\;,\\
\left(\Hcal_1\right)^{12}_{\;21}&=E^{(1)}_{213}F^{123}_{(1)}= - g_1^2\;,\left(\Hcal_1\right)^{21}_{\;21}=E^{(1)}_{213}F^{213}_{(1)}=g_1^2\;,
\end{split}
\ee
 while for the $XZ$ sector (free indices $1,3$):
\be
\begin{split}
\left(\Hcal_1\right)^{13}_{\;13}&=E^{(1)}_{132}F^{132}_{(1)}=g_2^2\;,\left(\Hcal_1\right)^{31}_{\;13}=E^{(1)}_{132}F^{312}_{(1)}=-g_1g_2\;,\\
\left(\Hcal_1\right)^{13}_{\;31}&=E^{(1)}_{312}F^{132}_{(1)}=-g_1g_2\;,\left(\Hcal_1\right)^{31}_{\;31}=E^{(1)}_{312}F^{312}_{(1)}=g_1^2\;,
\end{split}
\ee
and similarly for the $YZ$ sector. These reproduce the upper blocks of our Hamiltonians (\ref{XYHamiltonian}) and (\ref{XZHamiltonian}), respectively. Clearly the lower blocks $\Hcal_2$ arise in the same way using (\ref{HamEF}) for the second gauge index. 

Let us now consider the quantum plane structure of the theory which arises from the $F$-term relations. For the $XY$ sector these are 
\be
g_1Q_{12}\Qt_{21}=g_1\Qt_{12}Q_{21}\;\;,\quad g_2 Q_{21}\Qt_{12}=g_2\Qt_{21}Q_{12} \;,
\ee
while for the $XZ$ sector we have 
\be \label{XZqplane}
\phi_2 Q_{21}=\frac{1}{\kappa} Q_{21}\phi_1\;\;,\quad \phi_1Q_{12}=\kappa Q_{12}\phi_2\;,
\ee
and similarly for the  $YZ$ sector:
\be
\phi_2\Qt_{21}=\frac{1}{\kappa}\Qt_{21}\phi_1\;\;,\quad \phi_1\Qt_{12}=\kappa\Qt_{12}\phi_2\;.
\ee
We see that the $\SU(2)$ symmetry\footnote{
Note that this undeformed and unbroken  $\SU(2)$ symmetry acting on the $XY$ sector is precisely the extra  $\SU(2)_L$ global symmetry of the $\mathbb{Z}_2$ quiver we described in Section
\ref{subsec:QuiverTheory}.
} is undeformed in the $XY$ sector (as the factors of $g_1 g_2$ can be factored out), so the $R$-matrix in this sector is equivalent to the identity matrix $I$ (its classical value) in the quantum plane limit. 
However in the other two $XY$ and $YZ$ sectors the $\SU(2)$ symmetry of the $\Ncal=4$ theory (and of the orbifold point) becomes deformed as $\kappa$ moves away from its classical value of 1. 

To find the $R$-matrices corresponding to these quantum planes we will use the relation \cite{EwenOgievetsky94,Mansson:2008xv}
\be \label{qplaneR}
\Rhat^{i j}_{k l}  = \delta^i_k \delta^j_\ell - c \, E_{kl m}  F^{ijm}\;,
\ee
where  $c$ is a constant which in principle is arbitrary, as any choice will lead to the appropriate quantum plane relations. In this way we will obtain two $9\times 9$ $R$-matrices (one for each gauge group) which reproduce the full set of quantum plane relations above. However in the following, we will focus on each $\SU(2)$ subsector separately, to highlight the different features that they exhibit. It will also be convenient to rescale $E_{ijk}$ and $F^{ijk}$ in the following way, which leaves the Hamiltonians invariant:
\be
\begin{split}
E^{(1)}_{123}&=1\;,E^{(1)}_{231}=\kappa\;,E^{(1)}_{312}=1\;,E^{(1)}_{132}=-\kappa\;,E^{(1)}_{321}=-1\;,E^{(1)}_{213}=-1\;,\\
F_{(1)}^{123}&=\kappa^{-1}\;,F_{(1)}^{231}=1\;,F_{(1)}^{312}=\kappa^{-1}\;,F_{(1)}^{132}=-1\;,F_{(1)}^{321}=\kappa^{-1}\;,F_{(1)}^{213}=\kappa^{-1}\;, 
\end{split}
\ee
while those for the second gauge group are obtained as $\kappa\ra 1/\kappa$.

\mparagraph{XY sector}

We already stated that the $R$-matrix in this sector is the identity matrix $I$  in the quantum plane limit, however, let us explain how to see this.
Applying the definition (\ref{qplaneR}) with $c=1$, we obtain an $R$-matrix for this sector which, through the relation (\ref{qplane}) leads to the quantum deformations
\be
Q_{12}\Qt_{21}-\Qt_{12}Q_{21}\ra \frac{Q_{12}\Qt_{21}-\Qt_{12}Q_{21}}\kappa \;,\;\text{and}\;\; Q_{21}\Qt_{12}-\Qt_{21}Q_{12}\ra \kappa ( Q_{21}\Qt_{12}-\Qt_{21}Q_{12})\;.
\ee
as required. Note that this $R$-matrix is not the identity matrix, and it is not \emph{triangular}, i.e. it does not satisfy the condition $R_{21}=R_{12}^{-1}$. To obtain a triangular matrix, we need to choose $c=\kappa$ for the upper block, and $c=1/\kappa$ for the lower block, which reduces the $R$-matrix in each block to just the identity, and the quantum plane symmetry in this sector is simply the Lie algebra of $\SU(2)$.\footnote{Note that, as also discussed in \cite{Mansson:2008xv}, the choice of $c$ does not affect the $RTT$ relations defining the quantum algebra. } This is not surprising, as the orbifold treats the $X$ and $Y$ fields symmetrically. In order to obtain precisely the deformation of the superpotential above, one needs to rescale the $\epsilon_{ij}$ tensor of $\SU(2)$ by $1/\kappa$ or $\kappa$ depending on the gauge group.

\mparagraph{XZ sector}

The $R$-matrix in this sector can be obtained via (\ref{qplaneR})
with triangularity requiring the choice $ c = \frac{2 \kappa}{1+ \kappa^2}$. Explicitly, the $R$-matrix $R=P\Rhat$ reads
\be \label{Rmatrixkappa}
R
 =
 \begin{pmatrix}1&0&0&0&0&0&0&0\cr 0&{{2\,\kappa}\over{\kappa^2+1}}&-{{
 \kappa^2-1}\over{\kappa^2+1}}&0&0&0&0&0\cr 0&{{\kappa^2-1}\over{
 \kappa^2+1}}&{{2\,\kappa}\over{\kappa^2+1}}&0&0&0&0&0\cr 0&0&0&1&0&0
 &0&0\cr 0&0&0&0&1&0&0&0\cr 0&0&0&0&0&{{2\,\kappa}\over{\kappa^2+1}}&
 {{\kappa^2-1}\over{\kappa^2+1}}&0\cr 0&0&0&0&0&-{{\kappa^2-1}\over{
 \kappa^2+1}}&{{2\,\kappa}\over{\kappa^2+1}}&0\cr 0&0&0&0&0&0&0&1\cr 
 \end{pmatrix}
 =
 \begin{pmatrix}1&0&0&0&0&0&0&0\cr 0&{k}&k'&0&0&0&0&0
\cr 0&{-k'}&k&0&0&0&0&0\cr 0&0&0&1&0&0
 &0&0
 \cr 0&0&0&0&1&0&0&0
 \cr 0&0&0&0&0&k&
-k'&0\cr 0&0&0&0&0&k'&k&0\cr 0&0&0&0&0&0&0&1\cr 
 \end{pmatrix}
\ee
written in the same basis as (\ref{XZHamiltonian}), which we identify with the two-site basis $x^ix^j$ in (\ref{qplane}).
This $R$-matrix does not satisfy the usual YBE, apart from the special cases $\kappa=\pm1,0$. The $R$-matrix \eqref{Rmatrixkappa} can be seen to reduce to $I\oplus I$ in the $\kappa\ra 1$ limit, as expected.

It is very intriguing to observe that \eqref{Rmatrixkappa} is the direct sum of two Felder $\SU(2)$ dynamical $R$-matrices, as presented in (\ref{Rsym}),  for the special parameter values $\lambda=(1+\tau)/2$, $\eta=\pm\tau/4$ and with the rapidity fixed to $u=\half$. The choices  $\eta=\pm \tau/4$ are for the upper and lower block respectively. For these values, the coefficients of the $R$-matrix (\ref{Rsym}), normalised by dividing out by $\gamma$, reduce to the elliptic modulus $k$ and its complement $k'=\sqrt{1-k}$ related to the modular parameter $\tilde{m}=4\kappa^2/(1+\kappa^2)^2$ which, as we will see in Section \ref{sec:Elliptic}, can be read off from the 1-magnon dispersion relation. Note that the value $u=\half$ makes sense as the location of the quantum plane limit on the rapidity torus, being equal to one of the half-periods of the elliptic functions describing it. For the careful reader we note that for rational models, the quantum plane arises at $u\ra\infty$, however that is because the torus has been decompactified by sending the periods to infinity.

 Being triangular, the $R$-matrix can also be factorised as
\be
R=F_{21}F_{12}^{-1}=(F_{12})^{-2}\;.
\ee
with the twist being given by
\be \label{Ftwist}
F=\left(
\begin{array}{cccccccc}
 1 & 0 & 0 & 0 & 0 & 0 & 0 & 0 \\
 0 & \alpha & \beta & 0 & 0 & 0 & 0 & 0 \\
 0 & -\beta & \alpha & 0 & 0 & 0 & 0 & 0 \\
 0 & 0 & 0 & 1 & 0 & 0 & 0 & 0 \\
 0 & 0 & 0 & 0 & 1 & 0 & 0 & 0 \\
 0 & 0 & 0 & 0 & 0 & \alpha & -\beta & 0 \\
 0 & 0 & 0 & 0 & 0 & \beta & \alpha & 0 \\
 0 & 0 & 0 & 0 & 0 & 0 & 0 & 1 \\
\end{array}
\right) \,\quad \text{with:} \quad\begin{array}{c} \alpha=\frac{\kappa+1}{\sqrt{2}\sqrt{1+\kappa^2}}\;,\\
\\ \beta=\frac{\kappa-1}{\sqrt{2}\sqrt{1+\kappa^2}}\end{array}
\ee
We note that the twist is triangular, i.e. $F_{21}=F_{12}^{-1}$, is orthogonal as an $8\times 8$ matrix and has unit determinant. From this twist, taking appropriate care when working with our truncated tensor product on the three-site basis, one can obtain a coassociator with which the $R$-matrix (\ref{Rmatrixkappa}) satisfies the quasi-Hopf YBE (\ref{qhYBE}). 

Let us now consider the symmetries of the $XZ$ quantum plane, which in the undeformed theory are just an $SU(2)$ rotating $X$ with $Z$. This $\SU(2)$ is broken after the orbifold and the marginal deformation. However, from the above discussion we see that it will get uplifted to an  $SU(2)_\kappa$  quantum group defined by the RTT relations \cite{FRT90} with the $R$-matrix (\ref{Rmatrixkappa}):
\be
R^{i \, j}_{\, k\, l} \,  T^k\,_m  \,  T^l\,_n  =   T^j\,_l  \,  T^i\,_k \, R^{k \, l}_{\, m\, \, n}   \, .
\ee
Of course, given an arbitrary matrix $R$, the RTT relations by themselves are not sufficient to define a consistent quantum group. In particular, the consistency of  the higher order relations in the generators requires the YBE relation for $R$. Otherwise, requiring associativity can trivialise the relations. However, since our $R$-matrix is triangular and can be factorised via a Drinfeld twist, one can abandon associativity in a consistent way by working in a quasi-Hopf setting, as was done in \cite{Dlamini:2019zuk}. In such a setting, the action of the deformed generators on the quantum plane can be related to that of the undeformed ones using the twist, and the quantum algebra can be consistently defined (at the price of introducing a coassociator as in (\ref{coassociator}) which complicates the formalism somewhat.)

We thus see that, even though the orbifold and the marginal deformation superficially break part of the $PSU(2,2|4)$ global symmetry group of $\Ncal=4$ SYM, the broken generators are not really broken when one is willing to work in a quasi-Hopf setting. They are rather $\kappa$-deformed and should still be able to provide useful constraints on the theory. We will leave a more detailed study of the quantum group underlying the $XZ$ sector, as well as its generalisation to $SU(3)_\kappa$ and the full scalar sector, for future work.

The main complication in our setting, compared to the $\Ncal=1$ theories studied in \cite{Dlamini:2019zuk}, is the non-direct product nature of our state space, i.e. the fact that some combinations of the fields are not allowed due to the incompatibility of the gauge indices. One would thus need to work in a truncated state space which makes the construction less immediate. Work towards elucidating this structure is in progress. For the present work, we desire to avoid this issue and that is one of our motivations for introducing what we call the \emph{dynamical} notation in the next section.

Concluding this section we wish to emphasise the following points. Firstly our results in this section hold purely in the quantum plane limit.
We only obtained the $R$-matrix $R(u^* ; \kappa)$ in a special limit for the spectral parameter $u \to u^*$.
\footnote{In rational or trigonometric integrable systems, where the $R$-matrix is known as a function of the spectral parameter, one can obtain the quantum plane limit by taking the spectral parameter to infinity. In  elliptic models the dependence on the spectral parameter is such that the quantum plane limit corresponds to taking the spectral parameter to be equal one of the half-periods of the elliptic functions. Since the trigonometric and rational cases are reached by taking one or both (respectively) periods of the elliptic functions to infinity (thus decompactifying the rapidity torus in one or both directions) this leads to the statement above.}
To completely describe our model and attempt to solve it with the Algebraic Bethe ansatz approach  we need to first obtain the $R$-matrix as a function of the spectral parameter $R(u ; \kappa)$. In the conclusions we will comment on how one could go about doing that.
To fully understand the quasi-Hopf structure, it is also imperative to compute the coassociator, which will enable us to write down and check the quasi-Hopf YBE \eqref{qhYBE}.
Secondly, in order for a quantum system with quasi-Hopf symmetry to be  integrable, the Drinfeld twist $F$ needs to obey a specific cocycle condition. In particular, as reviewed in the next section,  if $F$ satisfies a shifted cocycle condition \cite{Babelon:1995rz,Enriquez_1998,Jimbo:1999zz} the corresponding $R$-matrix would satisfy the dynamical YBE. This is something we have not attempted to check in this current work. We hope to be able to report progress in this direction in a forthcoming paper.

\section{The dynamical spin chain} \label{sec:Dynamical}

Even though we do not (yet) have a spectral-parameter-dependent matrix $R(u ; \kappa)$,
we can still try to identify the main features it would need to have in order to reproduce our Hamiltonian. For this we find
it useful to introduce the notion of a dynamical parameter, which we will denote by $\lambda$. 
Apart from the fact that a dynamical description of our model will be much more elegant,  it also
plays an important role in the theory of elliptic quantum groups as introduced by Felder \cite{Felder:1994be,Felder:1994pb}. To understand the origins of this parameter, let us briefly review an important class of statistical mechanical models, the Solid-On-Solid (SOS) models \cite{Andrews:1984af}. For more details, we refer the reader to e.g. the review  \cite{Wadatietal89} or the book \cite{McCoyBook}. 

A SOS model is a statistical model defined by a set of Boltzmann face weights, depicted graphically as: 
\be
\begin{tikzpicture}[baseline={(0,1.5cm)}]
  \draw[-] (1,1)--(2,1)--(2,2)--(1,2)--(1,1);
  \node at (0.8,1){$a$};\node at(2.2,1){$b$};\node at (2.2,2){$c$};\node at (0.8,2){$d$};
  \node at(1.5,1.5){$u$};
  \node at (4.5,1.5){$=\BW{a}{b}{c}{d}{u}$};
\end{tikzpicture}
\ee
The parameter $u$ is the rapidity or spectral parameter, and $a,b,c,d$ are known as the heights. Here we are thinking of a square lattice with the weights associated to each unit face of the lattice, and their value determined by the heights associated with the four lattice points surrounding each face. Each model comes with a set of rules as to which heights are allowed to be adjacent. For instance, in the original Andrews-Baxter-Forrester (ABF) model \cite{Andrews:1984af} neighbouring heights can only differ by 1, i.e. we impose $|a-b|=1$ and similarly for $|b-c|$ etc.. To give a concrete example, let us write down the weights for this case, in symmetrised form
{\small
\be \label{Bweights}
\begin{split}
  \BW{a+1}{a+2}{a+1}{a}{u}&=\BW{a-1}{a-2}{a-1}{a}{u}=\frac{\theta_1(2\eta-u)}{\theta_1(2\eta)}\\
  \BW{a-1}{a}{a+1}{a}{u}&= \BW{a+1}{a}{a-1}{a}{u}=\frac{\sqrt{\theta_1(2\eta(a-1)+w_0)\theta_1(2\eta(a+1)+w_0)}}{\theta_1(2\eta a+w_0)}\frac{\theta_1(u)}{\theta_1(2\eta)}\\
  \BW{a+1}{a}{a+1}{a}{u}&=\frac{\theta_1(2\eta a+w_0+u)}{\theta_1(2\eta a+w_0)}\;\;,\quad
  \BW{a-1}{a}{a-1}{a}{u}=\frac{\theta_1(2\eta a+w_0-u)}{\theta_1(2\eta a+w_0)}
\end{split}
\ee
}
 We use the theta-function conventions of \cite{Thetavocabulary} which are also compatible e.g. with \cite{Gomez96}. The parameter $\eta$ is the deformation (or ``crossing'') parameter which appears in Baxter's elliptic $R$-matrix for the XYZ model, and $w_0$ is a tunable constant. Note that neighbouring heights indeed only differ by $\pm 1$. (The weights for all other possibilities are set to zero). The above weights satisfy the star-triangle relation
{\small
\be
\sum_g\BW{a}{b}{g}{f}{z-w}\BW{b}{c}{d}{g}{z}\BW{g}{d}{e}{f}{w}=\sum_g\BW{b}{c}{g}{a}{w}\BW{a}{g}{e}{f}{z}\BW{g}{c}{d}{e}{z-w}\;,
\ee
}
which can  be graphically represented as follows:
\be
\begin{tikzpicture}[baseline={(0,1.8cm)}]
\draw[-] (1,1)--(2,1)--(3,2)--(2,2)--(1,1);
\draw[-] (1,1)--(0,2)--(1,3)--(2,2);
\draw[-] (1,3)--(2,3)--(3,2);
\node at (4,2) {$=$};
\draw[-] (7,1)--(6,1)--(5,2)--(6,2)--(7,1);
\draw[-] (7,1)--(8,2)--(7,3)--(6,2);
\draw[-] (7,3)--(6,3)--(5,2);

\node at (-0.2,2) {$a$};\node at (1,0.8){$b$};\node at (2,0.8){$c$};\node at (3.2,2) {$d$};\node at (2,3.2) {$e$};\node at (1,3.2){$f$};
\node at (2.1,2.2){$g$};
\node at (4.8,2) {$a$};\node at (6,0.8){$b$};\node at (7,0.8){$c$};\node at (8.2,2) {$d$};\node at (7,3.2) {$e$};\node at (6,3.2){$f$};
\node at (6.3,2){$g$};

\node at (1,2){$z-w$};\node at (2,1.5) {$z$};\node at (2,2.5){$w$};
\node at (7,2){$z-w$};\node at (6,1.5) {$w$};\node at (6,2.5){$z$};

\end{tikzpicture}
\ee

The height $g$ is summed over all the values allowed by the $\pm1$ restriction and of course the equation is only non-trivial if the adjacent external heights also only differ by $\pm1$.

\mparagraph{The dynamical $R$-matrix}

For many applications, including to quantum groups, it is useful to have a description of the model as a vertex model with an $R$-matrix. For the ABF model, such a map was introduced by Felder \cite{Felder:1994be,Felder:1994pb}. As we are in an $\SU(2)$ setting we can define a 2-dimensional basis as
\be
e[1]=\doublet{1}{0} \;,\; e[-1]=\doublet{0}{1}
\ee
and construct an $R$-matrix $R^{ij}_{kl}$ on the product space $V\otimes V$ as
\be
\label{eq:ew}
R(u ; -2\eta d) e[c-d]\otimes e[b-c]=\sum_a \BW{a}{b}{c}{d}{u} e[b-a]\otimes e[a-d]
\, .
\ee
Graphically, we can represent this relation as 
\be
\begin{tikzpicture}[scale=0.7,baseline={(0,-0.2cm)}]
  \draw[-] (1-1,-1.4+0.5)--(1.9-1,-0.5+0.5)--(1-1,0.4+0.5)--(0.1-1,-0.5+0.5)--(1-1,-1.4+0.5);
  \draw[-] (0.13,.25-1) arc (45:135:0.2);
  \node at (0,-0.2-1){$a$};\node at (1.2,1-1){$b$};\node at (0,2.1-1){$c$};\node at (-1.2,1-1){$d$};

  \node at (2.5,0){$=$};

  \draw[->,blue,thick] (4,-1)--(6,1);
  \draw[->,blue,thick] (6,-1)--(4,1);
  \node at (4,-1.4) {$k$};
  \node at (6,-1.4) {$l$};
  \node at (4,1.4) {$i$};
  \node at (6,1.4) {$j$};

  \node at (3.5,0){$\lambda$};
\end{tikzpicture}
\ee
Note that the $R$-matrix depends on the height $d$, which is not fixed at a given site but dynamically determined by the configuration. We will call the combination $-2\eta d-w_0=\lambda$, where $\lambda$ is known as the dynamical parameter. Performing this vertex-face map, we find the $R$-matrix
\be \label{Rsym}
R(u ; \lambda)=\left(\begin{array}{cccc}
  \gamma& 0 & 0& 0\\
  0& \alpha & \beta_+ &0\\
  0&\beta_- & \alpha&0\\
  0&0&0&\gamma\end{array}\right)\;,
\ee
where
\be \label{eTLfin}
\begin{split}
  \gamma&=\frac{\theta_1(2\eta-u)}{\theta_1(2\eta)}\;,\\
  \alpha&=\frac{\sqrt{\theta_1(\lambda+2\eta)\theta_1(\lambda-2\eta)}}{\theta_1(-\lambda)}\frac{\theta_1(u)}{\theta_1(2\eta)}\;,\\
  \beta_\pm&=\frac{\theta_1(\lambda\pm u)}{\theta_1(\lambda)}\;.\\
\end{split}
\ee
This is Felder's dynamical $R$-matrix \cite{Felder:1994be,Felder:1994pb}, which has found extensive applications in the study of elliptic quantum groups.\footnote{However, as mentioned, we have performed a gauge transformation which has brought the $\alpha$ coefficients into symmetrised form \cite{Deguchi_2002}.} It should be clear from the construction that this $R$-matrix cannot satisfy the usual Yang-Baxter equation, since the weights in the star-triangle relation depended on the value of the heights, and in particular two adjacent weights will have different heights. Instead, as shown in \cite{Felder:1994be,Felder:1994pb}, this $R$-matrix satisfies the \emph{dynamical} Yang-Baxter equation :
  \be \label{dYBE1}
  \begin{split}
  R_{12}(u_1-u_2 ; \lambda+2\eta h^{(3)}) &R_{13}(u_1-u_3 ; \lambda)R_{23}(u_2-u_3 ; \lambda+2\eta h^{(1)}) 
  \\
    = \, 
  &R_{23}(u_2-u_3 ; \lambda)R_{13}(u_1-u_3 ; \lambda+2\eta h^{(2)}) R_{12}(u_1-u_2 ; \lambda)
\end{split}
 \ee
We see that the spectral parameter is shifted in consecutive $R$-matrices. The shift is proportional to $2\eta$, which is known as the \emph{step} and $h^{(i)}$ indicates the Cartan weight of each index line being crossed. In our $\SU(2)$ example these weights are $\pm 1$, depending on whether one crosses $e[1]$ or $e[2]$, so the shifts will be by $\pm 2\eta$. The dYBE can be represented graphically as follows:
\be
\begin{tikzpicture}[scale=0.8]
  \draw[->,blue,thick] (0,1)--(6,5);
  \draw[->,blue,thick] (2,0)--(2,6);
  \draw[->,blue,thick] (6,1)--(0,5);
  \node at (-0.2,0.5){$1$};\node at (2,-0.5){$2$};\node at (6.2,0.5){$3$};
  \node at (0,3) {$\scriptstyle\lambda$};\node at (0.8,5.7) {$\scriptstyle \lambda+2\eta h^{(3)}$};
  \node at (4,5) {$\scriptstyle\lambda+2\eta\!\sum\limits_{i=2,3}h^{(i)}$};\node at (1,1) {$\scriptstyle \lambda+2\eta h^{(1)}$};
  \node at (4,1) {$\scriptstyle\lambda+2\eta\sum\limits_{i=1,2}h^{(i)}$};\node at (5,3) {$\scriptstyle \lambda+2\eta \sum_ih^{(i)}$};
  \node at (2.4,3) {$\scriptstyle\lambda'$};
  
  \node at (8,3){$=$};
  \draw[->,blue,thick] (9,1)--(15,5);
  \draw[->,blue,thick] (13,0)--(13,6);
  \draw[->,blue,thick] (15,1)--(9,5);
  \node at (8.8,0.5){$1$};\node at (13,-0.5){$2$};\node at (15.4,1){$3$};
\node at (10,3) {$\scriptstyle\lambda$};\node at (12,4.7) {$\scriptstyle \lambda+2\eta h^{(3)}$};
  \node at (14.5,5.6) {$\scriptstyle\lambda+2\eta\!\sum\limits_{i=2,3}h^{(i)}$};\node at (12,1.5) {$\scriptstyle \lambda+2\eta h^{(1)}$};
  \node at (14.5,0.3) {$\scriptstyle\lambda+2\eta\sum\limits_{i=1,2}h^{(i)}$};\node at (15,3) {$\scriptstyle \lambda+2\eta \sum_ih^{(i)}$};
  \node at (12.6,3) {$\scriptstyle\lambda''$};
  
\end{tikzpicture}
\ee
where $\lambda'=\lambda+2\eta h^{(2)}$ and $\lambda''=\lambda+2\eta\sum\limits_{i=1,3}h^{(i)}$. Here we use the convention that (a) an $R$-matrix takes the value of the $\lambda$ parameter to its left (as determined by the arrows) and (b) the dynamical parameter is shifted with positive step as one crosses an index line from left to right (again as determined by the arrow). In pictures: 
\be
\begin{tikzpicture}[scale=0.6,baseline={(-0.2cm,1.4cm)}]

  \node at (-0.5,3) {$R^{i\;j}_{\;k\;l}(u ; \lambda) \;=$};
  \draw[->,blue,thick] (2,2)--(4,4);
  \draw[->,blue,thick] (4,2)--(2,4);
  \node at (2,1) {$k$};\node at (4,1) {$l$};
  \node at (2,4.5) {$i$};\node at (4,4.5){$j$};
  \node at (1.5,3) {$\lambda$};

  \node at (-3,3) {(a)};\node at (7,3){(b)};
  
  \draw[->,blue,thick] (1+8,3)--(5+8,3);
\node at (0.5+8,3) {$i$};
  \node at (3+8,3.5) {$\lambda$};
  \node at (3+8,2) {$\lambda+2\eta h^{i}$};

\end{tikzpicture}
\ee
So in the dynamical YBE, on the left $R_{12}$ and $R_{23}$ ``see'' the dynamical parameter $\lambda$ while $R_{13}$ sees $\lambda+2\eta h^{(2)}$, while on the right $R_{13}$ sees $\lambda$ while $R_{23}$,$R_{12}$ see $\lambda+2\eta h^{(1)}$ and $\lambda+2\eta h^{(3)}$, respectively. This explains the shifts appearing in (\ref{dYBE1}).

A very similar notation was also used recently in \cite{Yagi:2017hmj} following earlier work \cite{Yagi:2015lha,Maruyoshi:2016caf}. 
In these works, features of elliptic models, such as the transfer matrices, are mapped to surface defects in $\Ncal=1$ quiver theories, which, conversely, provide a construction of elliptic integrable models via branes. The setting of  \cite{Yagi:2017hmj,Yagi:2015lha,Maruyoshi:2016caf}  is  very different from ours. They study
 supersymmetric indices (which are partition functions on $\mathbb{S}^3 \times \mathbb{S}^1$ and count BPS objects) in the presence of  surface defects.
Our study aims to compute anomalous dimensions of non-BPS single-trace local operators in the planar limit.
Nonetheless, these works add to the conclusion that elliptic models have an important role to play in supersymmetric gauge theory.

Although clearly the dYBE does not define a completely new structure, as it is implied by the star-triangle relations of the corresponding SOS models, it brings with it several advantages, in particular if one is interested in more algebraic aspects. As mentioned, it allows for the definition of elliptic quantum groups. Also, and importantly for our purposes, it has been shown to be a special case of the quasi-Hopf YBE (\ref{qhYBE}), which arises when the twist from an initial Hopf algebra satisfies a so-called shifted cocycle relation \cite{Babelon:1995rz,Jimbo:1999zz}.
In this case the coassociator simplifies considerably as compared to the general case.
As already discussed in Section \ref{sec:QPlanes}, our system has a quasi-Hopf symmetry in the quantum plane limit, and it will be important to establish (possibly by a quasi-Hopf version of Baxterisation \cite{Jones1990}) whether this quasi-Hopf symmetry also holds away from that limit, i.e. in the presence of a spectral parameter. Although that would be important by itself, for practical applications it would be very desirable to bring the model into the dynamical framework above.

Another motivation for Felder's $R$-matrix (\ref{Rsym}) is that it is equivalent to Baxter's elliptic $R$-matrix for the 8-vertex model/XYZ model \cite{BaxterBook}, however it is of 6-vertex form, a feature typically associated with trigonometric models such as XXZ. Unlike the 8-vertex form, this allows for the definition of highest-weight states, which is a prerequisite for the application of the Algebraic Bethe Ansatz. Recall that Baxter, in his solution of the 8-vertex model, tackled this problem via a famous site-dependent change of basis, which locally brought the $R$-matrix into 6-vertex form. In the ABA context, this was used in \cite{Takhtajan81} for the diagonalisation of the XYZ Heisenberg chain. One can think of Felder's matrix as already being in the 6-vertex basis, making the application of the ABA straightforward \cite{Felder:1996xym,Deguchi_2002}, with the main difference to the usual case being the need to keep track of the dynamical parameter in the transfer matrix. 

\begin{figure}
\begin{center}
\begin{tikzpicture}[scale=0.5]
  \draw[-,blue,thick] (0,1)--(6,1);
  \filldraw[red,thick] (0,1) circle (5pt);  \filldraw[red,thick] (3,1) circle (5pt); \filldraw[red,thick] (6,1) circle (5pt);
\node at (0,0.3) {$1$};\node at(3,0.3) {$2$};\node at(6,0.3) {$3$};  
  \draw[-,blue,thick] (10,0)--(14,0)--(12,2)--(10,0);
  \filldraw[red,thick] (10,0) circle (5pt);  \filldraw[red,thick] (14,0) circle (5pt); \filldraw[red,thick] (12,2) circle (5pt);
\node at (9.5,-0.5) {$1$};\node at(14.5,-0.5) {$2$};\node at(12,2.7) {$3$};  
\end{tikzpicture}
\caption{\it Adjacency graphs of the $A_3$  (left) and $A_2^{(1)}$ models (right). The heights are labelled by 1,2,3. The graph of $A_3$ leads to the Ising model, as the $|a-b|=1$ rule means that the lattice contains a fixed (``frozen'') sublattice consisting of only the height 2, leaving only heights 1 and 3 as dynamical variables (the ``up'' and ``down'' spins). The $A_2^{(1)}$ diagram forms the basis for the adjacency graphs which are relevant to the dilute vertex models we are interested in.} \label{fig:Adjacency}
\end{center}
\end{figure}
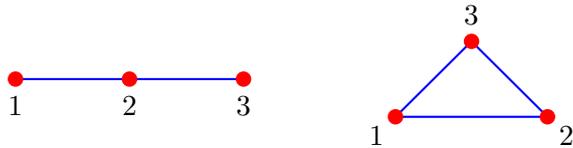

Let us now turn to generalisations of the ABF model. In the unrestricted case, the heights are allowed to span all integer values. However, it is interesting to also consider restricted situations (RSOS models), where the heights take values in a specific range. This can be achieved starting from (\ref{Bweights}) by tuning $\eta$ and $w_0$ so that the Boltzmann weights which connect heights inside the chosen interval to heights outside it vanish. It is useful to picture such a situation in term of an \emph{adjacency/incidence} graph, where one draws the allowed heights along with bonds indicating which heights can be adjacent to each other (in the ABF case, the $|a-b|=1$ rule requires them to be consecutive integers). Furthermore, one can also have cyclic (CSOS) models \cite{Pearce:1988en,Kuniba:1987yr,Pearce:1989ek} where, instead of vanishing for heights outside the interval, the weights are periodic, so that we can make an identification such as $a_{max}+1=a_{min}$. In this case the heights live on a circle rather than an interval. See Figure \ref{fig:Adjacency} for examples of RSOS and CSOS adjacency graphs for the simple case where there are only three heights.

 As shown in \cite{Pasquier:1986jc}, any classical or affine ADE Dynkin diagram provides a suitable adjacency/incidence graph for a critical RSOS model. See also e.g. \cite{DiFrancesco:1989ha,Roche90,Fendley:1989vt} for generalisations, including to adjacency graphs obtained by orbifolding of an initial graph. In the approach to $A_{n-1}^{(1)}$ RSOS models of \cite{JimboMiwaOkado87,Jimbo:1987mu}, the adjacency graph is given not by the Dynkin diagram itself, but by the set of dominant weights at level $k$ (the ``Weyl alcove'' at level $k$) of the affine algebra. So for fixed rank of the affine algebra, one can construct arbitrarily large adjacency graphs labelled by $k$. These can also be converted to CSOS-type models by suitably tuning the $w_0$ parameter to make the weights periodic \cite{Kuniba89}. We refer to the original literature, as well as to \cite{Dupic:2016ick,Morin-Duchesne:2018apx} for the details of the construction.

In the dynamical $R$-matrix picture, $\lambda$ is not restricted to be integer, and neither is the step $2\eta$. If $2\eta$ is generic we have the analogue of the SOS case. However, one can take the step to be an elliptic root of unity, $2N\eta=n_1+n_2\tau$, where we recall that the theta functions are antiperiodic under shifts of the argument by $1$ and quasiperiodic under shifts by $\tau$. Depending on the original value of $\lambda$, one can exploit the zeroes and periodicity of the theta functions to ensure that shifting $\lambda$ a certain number of times will either give zero weights (RSOS case) or give us back the same $R$-matrix (CSOS case).

It should be noted that choosing root-of-unity values for $2\eta$ leads to special features in the Bethe ansatz, similar to those discussed in \cite{Fabricius:2000yx} for the eight-vertex model. In the dynamical framework, these features, in particular degenerate eigenvectors and roots forming complete $N$-strings, are discussed in  \cite{Deguchi_2002}.

\mparagraph{Dilute RSOS models}

Let us now turn to a special class of $RSOS$ models, the \emph{dilute RSOS models} \cite{Nienhuis90,Kostov:1991cg,Warnaar:1992gj,Roche:1992ww,Warnaar:1993zn,Warnaar:1995law,BehrendPearce97}. These are models where adjacent heights are allowed to also take equal values, i.e. one modifies the adjacency condition to $|a-b|\leq 1$ (and similarly for $(b,c)$ etc). One can consider this as allowing paths from a given node that loop back to the same node (although this is not often indicated in the notation). This means that on top of the weights of ABF type (\ref{Bweights}) there will now exist Boltzmann weights of type
{\small
  \be
  \begin{split}
    &\BW{d\pm1}{d\pm1}{d\pm1}{d}{u}\;, \BW{d}{d\pm1}{d}{d}{u}\;, \BW{d}{d}{d\pm1}{d}{u}\;,\BW{d\pm1}{d}{d}{d}{u}\;,\\
    &\BW{d\pm1}{d\pm1}{d}{d}{u}\;, \BW{d}{d\pm1}{d\pm1}{d}{u}\;,\BW{d}{d}{d}{d}{u}
    \end{split}
 \ee
}
The name ``dilute'' comes through the link to loop models: Following \cite{Roche:1992ww}, one can represent the ABF Boltzmann weights in terms of the following two tiles:
\be
\begin{tikzpicture}[scale=0.7,baseline={(0,0.7cm)}]
  \node at (-3.6,1){$\BW{a}{b}{c}{d}{u}=$};

  \draw[-] (1-1,-1.4+1.5)--(1.9-1,-0.5+1.5)--(1-1,0.4+1.5)--(0.1-1,-0.5+1.5)--(1-1,-1.4+1.5);
  \draw[-] (0.13,.25) arc (45:135:0.2);
  \node at (0,-0.2){$a$};\node at (1.2,1){$b$};\node at (0,2.1){$c$};\node at (-1.2,1){$d$};
  \node at (0,1){$u$};
  \node at (2,1){$=$};

  \draw[-,black!70!green] (1+2.8,-1.4+1.5)--(1.9+2.8,-0.5+1.5)--(1+2.8,0.4+1.5)--(0.1+2.8,-0.5+1.5)--(1+2.8,-1.4+1.5);
  \draw[-,line width=0.5mm,black!60!red] (0.55+2.8,-0.95+1.5) arc (-45:45:0.65);
  \draw[-,line width=0.5mm,black!60!red] (1.45+2.8,-0.05+1.5) arc (135:225:0.65);

  \node at (6,1){$+$};
  
  \draw[-,black!70!green] (1+7,-1.4+1.5)--(1.9+7,-0.5+1.5)--(1+7,0.4+1.5)--(0.1+7,-0.5+1.5)--(1+7,-1.4+1.5);
  \draw[-,line width=0.5mm,black!60!red] (0.55+7,-0.05+1.5) arc (225:315:0.65);
  \draw[-,line width=0.5mm,black!60!red] (0.55+7,-0.95+1.5) arc (135:45:0.65);
\end{tikzpicture}
\ee
The lines on the tiles indicate domain walls separating different heights. Thus for the first tile we have $a=c$ and for the second $b=d$, and one can check that the ABF weights (\ref{Bweights}) fall under one of the two cases.\footnote{The relation is not 1--1, as there are weights where both $a=c$ and $b=d$. A notation for the tiles which makes them match the weights more precisely was used e.g. in \cite{Bianchini:2014bfa}.} If one now considers a two-dimensional lattice model with the two tiles above as building blocks, there will be regions of different height separated by domain walls. Each edge of the dual lattice will be part of a domain wall. Thus the ABF model is called \emph{dense}. On the other hand, the dilute Boltzmann weights above include the two ABF tiles but also tiles with one or no lines. The most general choice leads to the following set of 9 tiles:

  \be \label{looptiles}
  \begin{tikzpicture}[scale=0.6]
\node at (-0.4,-0.5){$\Bigg\{$};
  \draw[-,black!70!green] (1,-1.4)--(1.9,-0.5)--(1,0.4)--(0.1,-0.5)--(1,-1.4);
  \draw[-,line width=0.5mm,black!60!red] (0.55,-0.95) arc (-45:45:0.65);
  \draw[-,line width=0.5mm,black!60!red] (1.45,-0.05) arc (135:225:0.65);
\node at (2.2,-0.7){$,$};
  \draw[-,black!70!green] (1+2.5,-1.4)--(1.9+2.5,-0.5)--(1+2.5,0.4)--(0.1+2.5,-0.5)--(1+2.5,-1.4);
  \draw[-,line width=0.5mm,black!60!red] (0.55+2.5,-0.05) arc (225:315:0.65);
  \draw[-,line width=0.5mm,black!60!red] (0.55+2.5,-0.95) arc (135:45:0.65);
\node at (2.2+2.5,-0.7){$,$};
  \draw[-,black!70!green] (1+5,-1.4)--(1.9+5,-0.5)--(1+5,0.4)--(0.1+5,-0.5)--(1+5,-1.4);
\draw[-,line width=0.5mm,black!60!red] (0.55+5,-0.95) arc (-45:45:0.65);
\node at (2.2+5,-0.7){$,$};
  \draw[-,black!70!green] (1+7.5,-1.4)--(1.9+7.5,-0.5)--(1+7.5,0.4)--(0.1+7.5,-0.5)--(1+7.5,-1.4);
  \draw[-,line width=0.5mm,black!60!red] (1.45+7.5,-0.05) arc (135:225:0.65);
  \node at (2.2+7.5,-0.7){$,$};
  \draw[-,black!70!green] (1+10,-1.4)--(1.9+10,-0.5)--(1+10,0.4)--(0.1+10,-0.5)--(1+10,-1.4);
    \draw[-,line width=0.5mm,black!60!red] (0.55+10,-0.05) arc (225:315:0.65);
\node at (2.2+10,-0.7){$,$};
  \draw[-,black!70!green] (1+12.5,-1.4)--(1.9+12.5,-0.5)--(1+12.5,0.4)--(0.1+12.5,-0.5)--(1+12.5,-1.4);
  \draw[-,line width=0.5mm,black!60!red] (0.55+12.5,-0.95) arc (135:45:0.65);
\node at (2.2+12.5,-0.7){$,$};
  \draw[-,black!70!green] (1+15,-1.4)--(1.9+15,-0.5)--(1+15,0.4)--(0.1+15,-0.5)--(1+15,-1.4);
  \draw[-,line width=0.5mm,black!60!red] (0.55+15,-0.97)--(1.46+15,-0.05); 
  \node at (2.2+15,-0.7){$,$};
    \draw[-,black!70!green] (1+17.5,-1.4)--(1.9+17.5,-0.5)--(1+17.5,0.4)--(0.1+17.5,-0.5)--(1+17.5,-1.4);
  \draw[-,line width=0.5mm,black!60!red] (1.45+17.5,-0.98)--(0.53+17.5,-0.05); 
 \node at (2.2+17.5,-0.7){$,$};
 \draw[-,black!70!green] (1+20,-1.4)--(1.9+20,-0.5)--(1+20,0.4)--(0.1+20,-0.5)--(1+20,-1.4);
\node at (2.4+20,-0.5){$\Bigg\}$};
\end{tikzpicture}
\, .
\ee
See \cite{Dupic:2016ick,Morin-Duchesne:2020pbt} for recent discussions of the dilute loop models based on the tiles above and more details on their relations to the RSOS and vertex models.

After this rather lengthy review, let us now explain the relevance of RSOS models and the dynamical framework to the $\SU(3)$ sector of the spin chains arising from our quiver gauge theories, as discussed in \ref{sec:SU(3)}. More specifically, in the next section we will argue that one can obtain important insights about our Hamiltonians by considering the underlying vertex model as a dynamical vertex model (one where the couplings depend on the value of a dynamical parameter, as is the case for Felder's $R$-matrix), and that this vertex model can be mapped, at the level of the adjacency graph, to a dilute RSOS (or, more specifically, a CSOS) model. We do not address the question of whether suitable Boltzmann weights exist that realise this adjacency graph, but specifying the required features will provide important input in the search for such weights.

\subsection{The $\SU(3)$ sector as a dynamical 15-vertex model} \label{sec:15-vertex}

Let us summarise the main features of the one-loop spin chains coming from the holomorphic scalar sector of the marginally deformed $\mathbb{Z}_2$ quiver theory.  We saw that the three $\SU(2)$ sectors of the orbifold point look rather different after the deformation. In the $XY$ sector we obtain an alternating ferromagnetic Heisenberg spin chain, where the coupling is either $\kappa$ or $1/\kappa$ depending on whether the Hamiltonian is acting on even-odd or odd-even sites. These types of spin chains have been studied in the literature (e.g. \cite{Bell_1989,Medvedetal91}, see also section \ref{sec:XYsector} for more references). However, the spin chain in the $XZ$ (and equivalent $YZ$) sector is of a rather different type. Since the $Z$ field does not alter the node of the quiver, $Z$ insertions on the spin chain do not change the gauge coupling. In the language of the spin chain, crossing a $Z$ field does not change the Hamiltonian.  We will now explain why this makes this sector \emph{dilute}, where we use the term by analogy with the dilute RSOS models \cite{Warnaar:1993zn}, which we briefly reviewed above.

Following the intuition we have gained from the above discussion of the relation between dynamical spin chains and RSOS models, we will now describe why our spin chains coming from the $\Zset_2$ quiver theory should be understood as dynamical. Let us assume that a vertex model exists, whose $R$-matrix $R(u; \kappa) \equiv  R(u; \lambda)$ produces the Hamiltonian of our spin chain in the $\SU(3)$ sector. This $R$-matrix should depend on the ratio of the gauge couplings, which we have denoted $\kappa$. When crossing one of the bifundamental fields, $\kappa$ becomes exchanged with $1/\kappa$. In the dynamical spin chain language, crossing these bifundamental fields should take $\lambda\ra \lambda\pm 2\eta$. Thus, if $\kappa(\lambda\pm 2\eta)=1/\kappa(\lambda)$, the model has precisely the behaviour we require.
\be
R(u; \kappa) \equiv  R(u; \lambda) \quad  \Leftrightarrow   \quad R(u; \kappa^{-1}) \equiv  R(u; \lambda\pm 2\eta)
\ee
 What is more, our model is such that crossing two bifundamentals is equivalent to returning to the original coupling constant (and thus dynamical parameter $\lambda$). So one needs that $\lambda\pm 4\eta \sim \lambda$, implying that the $R$-matrix must have this periodicity
 \be
 R(u; \lambda)  =   R(u; \lambda \pm 4\eta)   \, .
\ee
  If we now assume that there exist a vertex model with this $R$-matrix, we can map it to an RSOS model which is necessarily cyclic (CSOS).

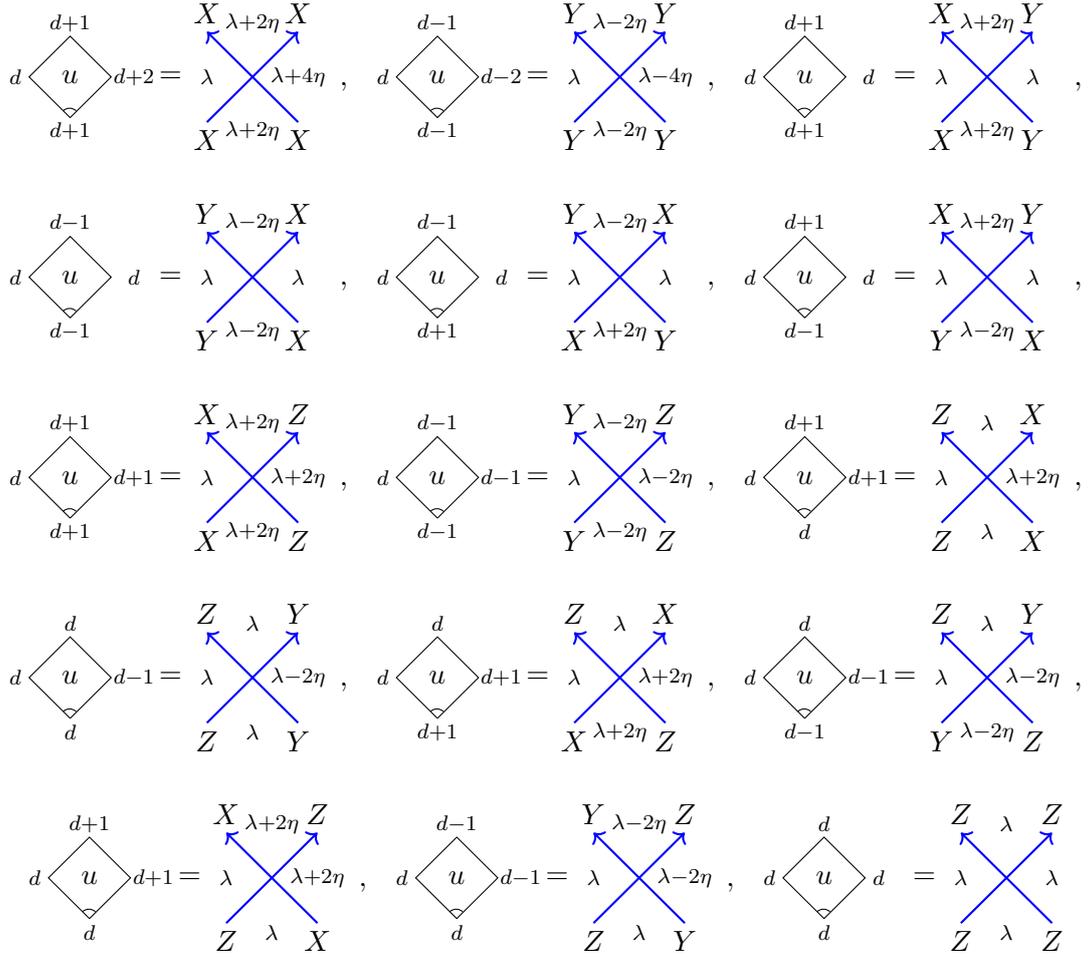
\begin{figure}[t]
\begin{center}
\begin{tikzpicture}[scale=0.6]
  \draw[-] (1-1,-1.4+0.5)--(1.9-1,-0.5+0.5)--(1-1,0.4+0.5)--(0.1-1,-0.5+0.5)--(1-1,-1.4+0.5);
  \draw[-] (0.13,.25-1) arc (45:135:0.2);
  \node at (0,-0.2-1){$\scriptstyle d+1$};\node at (1.4,1-1){$\scriptstyle d+2$};\node at (0,1.2){$\scriptstyle d+1$};\node at (-1.2,1-1){$\scriptstyle d$};
  \node at (0,0){$u$};
  
  \node at (2.2,0){$=$};

  \draw[->,blue,thick] (3,-1)--(5,1);
  \draw[->,blue,thick] (5,-1)--(3,1);
  \node at (3,-1.4) {$X$};
  \node at (5,-1.4) {$X$};
  \node at (3,1.4) {$X$};
  \node at (5,1.4) {$X$};

  \node at (3,0){$\scriptstyle\lambda$};
  \node at (4,-1.2){$\scriptstyle\lambda+2\eta$};
  \node at (4,1.2){$\scriptstyle\lambda+2\eta$};
  \node at (5,0){$\scriptstyle\lambda+4\eta$};
  \node at (6,-0.2){$,$};
\end{tikzpicture}
\begin{tikzpicture}[scale=0.6]
  \draw[-] (1-1,-1.4+0.5)--(1.9-1,-0.5+0.5)--(1-1,0.4+0.5)--(0.1-1,-0.5+0.5)--(1-1,-1.4+0.5);
  \draw[-] (0.13,.25-1) arc (45:135:0.2);
  \node at (0,-0.2-1){$\scriptstyle d-1$};\node at (1.4,1-1){$\scriptstyle d-2$};\node at (0,1.2){$\scriptstyle d-1$};\node at (-1.2,0){$\scriptstyle d$};
  \node at (0,0){$u$};
  
  \node at (2.2,0){$=$};

  \draw[->,blue,thick] (3,-1)--(5,1);
  \draw[->,blue,thick] (5,-1)--(3,1);
  \node at (3,-1.4) {$Y$};
  \node at (5,-1.4) {$Y$};
  \node at (3,1.4) {$Y$};
  \node at (5,1.4) {$Y$};

  \node at (3,0){$\scriptstyle\lambda$};
  \node at (4,-1.2){$\scriptstyle\lambda-2\eta$};
  \node at (4,1.2){$\scriptstyle\lambda-2\eta$};
  \node at (5,0){$\scriptstyle\lambda-4\eta$};
 \node at (6,-0.2){$,$};
  \end{tikzpicture}
\begin{tikzpicture}[scale=0.6]
  \draw[-] (1-1,-1.4+0.5)--(1.9-1,-0.5+0.5)--(1-1,0.4+0.5)--(0.1-1,-0.5+0.5)--(1-1,-1.4+0.5);
  \draw[-] (0.13,.25-1) arc (45:135:0.2);
  \node at (0,-0.2-1){$\scriptstyle d+1$};\node at (1.4,1-1){$\scriptstyle d$};\node at (0,1.2){$\scriptstyle d+1$};\node at (-1.2,0){$\scriptstyle d$};
  \node at (0,0){$u$};
  
  \node at (2.2,0){$=$};

  \draw[->,blue,thick] (3,-1)--(5,1);
  \draw[->,blue,thick] (5,-1)--(3,1);
  \node at (3,-1.4) {$X$};
  \node at (5,-1.4) {$Y$};
  \node at (3,1.4) {$X$};
  \node at (5,1.4) {$Y$};

  \node at (3,0){$\scriptstyle\lambda$};
  \node at (4,-1.2){$\scriptstyle\lambda+2\eta$};
  \node at (4,1.2){$\scriptstyle\lambda+2\eta$};
  \node at (5,0){$\scriptstyle\lambda$};
 \node at (6,-0.2){$,$};
  \end{tikzpicture}

\end{center}

\begin{center}
\begin{tikzpicture}[scale=0.6]
  \draw[-] (1-1,-1.4+0.5)--(1.9-1,-0.5+0.5)--(1-1,0.4+0.5)--(0.1-1,-0.5+0.5)--(1-1,-1.4+0.5);
  \draw[-] (0.13,.25-1) arc (45:135:0.2);
  \node at (0,-0.2-1){$\scriptstyle d-1$};\node at (1.4,1-1){$\scriptstyle d$};\node at (0,1.2){$\scriptstyle d-1$};\node at (-1.2,1-1){$\scriptstyle d$};
  \node at (0,0){$u$};
  
  \node at (2.2,0){$=$};

  \draw[->,blue,thick] (3,-1)--(5,1);
  \draw[->,blue,thick] (5,-1)--(3,1);
  \node at (3,-1.4) {$Y$};
  \node at (5,-1.4) {$X$};
  \node at (3,1.4) {$Y$};
  \node at (5,1.4) {$X$};

  \node at (3,0){$\scriptstyle\lambda$};
  \node at (4,-1.2){$\scriptstyle\lambda-2\eta$};
  \node at (4,1.2){$\scriptstyle\lambda-2\eta$};
  \node at (5,0){$\scriptstyle\lambda$};
  \node at (6,-0.2){$,$};
\end{tikzpicture}
\begin{tikzpicture}[scale=0.6]
  \draw[-] (1-1,-1.4+0.5)--(1.9-1,-0.5+0.5)--(1-1,0.4+0.5)--(0.1-1,-0.5+0.5)--(1-1,-1.4+0.5);
  \draw[-] (0.13,.25-1) arc (45:135:0.2);
  \node at (0,-0.2-1){$\scriptstyle d+1$};\node at (1.4,1-1){$\scriptstyle d$};\node at (0,1.2){$\scriptstyle d-1$};\node at (-1.2,0){$\scriptstyle d$};
  \node at (0,0){$u$};
  
  \node at (2.2,0){$=$};

  \draw[->,blue,thick] (3,-1)--(5,1);
  \draw[->,blue,thick] (5,-1)--(3,1);
  \node at (3,-1.4) {$X$};
  \node at (5,-1.4) {$Y$};
  \node at (3,1.4) {$Y$};
  \node at (5,1.4) {$X$};

  \node at (3,0){$\scriptstyle\lambda$};
  \node at (4,-1.2){$\scriptstyle\lambda+2\eta$};
  \node at (4,1.2){$\scriptstyle\lambda-2\eta$};
  \node at (5,0){$\scriptstyle\lambda$};
 \node at (6,-0.2){$,$};
  \end{tikzpicture}
\begin{tikzpicture}[scale=0.6]
  \draw[-] (1-1,-1.4+0.5)--(1.9-1,-0.5+0.5)--(1-1,0.4+0.5)--(0.1-1,-0.5+0.5)--(1-1,-1.4+0.5);
  \draw[-] (0.13,.25-1) arc (45:135:0.2);
  \node at (0,-0.2-1){$\scriptstyle d-1$};\node at (1.4,1-1){$\scriptstyle d$};\node at (0,1.2){$\scriptstyle d+1$};\node at (-1.2,0){$\scriptstyle d$};
  \node at (0,0){$u$};
  
  \node at (2.2,0){$=$};

  \draw[->,blue,thick] (3,-1)--(5,1);
  \draw[->,blue,thick] (5,-1)--(3,1);
  \node at (3,-1.4) {$Y$};
  \node at (5,-1.4) {$X$};
  \node at (3,1.4) {$X$};
  \node at (5,1.4) {$Y$};

  \node at (3,0){$\scriptstyle\lambda$};
  \node at (4,-1.2){$\scriptstyle\lambda-2\eta$};
  \node at (4,1.2){$\scriptstyle\lambda+2\eta$};
  \node at (5,0){$\scriptstyle\lambda$};
 \node at (6,-0.2){$,$};
  \end{tikzpicture}

\end{center}

\begin{center}
\begin{tikzpicture}[scale=0.6]
  \draw[-] (1-1,-1.4+0.5)--(1.9-1,-0.5+0.5)--(1-1,0.4+0.5)--(0.1-1,-0.5+0.5)--(1-1,-1.4+0.5);
  \draw[-] (0.13,.25-1) arc (45:135:0.2);
  \node at (0,-0.2-1){$\scriptstyle d+1$};\node at (1.4,1-1){$\scriptstyle d+1$};\node at (0,1.2){$\scriptstyle d+1$};\node at (-1.2,1-1){$\scriptstyle d$};
  \node at (0,0){$u$};
  
  \node at (2.2,0){$=$};

  \draw[->,blue,thick] (3,-1)--(5,1);
  \draw[->,blue,thick] (5,-1)--(3,1);
  \node at (3,-1.4) {$X$};
  \node at (5,-1.4) {$Z$};
  \node at (3,1.4) {$X$};
  \node at (5,1.4) {$Z$};

  \node at (3,0){$\scriptstyle\lambda$};
  \node at (4,-1.2){$\scriptstyle\lambda+2\eta$};
  \node at (4,1.2){$\scriptstyle\lambda+2\eta$};
  \node at (5,0){$\scriptstyle\lambda+2\eta$};
  \node at (6,-0.2){$,$};
\end{tikzpicture}
\begin{tikzpicture}[scale=0.6]
  \draw[-] (1-1,-1.4+0.5)--(1.9-1,-0.5+0.5)--(1-1,0.4+0.5)--(0.1-1,-0.5+0.5)--(1-1,-1.4+0.5);
  \draw[-] (0.13,.25-1) arc (45:135:0.2);
  \node at (0,-0.2-1){$\scriptstyle d-1$};\node at (1.4,1-1){$\scriptstyle d-1$};\node at (0,1.2){$\scriptstyle d-1$};\node at (-1.2,0){$\scriptstyle d$};
  \node at (0,0){$u$};
  
  \node at (2.2,0){$=$};

  \draw[->,blue,thick] (3,-1)--(5,1);
  \draw[->,blue,thick] (5,-1)--(3,1);
  \node at (3,-1.4) {$Y$};
  \node at (5,-1.4) {$Z$};
  \node at (3,1.4) {$Y$};
  \node at (5,1.4) {$Z$};

  \node at (3,0){$\scriptstyle\lambda$};
  \node at (4,-1.2){$\scriptstyle\lambda-2\eta$};
  \node at (4,1.2){$\scriptstyle\lambda-2\eta$};
  \node at (5,0){$\scriptstyle\lambda-2\eta$};
 \node at (6,-0.2){$,$};
  \end{tikzpicture}
\begin{tikzpicture}[scale=0.6]
  \draw[-] (1-1,-1.4+0.5)--(1.9-1,-0.5+0.5)--(1-1,0.4+0.5)--(0.1-1,-0.5+0.5)--(1-1,-1.4+0.5);
  \draw[-] (0.13,.25-1) arc (45:135:0.2);
  \node at (0,-0.2-1){$\scriptstyle d$};\node at (1.4,1-1){$\scriptstyle d+1$};\node at (0,1.2){$\scriptstyle d+1$};\node at (-1.2,0){$\scriptstyle d$};
  \node at (0,0){$u$};
  
  \node at (2.2,0){$=$};

  \draw[->,blue,thick] (3,-1)--(5,1);
  \draw[->,blue,thick] (5,-1)--(3,1);
  \node at (3,-1.4) {$Z$};
  \node at (5,-1.4) {$X$};
  \node at (3,1.4) {$Z$};
  \node at (5,1.4) {$X$};

  \node at (3,0){$\scriptstyle\lambda$};
  \node at (4,-1.2){$\scriptstyle\lambda$};
  \node at (4,1.2){$\scriptstyle\lambda$};
  \node at (5,0){$\scriptstyle\lambda+2\eta$};
 \node at (6,-0.2){$,$};
  \end{tikzpicture}

\end{center}

\begin{center}
\begin{tikzpicture}[scale=0.6]
  \draw[-] (1-1,-1.4+0.5)--(1.9-1,-0.5+0.5)--(1-1,0.4+0.5)--(0.1-1,-0.5+0.5)--(1-1,-1.4+0.5);
  \draw[-] (0.13,.25-1) arc (45:135:0.2);
  \node at (0,-0.2-1){$\scriptstyle d$};\node at (1.4,1-1){$\scriptstyle d-1$};\node at (0,1.2){$\scriptstyle d$};\node at (-1.2,1-1){$\scriptstyle d$};
  \node at (0,0){$u$};
  
  \node at (2.2,0){$=$};

  \draw[->,blue,thick] (3,-1)--(5,1);
  \draw[->,blue,thick] (5,-1)--(3,1);
  \node at (3,-1.4) {$Z$};
  \node at (5,-1.4) {$Y$};
  \node at (3,1.4) {$Z$};
  \node at (5,1.4) {$Y$};

  \node at (3,0){$\scriptstyle\lambda$};
  \node at (4,-1.2){$\scriptstyle\lambda$};
  \node at (4,1.2){$\scriptstyle\lambda$};
  \node at (5,0){$\scriptstyle\lambda-2\eta$};
  \node at (6,-0.2){$,$};
\end{tikzpicture}
\begin{tikzpicture}[scale=0.6]
  \draw[-] (1-1,-1.4+0.5)--(1.9-1,-0.5+0.5)--(1-1,0.4+0.5)--(0.1-1,-0.5+0.5)--(1-1,-1.4+0.5);
  \draw[-] (0.13,.25-1) arc (45:135:0.2);
  \node at (0,-0.2-1){$\scriptstyle d+1$};\node at (1.4,1-1){$\scriptstyle d+1$};\node at (0,1.2){$\scriptstyle d$};\node at (-1.2,0){$\scriptstyle d$};
  \node at (0,0){$u$};
  
  \node at (2.2,0){$=$};

  \draw[->,blue,thick] (3,-1)--(5,1);
  \draw[->,blue,thick] (5,-1)--(3,1);
  \node at (3,-1.4) {$X$};
  \node at (5,-1.4) {$Z$};
  \node at (3,1.4) {$Z$};
  \node at (5,1.4) {$X$};

  \node at (3,0){$\scriptstyle\lambda$};
  \node at (4,-1.2){$\scriptstyle\lambda+2\eta$};
  \node at (4,1.2){$\scriptstyle\lambda$};
  \node at (5,0){$\scriptstyle\lambda+2\eta$};
 \node at (6,-0.2){$,$};
  \end{tikzpicture}
\begin{tikzpicture}[scale=0.6]
  \draw[-] (1-1,-1.4+0.5)--(1.9-1,-0.5+0.5)--(1-1,0.4+0.5)--(0.1-1,-0.5+0.5)--(1-1,-1.4+0.5);
  \draw[-] (0.13,.25-1) arc (45:135:0.2);
  \node at (0,-0.2-1){$\scriptstyle d-1$};\node at (1.4,1-1){$\scriptstyle d-1$};\node at (0,1.2){$\scriptstyle d$};\node at (-1.2,0){$\scriptstyle d$};
  \node at (0,0){$u$};
  
  \node at (2.2,0){$=$};

  \draw[->,blue,thick] (3,-1)--(5,1);
  \draw[->,blue,thick] (5,-1)--(3,1);
  \node at (3,-1.4) {$Y$};
  \node at (5,-1.4) {$Z$};
  \node at (3,1.4) {$Z$};
  \node at (5,1.4) {$Y$};

  \node at (3,0){$\scriptstyle\lambda$};
  \node at (4,-1.2){$\scriptstyle\lambda-2\eta$};
  \node at (4,1.2){$\scriptstyle\lambda$};
  \node at (5,0){$\scriptstyle\lambda-2\eta$};
 \node at (6,-0.2){$,$};
  \end{tikzpicture}

\end{center}

\begin{center}
\begin{tikzpicture}[scale=0.6]
  \draw[-] (1-1,-1.4+0.5)--(1.9-1,-0.5+0.5)--(1-1,0.4+0.5)--(0.1-1,-0.5+0.5)--(1-1,-1.4+0.5);
  \draw[-] (0.13,.25-1) arc (45:135:0.2);
  \node at (0,-0.2-1){$\scriptstyle d$};\node at (1.4,1-1){$\scriptstyle d+1$};\node at (0,1.2){$\scriptstyle d+1$};\node at (-1.2,1-1){$\scriptstyle d$};
  \node at (0,0){$u$};
  
  \node at (2.2,0){$=$};

  \draw[->,blue,thick] (3,-1)--(5,1);
  \draw[->,blue,thick] (5,-1)--(3,1);
  \node at (3,-1.4) {$Z$};
  \node at (5,-1.4) {$X$};
  \node at (3,1.4) {$X$};
  \node at (5,1.4) {$Z$};

  \node at (3,0){$\scriptstyle\lambda$};
  \node at (4,-1.2){$\scriptstyle\lambda$};
  \node at (4,1.2){$\scriptstyle\lambda+2\eta$};
  \node at (5,0){$\scriptstyle\lambda+2\eta$};
  \node at (6,-0.2){$,$};
\end{tikzpicture}
\begin{tikzpicture}[scale=0.6]
  \draw[-] (1-1,-1.4+0.5)--(1.9-1,-0.5+0.5)--(1-1,0.4+0.5)--(0.1-1,-0.5+0.5)--(1-1,-1.4+0.5);
  \draw[-] (0.13,.25-1) arc (45:135:0.2);
  \node at (0,-0.2-1){$\scriptstyle d$};\node at (1.4,1-1){$\scriptstyle d-1$};\node at (0,1.2){$\scriptstyle d-1$};\node at (-1.2,0){$\scriptstyle d$};
  \node at (0,0){$u$};
  
  \node at (2.2,0){$=$};

  \draw[->,blue,thick] (3,-1)--(5,1);
  \draw[->,blue,thick] (5,-1)--(3,1);
  \node at (3,-1.4) {$Z$};
  \node at (5,-1.4) {$Y$};
  \node at (3,1.4) {$Y$};
  \node at (5,1.4) {$Z$};

  \node at (3,0){$\scriptstyle\lambda$};
  \node at (4,-1.2){$\scriptstyle\lambda$};
  \node at (4,1.2){$\scriptstyle\lambda-2\eta$};
  \node at (5,0){$\scriptstyle\lambda-2\eta$};
 \node at (6,-0.2){$,$};
  \end{tikzpicture}
\begin{tikzpicture}[scale=0.6]
  \draw[-] (1-1,-1.4+0.5)--(1.9-1,-0.5+0.5)--(1-1,0.4+0.5)--(0.1-1,-0.5+0.5)--(1-1,-1.4+0.5);
  \draw[-] (0.13,.25-1) arc (45:135:0.2);
  \node at (0,-0.2-1){$\scriptstyle d$};\node at (1.2,0){$\scriptstyle d$};\node at (0,1.2){$\scriptstyle d$};\node at (-1.2,1-1){$\scriptstyle d$};
  \node at (0,0){$u$};
  
  \node at (2.2,0){$=$};

  \draw[->,blue,thick] (3,-1)--(5,1);
  \draw[->,blue,thick] (5,-1)--(3,1);
  \node at (3,-1.4) {$Z$};
  \node at (5,-1.4) {$Z$};
  \node at (3,1.4) {$Z$};
  \node at (5,1.4) {$Z$};

  \node at (3,0){$\scriptstyle\lambda$};
  \node at (4,-1.2){$\scriptstyle\lambda$};
  \node at (4,1.2){$\scriptstyle\lambda$};
  \node at (5,0){$\scriptstyle\lambda$};

\end{tikzpicture}
\end{center}
\caption{\it The identification of the Boltzmann face weights of a dilute $A_2^{(1)}$ -type model with those of a dynamical 15-vertex model. For the specific $\mathbb{Z}_2$ quiver theory we are studying, the functional dependence of the weights on $d$ should be such that $d\pm2 \sim d$, or in terms of the $R$-matrix $\lambda \pm 4\eta \sim \lambda$.} \label{fig:15vertex}
\end{figure}

Furthermore, considering that crossing an adjoint $Z$ field does not alter the gauge group, the $R$-matrix should be such that $\lambda$ does not change when crossing a $Z$ field. In RSOS model language, this means that we should allow adjacent heights to be equal, and this means that the model is dilute, as reviewed above. All in all, in Figure \ref{fig:15vertex} 
we summarise what the properties of all the components of  the yet unknown  $R$-matrix $R(u; \lambda)$ should be with respect to
 how  $\lambda$ shifts across each field. In the language of vertex models we draw all the vertices of the $\SU(3)$ sector of the $\Zset_2$ quiver theory vertex model and explicitly draw their  $\lambda$ dependence.

 Moreover, in Figure \ref{fig:15vertex} 
 we show how our vertex model  
 (which should produce our  $R$-matrix $R(u; \lambda)$)
 can be put in correspondence with the Boltzmann weights of a dilute RSOS model (and thus also a dilute loop model as discussed above). Working in the $X,Y,Z$ basis, this identification is the unique one which respects the above periodicity and adjacency requirements as well as all the other symmetries of the problem.

Note that we have ordered the vertices so that they match the ordering in (\ref{looptiles}). In particular, the first 6 vertices (and corresponding Boltzmann weights) match with the first two ``dense'' model tiles (which can easily be seen to form a closed sector). In our model these would be the $R$-matrix components leading to the Hamiltonian of the $XY$ sector. Therefore, in the following we will call the $XY$ $\SU(2)$ sub-sector \emph{dense}. The remaining vertices (i.e. those of the $XZ$ and $YZ$ sectors) can be seen to correspond to the dilute model tiles. In particular, lines on the edges of tiles correspond to crossing an $X$ or $Y$ field, while absence of a line corresponds to the $Z$ field. What is more, in the RSOS language, edges with equal adjacent heights correspond to crossing a $Z$ field, while incrementing the height by $\pm 1$ along an edge of a weight corresponds to crossing an $X$ or $Y$ field, respectively. The empty tile corresponds to the four-$Z$ vertex. For this reason, we will call the $XZ$ and $YZ$ $\SU(2)$ sub-sector \emph{dilute}. 

In addition, note that our model cannot contain the vertices $XY\ra ZZ,YX\ra ZZ$ and their conjugates, which would correspond to the fourth and fifth tiles in (\ref{looptiles}), as they are not allowed by the $U(1)_r$ symmetry of the $\Ncal=2$ superconformal algebra.\footnote{Such vertices do appear in $\Ncal=1$ SCFT models. For instance, the $\sum_i (X^i)^3$ term appearing the general Leigh-Strassler superpotential (which played an important role in the study of \cite{Dlamini:2019zuk}) leads to vertices of this kind.} 
In the dilute model language, these tiles can be obtained from the second and third ones by crossing symmetry. We thus conclude that the corresponding loop/RSOS model is not crossing symmetric. However, this does not mean that the vertex model is not crossing symmetric. 

  The complete set of tiles (or the additional Boltzmann weights in RSOS language) in (\ref{looptiles}) would lead to a 19-vertex model, which is related to $A_2^{(2)}$ and the Izergin-Korepin model \cite{Izergin:1980pe}. That dilute model is known to have a solvable elliptic extension \cite{Warnaar:1992gj}. Our actual set of tiles/weights makes our model more like the 15-vertex model, which is related to the adjacency diagram of the $A_2^{(1)}$ algebra. For details of this model we refer to \cite{Dupic:2016ick,Morin-Duchesne:2018apx,Morin-Duchesne:2020pbt}. These references, however, are  mostly concerned with the trigonometric case. An elliptic extension should be obtainable using the relation to the $A_2^{(1)}$ RSOS models \cite{Jimbo:1987ra,Jimbo:1988gs,Kuniba89}. Whether such an extension has all the relevant features required to produce our specific Hamiltonians is the subject of ongoing study.

Having identified the overall structure of our model with that of a dilute RSOS model, the final step would be to choose the dynamical parameter and step $2\eta$ such that $\lambda\sim \lambda\pm4\eta$ (which also implies $\lambda\pm 2\eta=\lambda\mp 2\eta$). Then there are effectively only two values of the dynamical parameter, $\lambda$ and $\lambda'=\lambda+2\eta$. This means that there are two sets of vertices in the diagram above, one for $\lambda$ and one for $\lambda'$. If the parameter $\kappa$ is taken to depend on $\lambda$ such that $\kappa(\lambda')=1/\kappa(\lambda)$, we get exactly the needed structure, i.e. that the couplings in the Hamiltonian alternate depending on the gauge group to the left of each interaction vertex.\footnote{Note that with this restriction the fields $X$ and $Y$ play exactly the same role. However, the formalism can easily be generalised to e.g. $\Zset_k$ quivers, in which case there are $k$ nodes, and $X$ cycles through them in a clockwise fashion, while $Y$ goes anticlockwise. This corresponds to $X$ shifting the dynamical parameter upwards while $Y$ downwards, and the need for $k$ steps to return to the original node.}

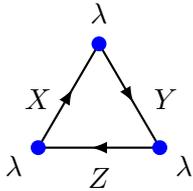
\begin{figure}[t]
\begin{center}
\begin{tikzpicture}[scale=0.8]
  \draw[thick] (0,0)--(1,2*0.86) node [sloped,pos=0.5,allow upside down,scale=1.5]{\arrowIn};
  \draw[thick] (1,2*0.86)--(2,0) node [sloped,pos=0.5,allow upside down,scale=1.5]{\arrowIn};
  \draw[thick] (2,0)--(0,0) node [sloped,pos=0.5,allow upside down,scale=1.5]{\arrowIn};
  \filldraw[blue,thick] (0,0) circle (3pt);  \filldraw[blue,thick] (1,2*0.86) circle (3pt); \filldraw[blue,thick] (2,0) circle (3pt);
   \node at (0,0.8){$X$}; \node at (2.1,0.8){$Y$}; \node at (1,-0.5){$Z$};
\node at (-0.4,-0.3){$\lambda$};\node at (1,2*0.86+0.5){$\lambda$};\node at (2.4,-0.3){$\lambda$};
\end{tikzpicture}
\end{center}
\caption{\it The adjacency graph of the $\Ncal=4$ theory seen as a (trivial) RSOS model. All vertices are identified.
  Here all fields are in the adjoint of the (single) gauge group and the model is not dynamical.}  \label{fig:N4adjacency} 
\end{figure}

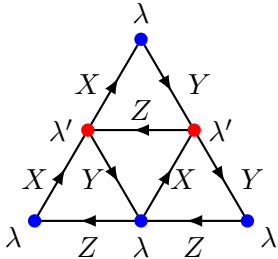
\begin{figure}[t]
\begin{center}
\begin{tikzpicture}[scale=0.7]
  \draw[thick] (0,0)--(1,2*0.86) node [sloped,pos=0.5,allow upside down,scale=1.5]{\arrowIn};
  \draw[thick] (1,2*0.86)--(2,0) node [sloped,pos=0.5,allow upside down,scale=1.5]{\arrowIn};
  \draw[thick] (2,0)--(0,0) node [sloped,pos=0.5,allow upside down,scale=1.5]{\arrowIn};
  \draw[thick] (1,2*0.86)--(2,4*0.86) node [sloped,pos=0.5,allow upside down,scale=1.5]{\arrowIn};
  \draw[thick] (2,4*0.86)--(3,2*0.86) node [sloped,pos=0.5,allow upside down,scale=1.5]{\arrowIn};
    \draw[thick] (3,2*0.86)--(1,2*0.86) node [sloped,pos=0.5,allow upside down,scale=1.5]{\arrowIn};
  \draw[thick] (2,0)--(3,2*0.86) node [sloped,pos=0.5,allow upside down,scale=1.5]{\arrowIn};
  \draw[thick] (3,2*0.86)--(4,0) node [sloped,pos=0.5,allow upside down,scale=1.5]{\arrowIn};
  \draw[thick] (4,0)--(2,0) node [sloped,pos=0.5,allow upside down,scale=1.5]{\arrowIn};
    
  \filldraw[blue,thick] (0,0) circle (3pt);  \filldraw[blue,thick] (2,0) circle (3pt);
  \filldraw[blue,thick] (4,0) circle (3pt);  \filldraw[red,thick] (1,2*0.86) circle (3pt);
   \filldraw[red,thick] (3,2*0.86) circle (3pt);  \filldraw[blue,thick] (2,4*0.86) circle (3pt);
   \node at (0,0.8){$X$};\node at (1,2.6){$X$};\node at (2.8,0.8){$X$};
   \node at (1.1,0.8){$Y$};\node at (4.1,0.8){$Y$};\node at (3.1,2.6){$Y$};
   \node at (1,-0.5){$Z$}; \node at (3,-0.5){$Z$};\node at (2,2.1){$Z$};
   \node at (-0.4,-0.3){$\lambda$};\node at (0.5,2*0.86){$\lambda'$};\node at (2,4*0.86+0.5){$\lambda$};
   \node at (2,-0.5){$\lambda$};\node at (4.4,-0.3){$\lambda$};\node at (3.5,2*0.86){$\lambda'$};
   \end{tikzpicture}
\end{center}
\caption{\it The adjacency diagram of the dilute RSOS model associated to the $\Zset_2$ quiver theory, in the dynamical $\Ncal=4$-like picture. Vertices of the same colour/height $\lambda$ are identified. This is the dual graph to the brane-tiling diagram of the quiver theory.} \label{fig:z2adjacency}  
\end{figure}

The next crucial step, before we are ready to write down the weights of our model, is to prescribe the adjacency graph of our RSOS model.
In particular, the specific set of Boltzmann weights should be precisely such that they capture the symmetries of the adjacency diagram of our $\Zset_2$ quiver.
We will explain why  this graph can be simply read off from the quiver. 
In $\Ncal=2$ language (where we have a 30-vertex model involving the six fields $\phi_1,\phi_2$, $Q_{12},Q_{21},\tQ_{12},\tQ_{21}$) the adjacency diagram is precisely our quiver depicted in Figure \ref{fig:z2}. 
This is because we can only  make single trace operators (which correspond to the spin chain states) following the arrows of the quiver.
In the dynamical $\Ncal=4$ language (where we have a 15-vertex model and the three fields $X,Y,Z$) we keep track of the gauge groups by explicitly denoting a blob in the quiver by its corresponding dynamical parameter $\lambda$.
In a picture more typical in the RSOS literature, the same quiver/adjacency diagram can be drawn as in Figure \ref{fig:z2adjacency}.  For comparison, in Figure \ref{fig:N4adjacency} we show the corresponding adjacency diagram for the $\Ncal=4$ SYM theory. For $\Ncal=4$ SYM the adjacency graph is a single triangle with the dynamical parameter $\lambda$ being the same at all the nodes. 
It is interesting to note that for both examples in the  language of brane tiling \cite{Hanany:2005ve,Franco:2005rj,Franco:2005sm,Yamazaki:2008bt}  the adjacency graph is the dual graph to the bipartite graph describing the quiver theory. 
   
To help clarify how the dynamical parameter, together with an appropriate choice of step, can lead to an alternating chain, in appendix
\ref{appendix:Alternating} we show how to obtain an alternating \emph{XX model} from Felder's dynamical $R$-matrix by choosing $2\eta=\tau/2$, with $\tau$ the imaginary period of the theta functions in the $R$-matrix. This is a dense-type model which is quite similar to our $XY$ sector, however, being a free-fermion-type model, it fails to capture the contribution of the $\sigma_z\otimes \sigma_z$ terms which are required for our XXX-type Hamiltonian.\footnote{We hope that no confusion will arise between our use of $XY$ and $XZ$ for the different $\SU(2)$ sectors of the gauge theory, and specific spin-chain models like $XXX$ or $XX$.}

\subsection{The dynamical Hamiltonian}

Answering the question of whether there exists a suitable $R$-matrix which is relevant to the models at hand is beyond the scope of this paper, but we hope to report on it in the near  future. Part of our motivation for discussing the dynamical parameter above is simply that it provides us with a very natural language with which to describe the spin chains under study, and, we expect, similar spin chains one might want to consider for larger sectors in the $\Zset_2$ quiver, or for ADE quivers in general. For alternating chains such as the $XY$ sector one can of course assign different Hamiltonians on even-odd and odd-even sites and study them without needing the dynamical parameter. But for dilute-type sectors where different fields affect the Hamiltonian in different ways (e.g. as in the $XZ$ sector where crossing a $Z$ field leaves $\kappa$ invariant while crossing an $X$ field takes $\kappa\ra1/\kappa$), the dynamical parameter appears to be very helpful in organising the computation. There is always the alternative of working in the $\Ncal=2$ picture, with the $8\times 8$ Hamiltonians (\ref{XYHamiltonian}) and (\ref{XZHamiltonian}) but this carries its own problems as the multiparticle basis is not a direct product of the single-particle basis (e.g. combinations such as $Q_{12}\tQ_{12}$ are not allowed by the gauge indices). Working in what we call the dynamical $\Ncal=4$ picture, with the dynamical parameter tracking the gauge group at each site of the chain, any combination one writes is automatically correct. 

So we can now summarise the Hamiltonians in the two sectors of relevance as follows:

\mparagraph{XY sector:}

\be
\label{eq:HXYdynamical}
\Hcal(\lambda)
=
\left(\begin{array}{cccc}
 0 & 0 & 0 & 0 \\
 0 & \kappa^{-1} & -\kappa^{-1} & 0 \\
 0 & -\kappa^{-1} & \kappa^{-1} & 0 \\
 0 & 0 & 0 & 0 \\
\end{array}
\right)
\quad
\mbox{and}
\quad
\Hcal(\lambda')
=
\left(\begin{array}{cccc}
 0 & 0 & 0 & 0 \\
 0 & \kappa & -\kappa & 0 \\
 0 & -\kappa & \kappa & 0 \\
 0 & 0 & 0 & 0 \\
\end{array}
\right)
\, .
\ee

\mparagraph{XZ sector:}

\begin{equation} \label{eq:HXZdynamical}
\Hcal(\lambda)
=
\begin{pmatrix}
0  & 0  & 0 & 0 \\
0  &  \kappa &  -1&  0 \\
 0 & -1  &  \kappa^{-1}  & 0  \\
0  & 0  & 0  & 0
\end{pmatrix}
\,
\quad
\mbox{and}
\quad
\Hcal(\lambda')
=
\begin{pmatrix}
0  & 0  & 0 & 0 \\
0  &  \kappa^{-1} &  -1 &  0 \\
 0 & -1   &  \kappa  & 0  \\
0  & 0  & 0  & 0
\end{pmatrix}
\, .
\end{equation}

It is worth noting here that a standard way to obtain alternating-type Hamiltonians in an Algebraic Bethe Ansatz framework is to introduce inhomogeneities (reference values of the spectral parameter at each site), and take their values to alternate as one moves along the chain. However, this typically leads to longer-range Hamiltonians (see e.g. \cite{deVega:1991rc}, or the recent work \cite{Bazhanov:2019xvy} and references therein) while our model is nearest-neighbour (at the one-loop level we are working with). This, combined with the fact that the $XZ$ sector Hamiltonian is not alternating, being determined by the number of $X$ fields one has crossed before reaching a given site, is what leads us to the dynamical picture described above. 

In the next section we will study the 2-magnon problem for the $XY$-sector Hamiltonian, and in Section \ref{sec:XZsector} that of the $XZ$-sector Hamiltonian above.

\section{The ``dense'' XY sector}
\label{sec:XYsector}

Let us now focus on the Heisenberg-like Hamiltonian (\ref{eq:HXYdynamical}) which describes the $XY$ sector. As discussed, we characterise this sector as ``dense'' since the dynamical parameter shifts every time we cross a field. Since the shift is the same regardless of whether the field is an $X$ or a $Y$, the dynamical parameter will alternate between two values $\lambda$ and $\lambda'$, and thus the Hamiltonian will be alternately $\Hcal(\lambda)$ or $\Hcal(\lambda')$ given in (\ref{eq:HXYdynamical}).\footnote{This behaviour is specific for the $\Zset_2$ quiver case. For $\Zset_k$ quivers, where we would expect $k-1$ parameters $\lambda_i$, the $X$ and $Y$ field would shift them differently.} So the net effect is that of an alternating bond Hamiltonian.

\begin{figure}[t]
  \begin{center}
  \begin{tikzpicture}[scale=0.6]
    \draw[-,thick,blue] (4,0)--(6,0);
    \draw[-,thick,red,dashed] (6,0)--(8,0);
    \draw[-,thick,blue] (8,0)--(10,0);
    \draw[-,thick,red,dashed] (10,0)--(12,0);
    \draw[-,thick,blue] (12,0)--(14,0);
    \draw[-,thick,red,dashed] (14,0)--(16,0);
    \draw[-,thick,blue] (16,0)--(18,0);
    
\draw[-,thick,dotted] (3,0)--(4,0);\draw[-,thick,dotted] (18,0)--(19,0);

\filldraw[red] (4,0) circle (5pt);
    \filldraw[blue] (6,0) circle (5pt);
        \filldraw[red] (8,0) circle (5pt);
        \filldraw[blue] (10,0) circle (5pt);
            \filldraw[red] (12,0) circle (5pt);
            \filldraw[blue] (14,0) circle (5pt);
                \filldraw[red] (16,0) circle (5pt);
        \filldraw[blue] (18,0) circle (5pt);

        \node at (5,0.7) {$\Hcal_{eo}$}; 
        \node at (7,0.7) {$\Hcal_{oe}$};
        \node at (9,0.7) {$\Hcal_{eo}$}; 
        \node at (11,0.7) {$\Hcal_{oe}$};
        \node at (13,0.7) {$\Hcal_{eo}$}; 
        \node at (15,0.7) {$\Hcal_{oe}$};
        \node at (17,0.7) {$\Hcal_{eo}$}; 

        \node at (4,-1) {$\scriptstyle 2s$};
        \node at (6,-1) {$\scriptstyle 2s+1$};

        \node at (8,-1) {$\scriptstyle 2s+2$};
        \node at (10,-1) {$\scriptstyle 2s+3$};
        \node at (12,-1) {$\scriptstyle 2s+4$};
        \node at (14,-1) {$\scriptstyle 2s+5$};
        \node at (16,-1) {$\scriptstyle 2s+6$};
        \node at (18,-1) {$\scriptstyle 2s+7$};

  \end{tikzpicture}
  \end{center}
      \caption{\it A section of the alternating bond spin chain for the XY sector. We emphasise the alternating nature by showing the bonds between lattice sites as a dashed line or a thick line.}\label{fig:alternating}
  \end{figure}
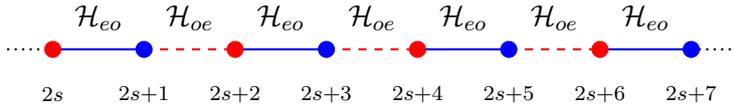

Choosing the convention that $\lambda'$ is on even sites and $\lambda$ on odd sites,  $H_{\lambda'}$ will act on even-odd sites while  $H_{\lambda}$ will act on odd-even sites (see Figure~\ref{fig:alternating}). 

\be \label{HamiltonianXY}
\Hcal_{eo}=\left(\begin{array}{cccc}
 0 & 0 & 0 & 0 \\
 0 & \kappa & -\kappa & 0 \\
 0 & -\kappa & \kappa & 0 \\
 0 & 0 & 0 & 0 \\
\end{array}
\right)\;,\;\;
\Hcal_{oe}=\left(\begin{array}{cccc}
 0 & 0 & 0 & 0 \\
 0 & 1/\kappa & -1/\kappa & 0 \\
 0 & -1/\kappa & 1/\kappa & 0 \\
 0 & 0 & 0 & 0 \\
\end{array}
\right)
\ee

Alternating nearest-neighbour chains such as this have been studied in \cite{Bell_1989, Medvedetal91}. In fact the discussion there was more general, as they considered not only bond alternation but also spin alternation (e.g. spin 1/2 alternating with spin 1). The dispersion relations and overall treatment of the chains are similar in both the bond and spin alternation case. There are also known materials which exhibit such alternating ferromagnetic behaviour, see e.g. \cite{Sirker_2008} for an example with bond alternation and \cite{Feyerhermetal01} for a case with spin alternation.\footnote{Let us remark that the literature on antiferromagnetic alternating chains is much more extensive, as well as that on chains with alternating antiferromagnetic/ferromagnetic bonds.}   

 The works \cite{Bell_1989, Medvedetal91} mostly studied the 2-magnon problem, however the 3- and multi-magnon problems were considered in \cite{Southern_1994,Southern98}, without, however, making use of Bethe-type ansatz but rather the recursion method (e.g. \cite{RecursionMethod}) which is applicable regardless of integrability. A long-wavelength approximation to this chain, which might be relevant for comparisons to the string sigma model side, is discussed in \cite{Huangetal91}.

\begin{figure}
\begin{center}
\includegraphics[width=12cm]{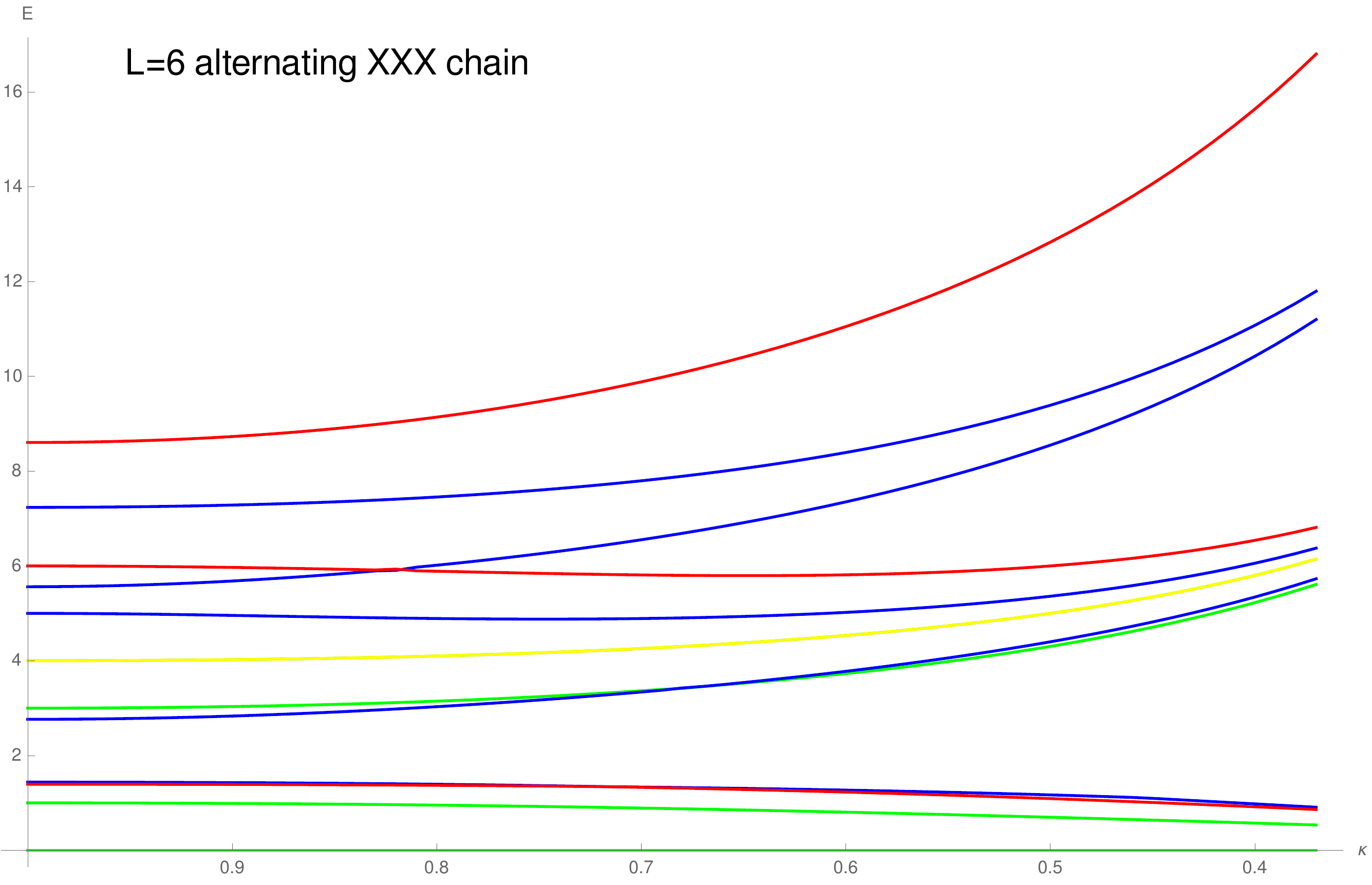}
\caption{\it The spectrum of the periodic (untwisted sector) $L=6$ alternating chain as a function of $\kappa$. $\kappa=1$ is the standard XXX Heisenberg model. Green denotes 1-magnon states, blue 2-magnon states and red 3-magnon states. The line starting at $E=4$ is a degenerate 1-magnon and 3-magnon state. There is also a 1-magnon state at E=0. We have plotted down to $\kappa=0.4$ for clarity. As $\kappa$ decreases further, the eigenvalues bunch into groups with limiting values of $6/\kappa,4/\kappa,2/\kappa$ and $0$.} \label{fig:L6periodic}
\end{center}
\end{figure}

Explicit diagonalisation of the alternating Hamiltonian for short chains reveals that the degeneracies of the eigenvalues are exactly the same as that for the XXX model at $\kappa=1$. Since the Hamiltonian is invariant under $\kappa\ra 1/\kappa$, without loss of generality we can take $\kappa\leq 1$. As $\kappa$ is decreased from $1$, the energies are modified but there is no splitting. This implies that the deformation preserves the symmetries of the XXX model. Some energy levels do cross,
however crossing only happens between
 states of different magnon number, thus they do not mix as the magnon number is a conserved quantity. An example of how the spectrum behaves as a function of $\kappa$ is shown for the $L=6$ periodic chain in Figure \ref{fig:L6periodic}.  As a check of our solution of the 2-magnon problem in the following, in Section \ref{sec:ClosedXY} we will compare the explicit diagonalisation to the 2-magnon energies and find exact agreement. 

We will now proceed to study the coordinate Bethe ansatz for this alternating chain. We start by noting that there are two equivalent pseudovacua, $\ket{\cdots XXXXX\cdots }$ and $\ket{\cdots YYYYY \cdots}$ which have zero energy. We will choose the $X$ vacuum as a reference and consider $Y$ magnon excitations on top of this reference state. Following in large part the techniques outlined in \cite{Bell_1989, Medvedetal91}, we will study the diagonalisation of the 1- and 2-magnon Hamiltonian for this alternating chain.

\subsection{One Magnon}

\begin{figure}[ht]
\begin{center}
\begin{tikzpicture}
  \begin{axis}[axis lines=left,axis line style={-},
    xtick={0,1.57,3.14},xticklabels={$0$,$\pi/2$,$\pi$},
    ytick={0,4,4.1},yticklabels={$0$,${}_{{}_4}$,${}^{{}^{2(\kappa+1/\kappa)}}$},x=3.2cm,y=1.3cm,xmax=3.3,ymax=4.2
    ]
    \addplot[orange,thick,domain=0:pi]{2*(1-cos(deg(x)))};
    \addplot[blue,thick,domain=0:pi/2] {(1/0.8+0.8-1/0.8*sqrt(1+0.8^4+2*0.8^2*cos(2*deg(x))))};
    \addplot[blue,thick,domain=pi/2:pi] {((1/0.8+0.8)+1/0.8*sqrt(1+0.8^4+2*0.8^2*cos(2*deg(x))))};
    \addplot[black!15,dashed,domain=0:pi] {4};
    \addplot[green,dashed,domain=0:pi] {2/0.8+2*0.8};
  \end{axis};
  \node at (3,2.6) {$\scriptstyle2\,(\kappa^{-1}-\kappa)\Bigg\{$};
 \node at (4.8,0.8) {Acoustic branch};\node at (5,4.5){Optical branch};
\end{tikzpicture}
  \caption{\it A plot of the 1-magnon energy $E_1$ (blue line) for $\kappa<1$, as compared to the XXX energy $E_1^{XXX}=2(1-\cos p)$ at $\kappa=1$. The gap between the branches is $2(1/\kappa-\kappa)$, and the maximum of the energy is $2(\kappa+\kappa^{-1})$.}\label{Gap}
\end{center}
\end{figure}
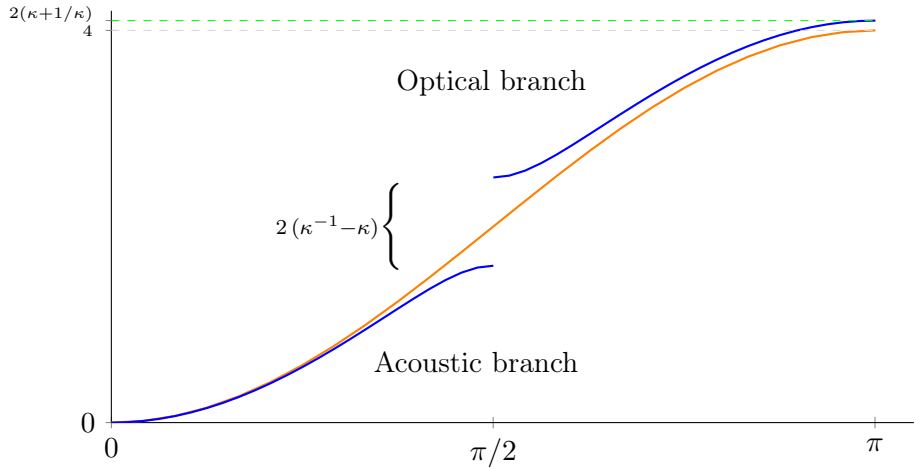

Let us start by considering a single $Y$ magnon in the $X$ vacuum and proceed to solve the one magnon problem $H|p\rangle = E_1(p)|p\rangle$. Due to the alternating-bond nature of the spin chain, there will be two equations to solve, namely, for $Y$ on even sites and for $Y$ on odd sites. Denoting these states by $\ket{2r}$ and $\ket{2s+1}$, the equations are
\be
\begin{split}
&\left(\kappa+\kappa^{-1}\right) \ket{2r}-\kappa^{-1}\ket{2r-1}-\kappa \ket{2r+1}=E \ket{2r} \;\; \text{and}\\
&\left(\kappa+\kappa^{-1}\right) \ket{2s+1}-\kappa \ket{2s}-\kappa^{-1} \ket{2s+s}=E \ket{2s+1} \;.
\end{split}
\ee
We will therefore take for our ansatz a superposition of a single $Y$ excitation on odd and even sites
\begin{equation}
\label{eq:1magnonXY}
    |p\rangle = \displaystyle\sum_{\ell \in 2\mathbb{Z}}\ \psi_{e}(\ell)|\ell\rangle + \displaystyle\sum_{\ell \in 2\mathbb{Z}+1}\ \psi_{o}(\ell)|\ell\rangle.
\end{equation}
This leads to the following equations:
\begin{equation}
    \begin{aligned}
        &\kappa^{-1}\big(\psi_e(2r) - \psi_o(2r-1)\big) + \kappa\big(\psi_e(2r) -\psi_o(2r+1)\big) = E_1(p)\psi_e(2r),
    \end{aligned}
\end{equation}
and
\begin{equation}
    \begin{aligned}
        & \kappa\big(\psi_o(2s+1)- \psi_e(2s)\big) + \kappa^{-1}\big(\psi_o(2s+1) - \psi_e(2s+2)\big)= E_1(p)\psi_o(2s+1)  \, ,
    \end{aligned}
\end{equation}
which can  be solved easily by the Bethe-type ansatz
\begin{equation}
    \begin{aligned}
        \psi_{e}(\ell) = A_e(p)e^{ip\ell}, \quad \psi_{o}(\ell) = A_o(p)e^{ip\ell} \, ,
    \end{aligned}
\end{equation}
where the ratio between the even and odd amplitudes is fixed to be
\begin{equation}\label{eq:ratio}
    r(p;\kappa) = \frac{A_o(p)}{A_e(p)} = \mp\frac{e^{ip}\sqrt{1 +\kappa^{2} e^{-2ip}}}{\sqrt{1+\kappa^{2}e^{2ip}}} \, .
\end{equation}
The eigenvalue of the eigenvector \eqref{eq:1magnonXY} is
\be \label{XYdispersion1}
E_1(p)= E_1(p;\kappa)= \frac{1}{\kappa}+\kappa \pm\frac{1}{\kappa}\sqrt{(1+\kappa^2)^2-4\kappa^2\sin^2p} \, .
\ee
Similarly to \cite{Bell_1989,Medvedetal91}, we will call the negative branch of the square root the \emph{acoustic branch} and the positive one the \emph{optical branch}. The acoustic branch is the one which includes the zero-energy state $E_1(0)=0$. As can be seen in Figure \ref{Gap}, as $\kappa$ is tuned away from 1, a gap of magnitude $2(1/\kappa-\kappa)$ develops between the branches at the boundary of the Brillouin zone, which is at $p=\pi/2$. Therefore, scattering states are confined either to the lower (``acoustic'') or upper (``optical'') branch.

Note that by a choice of branch cut we can also bring the energy eigenvalue to the form
\begin{equation} \label{XYdispersion2}
    E_1(p;\kappa) = \kappa + \frac{1}{\kappa} \pm\frac{1}{\kappa}\sqrt{1+\kappa^{2}e^{-2ip}}\sqrt{1+\kappa^{2}e^{2ip}}\;. 
\end{equation}
In this section we will choose to use the dispersion relation in this form. Some motivation for this will be discussed in Section \ref{sec:Elliptic}. Without loss of generality we will work with the acoustic branch, however when considering specific solutions (such as in section \ref{sec:ClosedXY}) magnons belonging to both branches need to be considered (as well as solutions with complex momenta which can lead to energies between the branches.)

We observe that the energy is even under reflection of the momentum, while the ratio function is inverted:
\be
E_1(-p ; \kappa)=E_1(p ; \kappa) \;\;,\;\;r(-p ; \kappa)=\frac{1}{r(p ; \kappa)}
\ee
Similarly, the energy is invariant under the $\Zset_2$ transformation $\kappa\ra 1/\kappa$, while the ratio is again inverted:
\be
E_1(p ; 1/\kappa) = E_1(p ; \kappa) \; \;,\; \; r(p ; 1/\kappa)=\frac{1}{r(p ; \kappa)} \, .
\ee
This behaviour of the ratio function is natural, since the $\Zset_2$ transformation exchanges the gauge groups and thus the even and odd sites of the chain.
Thus, the energy eigenfunction is also an eigenfunction of the $\mathbb{Z}_2$ symmetry $\kappa\ra 1/\kappa$:
\be
\label{eq:Z2EigenvalueXY}
\mathbb{Z}_2      |p\rangle   =    \frac{1}{r(p ; \kappa)}    |p\rangle  \, .
\ee
In the following, we will work with a fixed value of $\kappa$, so where no confusion can arise we will simply write $r(p)=r(p ; \kappa)$.

In the orbifold limit $\kappa \rightarrow 1$, the alternating-bond behaviour disappears, as both  Hamiltonians reduce to the well-known ferromagnetic XXX Hamiltonian. This fact is also reflected in the dispersion relation reducing to the standard XXX form and the ratio between even and odd sites becoming trivial:
\begin{equation}
    E_1(p;1) = 2 (1-\cos(p))=4\sin^{2}(p/2) \;,\; \; r(p;1) = 1 \;.
\end{equation}
Note, however, that even though the Hamiltonian is that of the XXX model, the orbifold limit is not equivalent to $\Ncal=4$ SYM as one also needs to include the twisted sector. The orbifold limit and the corresponding Bethe ansatz have been discussed in detail in \cite{Wang:2003cu,Ideguchi:2004wm,Beisert:2005he,Solovyov:2007pw}.

We will now proceed to the 2-magnon problem which, as we will see, exhibits several novel features compared to the XXX case. 

\subsection{Two Magnons}

Following the treatment of \cite{Bell_1989,Medvedetal91}, we will organise the eigenvalue problem for two magnons into \emph{non-interacting} and \emph{interacting equations}. The difference from the standard analysis is that the equations depend on whether the magnons are at even-even, even-odd, odd-even or odd-odd sites. The non-interacting equations for even-even and even-odd sites are

\be \label{NonInt2magXY}
\begin{split}
&2(\kappa+\kappa^{-1})\sket{2r,2s}-\kappa^{-1}(\sket{2r-1,2s}+\sket{2r,2s-1})-\kappa(\sket{2r+1,2s}+\sket{2r,2s+1})=E\sket{2r,2s} \;\;\text{and}\\
&2(\kappa+\kappa^{-1})\sket{2r,2s+1}-\kappa^{-1}(\sket{2r-1,2s+1}+\sket{2r,2s+2})-\kappa(\sket{2r+1,2s+1}+\sket{2r,2s})=E\sket{2r,2s+1} \;,
\end{split}
\ee
and the odd-even and odd-odd cases can be obtained from these by taking $\kappa\ra \kappa^{-1}$. There are two interacting equations, depending on whether the neighbouring magnons are at even-odd or odd-even sites:
\be \label{Int2magXY}
\begin{split}
&2\kappa^{-1}\ket{2r,2r\!+\!1}-\kappa^{-1}\ket{2r\!-\!1,2r\!+\!1}-\kappa^{-1}\ket{2r,2r\!+\!2}=E\ket{2r,2r\!+\!1}\; \;\text{and}\\
&2\kappa\ket{2r\!-\!1,2r}-\kappa\ket{2r\!-\!2,2r}-\kappa\ket{2r\!-\!1,2r\!+\!1}=E\ket{2r\!-\!1,2r}\;
\end{split}
\ee

Since the non-interacting equations are simply sums of the 1-magnon equations, it is easy to check that an ansatz of the type
\begin{equation} \label{XY2magnonnaive}
 \begin{split}  
   \ket{p_1,p_2}=&\sum_{r<s} \psi_{ee}(2r, 2s)\ket{2r,2s} \!+\! \sum_{r<s\!+\!1} \psi_{eo}(2r,2s\!+\!1)\ket{2r,2s\!+\!1}\\
        &\!+\!\sum_{r<s\!+\!1} \psi_{oe}(2r\!-\!1, 2s)\ket{2r\!-\!1,2s}\!+\!\sum_{r<s} \psi_{oo}(2r\!+\!1, 2s\!+\!1)\ket{2r\!+\!1,2s\!+\!1}\;,
   \end{split}
\end{equation}
where (with $\ell_1,\ell_2$ being even or odd as specified by the indices on $\psi$)
\begin{equation}
    \begin{aligned}
        &\psi_{ee}(\ell_1, \ell_2) = A_{ee}(p_1,p_2)e^{ip_1\ell_1+ip_2\ell_2},\\
        &\psi_{eo}(\ell_1, \ell_2) = A_{eo}(p_1,p_2)e^{ip_1\ell_1+ip_2\ell_2},\\
        &\psi_{oe}(\ell_1, \ell_2) = A_{oe}(p_1,p_2)e^{ip_1\ell_1+ip_2\ell_2},\\
        &\psi_{oo}(\ell_1, \ell_2) = A_{oo}(p_1,p_2)e^{ip_1\ell_1+ip_2\ell_2}\;,
    \end{aligned}
\end{equation}
and where the coefficients are fixed in terms of $A_{ee}(p_1,p_2)$ as 
\begin{equation}\label{eq:ratios}
  \begin{split}
    A_{eo}(p_1,p_2) &=r(p_2)A_{ee}(p_1,p_2)\;,\\
    A_{oe}(p_1,p_2) &=r(p_1)A_{ee}(p_1,p_2) \;,\\
    A_{oo}(p_1,p_2) &=r(p_1)r(p_2)A_{ee}(p_1,p_2)\;,
\end{split}
  \end{equation}
solves the non-interacting equations, with the additive eigenvalue $E_2(p_1,p_2) = E_1(p_1) + E_1(p_2)$. To solve the interacting equations, the usual Bethe approach would be to add to the above terms also those with swapped momenta, i.e. with $e^{i\ell_1 p_2 +i\ell_2p_1}$ in the exponent. However, in this case this does not lead to a solution, and we will need to enhance the Bethe ansatz (\ref{XY2magnonnaive}) to obtain a solution.

\subsubsection{Centre-of-mass frame} \label{sec:CoMXY}

One way to improve on (\ref{XY2magnonnaive}) is by adding, apart from swapped terms, also contact terms to the wavefunction. As we will see, this will indeed provide a solution, however only in the centre-of-mass frame where $p_1+p_2=0$. The origin of these contact terms will become clearer in the next section, after we solve the more general case of $p_1+p_2\neq 0$.

So we will now write 
\begin{equation} \label{XY2magnonnonint}
 \begin{split}  
   \ket{p_1,p_2}=&\sum_{r<s} \psi_{ee}(2r, 2s)\ket{2r,2s} \!+\! \sum_{r<s} \psi_{eo}(2r,2s\!+\!1)\ket{2r,2s\!+\!1}\\
        &\!+\!\sum_{r<s} \psi_{oe}(2r\!-\!1, 2s)\ket{2r\!-\!1,2s}\!+\!\sum_{r<s} \psi_{oo}(2r\!+\!1, 2s\!+\!1)\ket{2r\!+\!1,2s\!+\!1}\;,
   \end{split}
\end{equation}
where
\begin{equation} \label{Bethecontact}
    \begin{aligned}
        \psi_{ee}(\ell_1,\ell_2) &= A_{ee}(p_1,p_2)e^{ip_1\ell_1+ip_2\ell_2} + A_{ee}(p_2,p_1)e^{ip_2\ell_1+ip_1\ell_2}\\
        \psi_{oo}(\ell_1,\ell_2) &= A_{oo}(p_1,p_2)e^{ip_1\ell_1+ip_2\ell_2} + A_{oo}(p_2,p_1)e^{ip_2\ell_1+ip_1\ell_2}\\
        \psi_{eo}(\ell_1,\ell_2) &= A_{eo}(p_1,p_2)e^{ip_1\ell_1+ip_2\ell_2}(1+\delta_{\ell_1+1,\ell_2}\mathcal{A}(p_1,p_2)) \\
        &+ A_{eo}(p_2,p_1)e^{ip_2\ell_1+ip_1\ell_2}(1+\delta_{\ell_1+1,\ell_2 }\mathcal{A}(p_2,p_1))\\
        \psi_{oe}(\ell_1,\ell_2) &= A_{oe}(p_1,p_2)e^{ip_1\ell_1+ip_2\ell_2}(1+\delta_{\ell_1+1,\ell_2 }\mathcal{B}(p_1,p_2)) \\
        &+ A_{oe}(p_2,p_1)e^{ip_2\ell_1+ip_1\ell_2}(1+\delta_{\ell_1+1,\ell_2}\mathcal{B}(p_2,p_1)).
    \end{aligned}
\end{equation}
We have locally enhanced the wavefunctions for $\psi_{eo}$, $\psi_{oe}$ for nearest-neighbour terms by adding general coefficients $\mathcal{A}$, $\mathcal{B}$ which play the role of contact terms. Note that we are not assuming that $\mathcal{A}(p_2,p_1),\mathcal{B}(p_2,p_1)$ are related to $\mathcal{A}(p_1,p_2),\mathcal{B}(p_1,p_2)$ by permuting the momenta, so we have four independent contact terms. At this stage we have not imposed the CoM frame yet (so we keep $p_2$ in the notation for now) but we will soon see that the condition $p_2=-p_1$ is required for consistency of the ansatz.

Factoring out the overall coefficient $A_{ee}(p_1,p_2)$, we can write the wavefunctions as
\begin{equation} \label{BAnsatzSXY}
    \begin{aligned}
        \psi_{ee}(\ell_1,\ell_2) &= e^{ip_1\ell_1+ip_2\ell_2} + S(p_1,p_2)e^{ip_2\ell_1+ip_1\ell_2}\\
        \psi_{oo}(\ell_1,\ell_2) &= r(p_1)r(p_2)e^{ip_1\ell_1+ip_2\ell_2} + r(p_1)r(p_2)S(p_1,p_2)e^{ip_2\ell_1+ip_1\ell_2}\\
        \psi_{eo}(\ell_1,\ell_2) &= r(p_2)e^{ip_1\ell_1+ip_2\ell_2}(1+\delta_{\ell_1+1,\ell_2}\mathcal{A}(p_1,p_2)) \\
        &+ r(p_1)S(p_1,p_2)e^{ip_2\ell_1+ip_1\ell_2}(1+\delta_{\ell_1+1,\ell_2}\mathcal{A}(p_2,p_1))\\
        \psi_{oe}(\ell_1,\ell_2) &= r(p_1)e^{ip_1\ell_1+ip_2\ell_2}(1+\delta_{\ell_1+1,\ell_2 }\mathcal{B}(p_1,p_2)) \\
        &+ r(p_2)S(p_1,p_2)e^{ip_2\ell_1+ip_1\ell_2}(1+\delta_{\ell_1+1,\ell_2 }\mathcal{B}(p_2,p_1)).
    \end{aligned}
\end{equation}
where we have defined $S(p_1,p_2)=A_{ee}(p_2,p_1)/A_{ee}(p_1,p_2)$. We thus need to determine $S(p_1,p_2)$ and the four contact terms $\mathcal{A}(p_1,p_2),\mathcal{A}(p_2,p_1),\mathcal{B}(p_1,p_2),\mathcal{B}(p_2,p_1)$ in order to obtain a solution. 

\mparagraph{Next-to-nearest neighbour magnons}

The first thing to consider when adding contact terms that only arise at nearest-neighbour sites is that they will jump out to the next-to-nearest neighbour terms and enter the even-even and odd-odd non-interacting equations (\ref{NonInt2magXY}) when $s=r+1$. However, we have already computed the eigenvalue for the non-interacting equations, which of course includes this case. To not spoil the eigenvalue that we found, we will need to constrain  the contact terms. Looking at the  $(2r,2r+2)$ equation, we find the constraint
\begin{equation}\label{eq:constraint1}
   \mathcal{A}(p_1,p_2) = -\kappa^{2}\frac{r(p_1)}{r(p_2)}e^{i(p_1+p_2)}\mathcal{B}(p_1,p_2)\;.
\end{equation}
 Similarly, from the $(2r+1,2r+3)$ equation, we have
\begin{equation} \label{eq:constraint2}
        \mathcal{A}(p_1,p_2) = -\kappa^{2}\frac{r(p_1)}{r(p_2)}e^{-i(p_1+p_2)}\mathcal{B}(p_1,p_2)\;.
\end{equation}
These equations can only agree if $K = p_1+p_2 = n\pi,\ n \in \mathbb{Z}$. We will choose $n = 0$ to keep $K$ in the first Brillouin zone $K \in (-\pi/2,\pi/2]$. This leads to a single constraint:
\begin{equation}\label{Contactconstraint}
    \begin{aligned} 
        \mathcal{A}(p,-p) = -\kappa^{2}\frac{r(p)}{r(-p)}\mathcal{B}(p,-p).
    \end{aligned}
\end{equation}
We will thus restrict to the centre-of-mass case for the rest of this section.

\mparagraph{Interacting Equations} 

Substituting the ansatz (\ref{Bethecontact}) together with (\ref{Contactconstraint}) in the interacting equations (\ref{Int2magXY}), we can solve for $A_{ee}(-p,p)$ in terms of $A_{ee}(p,-p)$ in each case to find an expression for the $S$-matrix $S(p,-p)$. We find
\be
S^{(eo)}(p,-p)=\frac{e^{-2 i p} \left(\mathcal{B}(p,-p) \kappa ^2 (2-E_2 \kappa )-(\mathcal{B}(p,-p)-1) e^{2 i p} (E_2 \kappa -2)+2\right)}{\mathcal{B}(-p,p) (E_2 \kappa -2) \left(1+\kappa ^2 e^{2 i p}\right)-E_2 \kappa -2 e^{2 i p}+2}\;,
\ee
from the even-odd equation, while the odd-even equation gives 
\be
S^{(oe)}(p,-p)=\frac{e^{-2ip}\left(-(1+\mathcal{B}(p,-p)) e^{2ip} (E_2-2\kappa)+\kappa(-2+\mathcal{B}(p,-p)\kappa(2\kappa-E_2))\right)}{
  E_2\left(1+\mathcal{B}(-p,p)(1+e^{2ip}\kappa^2)-2\kappa(1-e^{2ip}+\mathcal{B}(-p,p)(1+e^{2ip}\kappa^2))\right)}\;.
\ee
 To keep the expressions compact we write $E_2$ for the centre-of-mass energy $E_2=2(\kappa+1/\kappa)-2\sqrt{1+\kappa^2e^{2ip}}\sqrt{1+\kappa^2e^{-2ip}}$.

 Of course, the two expressions for the $S$-matrix need to agree. Setting them equal we can solve for  $\mathcal{B}(-p,p)$ to find
 \be
 \mathcal{B}(-p,p)=F(p) \mathcal{B}(p,-p)+\mathcal{G}(p)
 \ee
where 
 \be
F(p)=-\frac{\left(\kappa ^2+e^{2 i p}\right) \left(E_2^2 \kappa +E_2 \left(\kappa ^2+1\right) \left(-2+e^{2 i p}\right)-4 \kappa  \left(-1+e^{2 i p}\right)\right)}{\left(1+\kappa ^2 e^{2 i p}\right) \left(E_2^2 \kappa  \left(-e^{2 i p}\right)+E_2 \left(\kappa ^2+1\right) \left(-1+2 e^{2 i p}\right)-4 \kappa  \left(-1+e^{2 i p}\right)\right)}
\ee
and
\be \label{contact}
\mathcal{G}(p)=\frac{E_2 \left(\kappa ^2-1\right) \left(-1+e^{4 i p}\right)}{\left(1+\kappa ^2 e^{2 i p}\right) \left(E_2^2 \kappa  e^{2 i p}-E_2 \left(\kappa ^2+1\right) \left(-1+2 e^{2 i p}\right)+4 \kappa  \left(-1+e^{2 i p}\right)\right)}
\ee
Substituting this solution for $\mathcal{B}(-p,p)$ into either of the $S$-matrices above we find that the dependence on $\mathcal{B}(p,-p)$ cancels out and we obtain our final $S$-matrix in the centre-of-mass frame:
\be \label{SCoMXY}
S(p,-p)=\frac{e^{-2 i p} \left(E_2^2 \kappa  \left(-e^{2 i p}\right)+E_2 \left(\kappa ^2+1\right) \left(-1+2 e^{2 i p}\right)-4 \kappa  \left(-1+e^{2 i p}\right)\right)}{E_2^2 \kappa +E_2 \left(\kappa ^2+1\right) \left(-2+e^{2 i p}\right)-4 \kappa  \left(-1+e^{2 i p}\right)}
\ee
In Section \ref{sec:ClosedXY} we will show, for short closed chains, that the energies obtained by solving the centre-of-mass Bethe ansatz with this $S$-matrix correctly reproduce the energies determined by explicit diagonalisation.

\mparagraph{Final form of the contact terms}

At this stage, looking at our ansatz (\ref{BAnsatzSXY}) evaluated in the $K=0$ frame, the only term left to determine is $\mathcal{B}(p,-p)$
which appears in the direct even-odd (after using (\ref{Contactconstraint}))and odd-even terms, as everything else is expressed in terms of it (and $\mathcal{G}(p)$ which was fixed above). However, now that we have $S(p,-p)$ we can check that \be \label{Bmpp}
\mathcal{B}(-p,p)=-\frac{1+\kappa^2 e^{-2ip}}{1+\kappa^2 e^{2ip}} S(p,-p)^{-1}\mathcal{B}(p,-p)+\mathcal{G}(p)
\ee
Noting that the ratio between the swapped and direct terms, for $\ell_2=\ell_1+1$, is
\be
\frac{r(p_2) S(p_1,p_2) e^{ip_2\ell_1+ip_1\ell_2}}{r(p_1) e^{ip_1\ell_1+ip_2\ell2}}\ra \frac{1+\kappa^2 e^{2ip}}{1+\kappa^2 e^{-2ip}}S(p,-p) \quad \text{as} \;p_1\ra p,\;p_2\ra -p
\ee
we find that the $\mathcal{B}(p,-p)$ term coming from (\ref{Bmpp}) precisely cancels the $\mathcal{B}(p,-p)$ term in both the even-odd and odd-even wavefunctions.\footnote{Of course, this cancellation means that the ansatz (\ref{BAnsatzSXY}) was too permissive and we could have started with a more restrictive ansatz, with a contact term only in the direct or only on the swapped wavefunctions.} We are then left with only the $\mathcal{G}(p)$ contact term. Recalling also that $r(-p)=1/r(p)$, we can write the final solution of the XY 2-magnon problem in the centre-of-mass frame:
\begin{equation}\label{eq:CoMwav1}
    \begin{aligned}
        &\psi_{ee}(\ell_1,\ell_2) = e^{ip\ell_1+i(-p)\ell_2} + S(p,-p)e^{i(-p)\ell_1+ip\ell_2}\\
        &\psi_{oo}(\ell_1,\ell_2) = e^{ip\ell_1+i(-p)\ell_2} + S(p,-p)e^{i(-p)\ell_1+ip\ell_2)}\\
        &\psi_{oe}(\ell_1,\ell_2) = r(p)e^{ip\ell_1+i(-p)\ell_2} + \frac{1}{r(p)}S(p,-p)e^{i(-p)\ell_1+ip\ell_2}(1+\delta_{\ell_1+1,\ell_2}\mathcal{G}(p))\\
        &\psi_{eo}(\ell_1,\ell_2) = \frac{1}{r(p)}e^{ip\ell_1+i(-p)\ell_2} + r(p)S(p,-p)e^{i(-p)\ell_1+ip\ell_2}(1-\delta_{\ell_1+1,\ell_2} \frac{\kappa^{2}}{r(p)^{2}}\mathcal{G}(p))
    \end{aligned}
\end{equation}
Note that we now have a contact term only in the swapped parts of $\psi_{oe}$ and $\psi_{eo}$. However, by their nature the placement of contact terms is ambiguous and we could for instance move them to the direct terms. Noting the identity
\be
S(p,-p)\mathcal{G}(p)/r(p)=\mathcal{G}(-p) e^{-2ip} r(p)
\ee
an equivalent form of writing the solution is as follows:
\begin{equation}\label{eq:CoMwav2}
    \begin{aligned}
        &\psi_{ee}(\ell_1,\ell_2) = e^{ip\ell_1+i(-p)\ell_2} + S(p,-p)e^{i(-p)\ell_1+ip\ell_2}\\
        &\psi_{oo}(\ell_1,\ell_2) = e^{ip\ell_1+i(-p)\ell_2} + S(p,-p)e^{i(-p)\ell_1+ip\ell_2)}\\
        &\psi_{oe}(\ell_1,\ell_2) = r(p)e^{ip\ell_1+i(-p)\ell_2}(1-\delta_{\ell_1+1,\ell_2}G(-p)) + \frac{1}{r(p)}S(p,-p)e^{i(-p)\ell_1+ip\ell_2}\\
        &\psi_{eo}(\ell_1,\ell_2) = \frac{1}{r(p)}e^{ip\ell_1+i(-p)\ell_2}(1+\delta_{\ell_1+1,\ell_2}\ \kappa^{2}r(p)^{2}G(-p)) + r(p)S(p,-p)e^{i(-p)\ell_1+ip\ell_2}
    \end{aligned}
\end{equation}
This final form will be useful when recovering the contact terms from the general solution in appendix \ref{XYCoMlimit}.

At this point, if we were only interested in 2-magnon excitations we would be done, since in string theory we need to impose the zero
momentum condition. However, we wish to solve the chain for arbitrary 2-magnon centre-of-mass momentum $K$, as that would be necessary if we were to feed the solution into the three-magnon problem at a later stage. Therefore, starting from the next section we will study how to solve the 2-magnon problem for $K\neq 0$. Apart from exhibiting several interesting features, our solution will also shed light on the origin of the centre-of-mass contact terms, which were introduced by hand.

\subsubsection{General solution} \label{sec:TwomagXY}

To solve the 2-magnon problem beyond the CoM frame we will follow the work of \cite{Bell_1989,Medvedetal91} on alternating chains. 
As usual for the coordinate Bethe ansatz, one starts by splitting the eigenvalue equations into non-interacting and interacting ones.
The non-interacting ones are those involving states where no two magnons are next to each other. First one finds all solutions
of the non-interacting equations (all the values of the momenta $p_1$ and $p_2$ that solve all those equations
with given energy $E$ and total momentum $K$). One then combines them appropriately to solve the interacting
equations. The new feature in \cite{Bell_1989,Medvedetal91} compared to more standard spin chains is that there is more than one
set of momenta $p_1,p_2$ giving the same $E$ and $K$. These additional momenta also need to be added in order to obtain a solution
of the interacting equations.

One can also think of the above in terms of the formalism of \cite{Bibikov_2016}, where apart from the usual Bethe swap of momenta one allows a discrete number of additional momenta. (In that work the additional momenta were needed for the 3-magnon problem, but our case we see the need already at the 2-magnon level). 

For the non-interacting equations (\ref{NonInt2magXY}) we make the same ansatz as (\ref{XY2magnonnonint}), including the relations (\ref{eq:ratios}). The difference is in treating the interacting equations (\ref{Int2magXY}).  Following the treatment in \cite{Bell_1989, Medvedetal91}, instead of contact terms as in (\ref{Bethecontact}) we will add an extra set of momenta $\{k_1,k_2 \}$ which also correspond to the same total momentum $K$ and energy $E_2$. By adding these terms (and their permutations) to our ansatz, we will show that the interacting equations are satisfied without any constraints on the momenta.

\mparagraph{The additional momenta}

\mbox{}

The main important feature of the type of dispersion relation (\ref{XYdispersion1}) is that one can achieve the same 2-magnon energy with more than one set of momenta. To see this, let us consider the solutions of the equations
\be \label{KEconstraints}
K=p_1+p_2 \;\;\text{and} \;\;E_2(p_1,p_2)=E_1(p_1)+E_2(p_2)
\ee
for given total momentum $K$ and total energy $E_2$. Rewriting (\ref{XYdispersion1}) slightly, we have
\begin{equation}
    \begin{aligned}
        E_2(p_1,p_2) = 2(\kappa + 1/\kappa)-\sqrt{1+\kappa^4+2\kappa^2\cos(2p_1)}-\sqrt{1+\kappa^4+2\kappa^2\cos(2p_1)} \;.
    \end{aligned}
\end{equation}
For concreteness we focus on the lower branch for both magnons, however the discussion below applies to all choices of branch (as we will be removing the square roots by squaring repeatedly). Following the discussion in \cite{Bell_1989, Medvedetal91}, in order to solve (\ref{KEconstraints}) we first switch from $p_1,p_2$ to the total momentum $K$ and relative momentum $q$,
\be
K=p_1+p_2\;\;,\;\;q=\frac{p_1-p_2}{2}\;.
\ee
Writing $p_{1,2}=K/2\pm q$ and expanding $\cos(2p_{1,2})=\cos K\cos2q \mp \sin K\sin2q$, and removing the square roots by squaring twice,
we find that there are two solutions (beyond the usual swapped solutions $q\ra -q$) for $\cos(2q)$ given a fixed $E_2$ and $K$. The sum of these two solutions is:
\be \label{cosqsum}
\boxed{
\cos(2q_1)+\cos(2q_2)=-\frac{(E_2-2(\kappa+1/\kappa))^2\cos K}{2\sin^2K}}\;.
\ee
Let us assume an initial set of momenta $(p_1,p_2)$ with sum $K=p_1+p_2$ and difference $2q_1=p_1-p_2$. Then there will be another solution which we will call $(k_1,k_2)$, with the same $K=k_1+k_2$ but with difference $2q_2=k_1-k_2$, given by
\be \label{ksols}
k_{1,2}=\frac{K}{2}\pm \frac{\pi}{2}\mp\half\arccos\left(\cos(p_1-p_2)+\frac{(E_2-2(\kappa+1/\kappa))^2\cos K}{2\sin^2K}\right)
\ee
where we used that $\arccos(-x)=\pi-\arccos(x)$. For real $p_1,p_2$, this new set of momenta is generally complex-valued with equal but opposite imaginary parts (since we always take $K$ to be real valued). However, depending on the values of the initial $p$ momenta and of $\kappa$, it is possible for the $k$ momenta to also be real. In the case where the $k$ momenta are complex, their real parts differ by $\pi$. This makes them different from a typical bound state which has equal real parts given by $K/2$.

Let us remark that the possibility of additional momenta is not there for the XXX dispersion relation $E_2=2(1-\cos(p_1))+2(1-\cos(p_2))$, where the only solutions (up to periodicity) are the original $p_1,p_2$ and their permutation $p_2,p_1$. This feature is characteristic of staggered-type chains such as the one under study (see e.g. a comment in \cite{BaxterBook}, p.137). Of course, the swapped solutions $q_1\ra -q_1$ and $q_2 \ra -q_2$ are still there in (\ref{cosqsum}), so we find a total of four solutions of our 2-magnon dispersion relation:

\begin{center}
\begin{tabular}{|c|c|c|c|} \hline
Direct $p$ &  Swapped $p$& Direct $k$ &Swapped $k$ \\ \hline
$(p_1,p_2)$ &  $(p_2,p_1)$ & $(k_1,k_2)$ & $(k_2,k_1)$\\ \hline
\end{tabular}
\end{center}

As all these momenta have the same total energy and momentum, the most general wavefunction for fixed $K$ and $E_2$ will be a superposition of all of them. As expected from the discussion in \cite{Bell_1989, Medvedetal91}, generalising the Bethe ansatz to include the $k$ momenta will indeed lead to a solution of the interacting equations.

\mparagraph{Generalised Bethe ansatz}

\mbox{}

Given the above discussion, we will now update the wavefunction (\ref{XY2magnonnaive}) to include all the four sets of momenta. We have:

\begin{equation} \label{XY2magnonadditional}
 \begin{split}  
   \ket{p_1,p_2,k_1,k_2}=&\sum_{r<s} \Big(\psi_{ee}(2r, 2s)\ket{2r,2s} \!+\! \psi_{eo}(2r,2s\!+\!1)\ket{2r,2s\!+\!1}\\
  &\!+\! \psi_{oe}(2r\!-\!1, 2s)\ket{2r\!-\!1,2s}\!+ \psi_{oo}(2r\!+\!1, 2s\!+\!1)\ket{2r\!+\!1,2s\!+\!1}\Big)\;,
   \end{split}
\end{equation}
with
\begin{equation}
    \begin{aligned}
        \psi_{ee}(\ell_1, \ell_2) &= A_{ee}(p_1,p_2)e^{ip_1\ell_1+ip_2\ell_2} + A_{ee}(p_2,p_1)e^{ip_2\ell_1+ip_1\ell_2}\\ &+A_{ee}(k_1,k_2)e^{ik_1\ell_1+ik_2\ell_2} + A_{ee}(k_2,k_1)e^{ik_2\ell_1+ik_1\ell_2},\\
        \psi_{eo}(\ell_1, \ell_2) &= A_{eo}(p_1,p_2)e^{ip_1\ell_1+ip_2\ell_2} + A_{eo}(p_2,p_1)e^{ip_2\ell_1+ip_1\ell_2} \\ &+A_{eo}(k_1,k_2)e^{ik_1\ell_1+ik_2\ell_2} + A_{eo}(k_2,k_1)e^{ik_2\ell_1+ik_1\ell_2},\\
        \psi_{oe}(\ell_1, \ell_2) &= A_{oe}(p_1,p_2)e^{ip_1\ell_1+ip_2\ell_2} + A_{oe}(p_2,p_1)e^{ip_2\ell_1+ip_1\ell_2} \\ &+A_{oe}(k_1,k_2)e^{ik_1\ell_1+ik_2\ell_2} + A_{oe}(k_2,k_1)e^{ik_2\ell_1+ik_1\ell_2},\\
        \psi_{oo}(\ell_1, \ell_2) &= A_{oo}(p_1,p_2)e^{ip_1\ell_1+ip_2\ell_2} + A_{oo}(p_2,p_1)e^{ip_2\ell_1+ip_1\ell_2} \\ &+ A_{oo}(k_1,k_2)e^{ik_1\ell_1+ik_2\ell_2} + A_{oo}(k_2,k_1)e^{ik_2\ell_1+ik_1\ell_2},
    \end{aligned}
\end{equation}
Each of these extra terms are solutions of the non-interacting equations and therefore their sum is also a solution of the non-interacting equations. From the non-interacting equations, the coefficients for the direct and swapped $p$ momenta are fixed in terms of $A_{ee}(p_1,p_2)$and $A_{ee}(p_2,p_1)$ as in equation (\ref{eq:ratios}), of course with $p_1\leftrightarrow p_2$ in the swapped case. In exactly the same manner, the coefficients for the direct $k$ momenta and swapped $k$ momenta are fixed, through the appropriate replacements in equation (\ref{eq:ratios}), in terms of the coefficients $A_{ee}(k_1,k_2)$ and $A_{ee}(k_2,k_1)$, respectively. Explicitly, we have (with $\ell_1,\ell_2$ even or odd as required by the labels):
\begin{equation} \label{GeneralisedBethe}
    \begin{split}
        \psi_{ee}(\ell_1, \ell_2) &= A_{ee}(p_1,p_2)e^{ip_1\ell_1+ip_2\ell_2} + A_{ee}(p_2,p_1)e^{ip_2\ell_1+ip_1\ell_2} \\ &+ A_{ee}(k_1,k_2)e^{ik_1\ell_1+ik_2\ell_2} + A_{ee}(k_2,k_1)e^{ik_2\ell_1+ik_1\ell_2},\\
        \psi_{eo}(\ell_1, \ell_2) &= r(p_2)\ A_{ee}(p_1,p_2)e^{ip_1\ell_1+ip_2\ell_2} + r(p_1)\ A_{ee}(p_2,p_1)e^{ip_2\ell_1+ip_1\ell_2} \\ &+ r(k_2)\ A_{ee}(k_1,k_2)e^{ik_1\ell_1+ik_2\ell_2} + r(k_1)\ A_{ee}(k_2,k_1)e^{ik_2\ell_1+ik_1\ell_2},\\
        \psi_{oe}(\ell_1, \ell_2) &= r(p_1)\ A_{ee}(p_1,p_2)e^{ip_1\ell_1+ip_2\ell_2} + r(p_2)\ A_{ee}(p_2,p_1)e^{ip_2\ell_1+ip_1\ell_2} \\ &+ r(k_1)\ A_{ee}(k_1,k_2)e^{ik_1\ell_1+ik_2\ell_2} + r(k_2)\ A_{ee}(k_2,k_1)e^{ik_2\ell_1+ik_1\ell_2},\\
        \psi_{oo}(\ell_1, \ell_2) &= r(p_1)r(p_2)\ A_{ee}(p_1,p_2)e^{ip_1\ell_1+ip_2\ell_2} + r(p_1)r(p_2)\ A_{ee}(p_2,p_1)e^{ip_2\ell_1+ip_1\ell_2} \\ &+ r(k_1)r(k_2)\ A_{ee}(k_1,k_2)e^{ik_1\ell_1+ik_2\ell_2} +r(k_1)r(k_2)\ A_{ee}(k_2,k_1)e^{ik_2\ell_1+ik_1\ell_2},
    \end{split}
\end{equation}
Imposing the non-interacting equations has left us with four remaining coefficients: $A_{ee}(p_1,p_2)$, $A_{ee}(p_2,p_1)$, $A_{ee}(k_1,k_2)$ and $A_{ee}(k_2,k_1)$. These coefficients will be fixed in terms of $A_{ee}(p_1,p_2)$ through the interacting equations and will give us our final S-matrices.\\

\mparagraph{Interacting equations}

It is useful to simplify the interacting equations (\ref{Int2magXY}) by noting the fact that the even-odd and odd-even non-interacting equations (\ref{NonInt2magXY}) also hold if we set $r=s$ (even though these equations are unphysical). So we can subtract the even-odd equation interacting equation in (\ref{Int2magXY}) from the corresponding one in (\ref{NonInt2magXY}) and similarly for the odd-even one, to obtain
the two equations:
\begin{equation}\label{eq:con1}
        \frac{1}{\kappa}\psi_{ee}(2r,2r) - \frac{2}{\kappa}\psi_{oe}(2r-1,2r) + \frac{1}{\kappa}\psi_{oo}(2r-1,2r-1) = 0\;,
\end{equation}
and 
\begin{equation}\label{eq:con2}
        \kappa\ \psi_{ee}(2s,2s) - 2\kappa\ \psi_{eo}(2s, 2s+1) + \kappa\ \psi_{oo}(2s+1,2s+1) = 0\;.
\end{equation}
Now we apply the following procedure to solve these equations:
\begin{itemize}
    \item We take equation~(\ref{eq:con1}) and disregard the coefficients $A_{ee}(k_1,k_2)$ and $A_{ee}(k_2,k_1)$ momentarily. We then solve equation~(\ref{eq:con1}) for $A_{ee}(p_2,p_1)$ in terms of $A_{ee}(p_1,p_2)$. This gives us the S-matrix for $(p_1,p_2) \rightarrow (p_{2},p_{1})$ scattering.
    \item Next, we disregard the coefficients $A_{ee}(p_1,p_2)$ and $A_{ee}(p_2,p_1)$ momentarily and solve again equation~(\ref{eq:con1}) but this time for $A_{ee}(k_2,k_1)$ in terms of $A_{ee}(k_1,k_2)$. This gives us the S-matrix for $(k_1,k_2) \rightarrow (k_{2},k_{1})$ scattering. 
    \item Since each scattering sector is solved, their sum also solves equation~(\ref{eq:con1}). It does not however solve equation~(\ref{eq:con2}), but we still have two coefficients remaining: $A_{ee}(p_1,p_2)$ and $A_{ee}(k_1,k_2)$ (after substituting the results from the previous point). We can therefore solve for $A_{ee}(k_1,k_2)$ in terms of $A_{ee}(p_1,p_2)$ which gives us the S-matrix for $(p_1,p_2) \rightarrow (k_1,k_2)$ scattering. Thus, equation~(\ref{eq:con2}) is also solved.
\end{itemize}
It should be noted that we could also start this procedure with equation~(\ref{eq:con2}) first and then end with equation~(\ref{eq:con1}). The S-matrices will be different to the first procedure but are related by inversion of $\kappa$, except for the S-matrices where we go from $p$'s to $k$'s.\\

Following the above procedure, in the end we find two $\mathbb{Z}_2$-conjugate solutions for the remaining coefficients. The first solution is given by
\begin{equation} \label{FourtermA}
    \begin{aligned} 
        &A_{ee}(p_1,p_2) = \big(a(k_2,k_1)b(k_1,k_2) - b(k_2,k_1)a(k_1,k_2)\big)a(p_1,p_2)\\
        &A_{ee}(p_2,p_1) = -\big(a(k_2,k_1)b(k_1,k_2) - b(k_2,k_1)a(k_1,k_2)\big)a(p_2,p_1)\\
        &A_{ee}(k_1,k_2) = -\big(a(p_2,p_1)b(p_1,p_2) - b(p_2,p_1)a(p_1,p_2)\big)a(k_1,k_2)\\
        &A_{ee}(k_2,k_1) = -\big(a(p_1,p_2)b(p_2,p_1) - b(p_1,p_2)a(p_2,p_1)\big)a(k_2,k_1)
    \end{aligned}
\end{equation}
and the second solution (for which we will use the letter $B$) is given by
\begin{equation} \label{FourtermB}
    \begin{aligned}
        &B_{ee}(p_1,p_2) = \big(a(k_2,k_1)b(k_1,k_2) - b(k_2,k_1)a(k_1,k_2)\big)b(p_1,p_2)\\
        &B_{ee}(p_2,p_1) = -\big(a(k_2,k_1)b(k_1,k_2) - b(k_2,k_1)a(k_1,k_2)\big)b(p_2,p_1)\\
        &B_{ee}(k_1,k_2) = -\big(a(p_2,p_1)b(p_1,p_2) - b(p_2,p_1)a(p_1,p_2)\big)b(k_1,k_2)\\
        &B_{ee}(k_2,k_1) = -\big(a(p_1,p_2)b(p_2,p_1) - b(p_1,p_2)a(p_2,p_1)\big)b(k_2,k_1)
    \end{aligned}
\end{equation}
where the coefficients $a,b$ are 
\begin{equation} \label{abcoeffsXY}
   \boxed{ \begin{split}
        a(p_1,p_2,\kappa) &= e^{i(p_1+p_2)}-2e^{ip_1}r(p_2,\kappa)+r(p_1,\kappa)r(p_2,\kappa)\\
        b(p_1,p_2,\kappa) &= 1 - 2e^{ip_1}r(p_1,\kappa)+r(p_1,\kappa)r(p_2,\kappa)e^{i(p_1+p_2)}
    \end{split}}
\end{equation}
Clearly these coefficients are deformations of the terms appearing in the XXX $S$-matrix, to which they reduce in the $\kappa=1$ limit. That is, $a(p_1,p_2,1)=b(p_1,p_2,1)=1-2e^{ip_1}+e^{i(p_1+p_2)}$. 

For the wavefunction $\psi_{ee}(\ell_1,\ell_2)$, we can therefore combine the two solutions with general coefficients $\alpha,\beta$
\begin{equation}
    \psi_{ee}(\ell_1,\ell_2) = \alpha\ \psi^{A}_{ee}(\ell_1,\ell_2) + \beta\ \psi^{B}_{ee}(\ell_1,\ell_2),
\end{equation}
where the notation is that $\psi^A_{ee}$ has the $A$-coefficients and $\psi^B_{ee}$ has the $B$-coefficients. Note that $A,B$ are not indices, just different labels for the different wavefunctions. The even-odd,odd-even and odd-odd parts of the wavefunction are as above 
but with appropriate placements of $r(p)$. As we will show later, combining $\psi^{A}$ and $\psi^{B}$ in this way is necessary in order to get an eigenstate of the $\mathbb{Z}_2$ symmetry.\\

In summary, the total eigenstate that solves the two magnon problem is given by
\begin{equation}\label{eq:XY4lincom}
    \ket{\psi}_{\text{tot}} = \alpha\ |p_1,p_2,k_1,k_2 \rangle_{A} +\beta\ |p_1,p_2,k_1,k_2 \rangle_{B},
\end{equation}
Let us note that the structure of the solution bears a strong similarity to approach of \cite{Bibikov_2016}. In that work, an extra set of momenta was used to solve the three-magnon problem for a spin-1 non-integrable system (for which there exists a scattering channel that exhibits diffractive scattering) by introducing a finite number of extra \emph{discrete diffractive terms} to the usual Bethe ansatz. The main difference here is that the additional set of momenta already appears at the 2-magnon level. Writing e.g. $\psi_{ee}^A(\ell_1,\ell_2)$ as:
\begin{equation}
    \psi_{ee}^A(\ell_1,\ell_2) = \Psi(k_1,k_2)\Omega^A(p_1,p_2; \ell_1,\ell_2) - \Psi(p_1,p_2)\Omega^A(k_1,k_2; \ell_1,\ell_2), 
\end{equation}
where (here $q_i$ stands for either $p_i$ or $k_i$)
\begin{equation}
    \Psi(q_1,q_2) = a(q_2,q_1)b(q_1,q_2) - b(q_2,q_1)a(q_1,q_2),
\end{equation}
and
\begin{equation}
    \Omega^A(q_1,q_2; \ell_1,\ell_2) = a(q_1,q_2)e^{iq_1\ell_1+iq_2\ell_2} - a(q_2,q_1)e^{iq_2\ell_1+iq_1\ell_2}\;,
\end{equation}
the similarity to \cite{Bibikov_2016} is apparent.

\mparagraph{Properties of the $S$-matrices}

\mbox{}

Factoring  $A_{ee}(p_1,p_2)$ out from the above wavefunctions, we have
{\small
 \begin{equation} 
    \begin{split}
      \psi_{ee}(\ell_1, \ell_2) &= e^{ip_1\ell_1+ip_2\ell_2} + S(p_1,p_2)e^{ip_2\ell_1+ip_1\ell_2} \\ &+ T(p_1,p_2,k_1,k_2)e^{ik_1\ell_1+ik_2\ell_2} + T(p_1,p_2,k_1,k_2)S(k_1,k_2)e^{ik_2\ell_1+ik_1\ell_2}\;,\\
        \psi_{eo}(\ell_1, \ell_2) &= r(p_2) e^{ip_1\ell_1+ip_2\ell_2} + r(p_1) S(p_1,p_2)e^{ip_2\ell_1+ip_1\ell_2} \\ &+r(k_2)T(p_1,p_2,k_1,k_2)e^{ik_1\ell_1+ik_2\ell_2} + r(k_1) T(p_1,p_2,k_1,k_2)S(k_1,k_2)e^{ik_2\ell_1+ik_1\ell_2}\;,\\
        \psi_{oe}(\ell_1, \ell_2) &= r(p_1) e^{ip_1\ell_1+ip_2\ell_2} + r(p_2) S(p_1,p_2)e^{ip_2\ell_1+ip_1\ell_2} \\ &+r(k_1) T(p_1,p_2,k_1,k_2)e^{ik_1\ell_1+ik_2\ell_2} + r(k_2)T(p_1,p_2,k_1,k_2)S(k_1,k_2)e^{ik_2\ell_1+ik_1\ell_2}\;,\\
        \psi_{oo}(\ell_1, \ell_2) &= r(p_1)r(p_2) e^{ip_1\ell_1+ip_2\ell_2} + r(p_1)r(p_2) S(p_1,p_2)e^{ip_2\ell_1+ip_1\ell_2} \\ &+ r(k_1)r(k_2)T(p_1,p_2,k_1,k_2)e^{ik_1\ell_1+ik_2\ell_2} +r(k_1)r(k_2)T(p_1,p_2,k_1,k_2)S(k_1,k_2)e^{ik_2\ell_1+ik_1\ell_2},
    \end{split}
 \end{equation}
 }
where we have defined three S-matrices:
\begin{equation}
        S(p_1,p_2) = \frac{A_{ee}(p_2,p_1)}{A_{ee}(p_1,p_2)}\;,\; S(k_1,k_2) = \frac{A_{ee}(k_2,k_1)}{A_{ee}(k_1,k_2)}\;,
        \;T(p_1,p_2,k_1,k_2) = \frac{A_{ee}(k_1,k_2)}{A_{ee}(p_1,p_2)}.
\end{equation}
The $S$-matrix $S(p_1,p_2)$ is the usual Bethe ansatz $S$-matrix, while $S(k_1,k_2)$ plays the same role for the $k$ momenta. However $T(p_1,p_2,k_1,k_2)$ is a new object, which describes the scattering where an initial state with $p_1,p_2$ momenta becomes a final state with $k_1,k_2$ momenta.

For the $A$ solution, these $S$-matrices take the form
\begin{equation} \label{SAmatrixXY}
    S^A(p_1,p_2;\kappa) =-\frac{a(p_2,p_1)}{a(p_1,p_2)},\quad S^A(k_1,k_2;\kappa) =-\frac{a(k_2,k_1)}{a(k_1,k_2)}.
\end{equation}
while $T$ is
\begin{equation}
\label{TAmatrixXY}
        T^{A}(p_1,p_2;k_1,k_2,\kappa) = -\frac{a(p_2,p_1)b(p_1,p_2) - b(p_2,p_1)a(p_1,p_2)}{a(k_2,k_1)b(k_1,k_2) - b(k_2,k_1)a(k_1,k_2)}\frac{a(k_1,k_2)}{a(p_1,p_2)}.
\end{equation}
Similarly, for the solution with $B$ coefficients we have,
\begin{equation} \label{SBmatrixXY}
    S^B(p_1,p_2;\kappa) =-\frac{b(p_2,p_1)}{b(p_1,p_2)},\quad S^B(k_1,k_2;\kappa) =-\frac{b(k_2,k_1)}{b(k_1,k_2)},
\end{equation}
and 
\begin{equation}
\label{TBmatrixXY}
        T^{B}(p_1,p_2;k_1,k_2;\kappa)
        = -\frac{a(p_2,p_1)b(p_1,p_2) - b(p_2,p_1)a(p_1,p_2)}{a(k_2,k_1)b(k_1,k_2) - b(k_2,k_1)a(k_1,k_2)}\frac{b(k_1,k_2)}{b(p_1,p_2)}.
\end{equation}
It is intriguing to note that we can write all the $S$-matrices above \eqref{SAmatrixXY}, \eqref{SBmatrixXY}, \eqref{TAmatrixXY}  and \eqref{TBmatrixXY} in a unified form by including the ratio of the Wronskian-type terms even  for the  $S^A$, $S^B$ where the Wronskian-type factor   becomes trivial. Using generic momenta $q_1,q_2\ra q_1',q_2'$ for the scattering process, we obtain the unifying formula:
\be
   \boxed{
S(q_1,q_2;q_1',q_2')=-\frac{a(q_2,q_1)b(q_1,q_2) - b(q_2,q_1)a(q_1,q_2)}{a(q_2',q_1')b(q_1',q_2') - b(q_2',q_1')a(q_1',q_2')}\frac{b(q_1',q_2')}{b(q_1,q_2)}
}
\ee
and we see that $S(p_1,p_2;p_2,p_1)=-S(p_1,p_2), S(k_1,k_2;k_2,k_1)=-S(k_1,k_2)$ and $ S(p_1,p_2;k_1,k_2)=T(p_1,p_2;k_1,k_2)$. 

The $S$ matrices for the $p$ and $k$ momenta satisfy the unitary condition
\be
S^{A}(p_1,p_2)S^{A}(p_2,p_1)=1\;\;,\;\; S^{B}(p_1,p_2)S^{B}(p_2,p_1)=1\;
\ee
as well as
\be
S^A(p,p)=-1\;\;,\;\;S^B(p,p)=-1 .
\ee
They also smoothly reduce to the XXX $S$-matrix as $\kappa\ra 1$, e.g
\be
S^A(p_1,p_2)=-\frac{e^{i(p_1+p_2)}-2e^{ip_2}r(p_1,\kappa)+r(p_1,\kappa)r(p_2,\kappa)}{e^{i(p_1+p_2)}-2e^{ip_1}r(p_2,\kappa)+r(p_1,\kappa)r(p_2,\kappa)} \overset{\kappa\ra 1}{\longrightarrow}
-\frac{e^{i(p_1+p_2)}-2e^{ip_2}+1}{e^{i(p_1+p_2)}-2e^{ip_1}+1}\;.
\ee
Using  that $r(0,\kappa)=1$, it can also be easily seen that
\be
S^{A,B}(0,p)=S^{A,B}(p,0)=1
\ee
On the other hand, $|T| \neq 1$ and therefore it is not a phase.\footnote{However we expect that a process where $p_1+p_2\ra k_1+k_2$ is compensated by the inverse process.} There is an interesting identity that one can derive by thinking physically about scattering from $\{p_1,p_2\}$ to $\{k_2,k_1\}$: one can first send $\{p_1,p_2\} \rightarrow \{k_1,k_2\}$ and then $\{k_1,k_2\} \rightarrow \{k_2,k_1\}$ or one can first send $\{p_1,p_2\} \rightarrow \{p_2,p_1\}$ and then $\{p_2,p_1\} \rightarrow \{k_2,k_1\}$. This gives
\begin{equation}
    S^{A}(k_1,k_2)T^{A}(p_1,p_2,k_1,k_2) = T^{A}(p_2,p_1,k_2,k_1)S^{A}(p_1,p_2),
\end{equation}
and a similar identity for the $B$ coefficient solutions.

\mparagraph{Symmetries}

\mbox{}

The two solutions $\ket{p_1,p_2,k_1,k_2}_A$ and $\ket{p_1,p_2,k_1,k_2}_B$ are related by $\mathbb{Z}_2$ symmetry. Recalling that $r(p,1/\kappa) = 1/r(p,\kappa)$, we can see that
\begin{equation} \label{abZ2}
    \begin{aligned}
        a(p_1,p_2,1/\kappa) =\frac{1}{r(p_1)r(p_2)}b(p_1,p_2,\kappa), \quad b(p_1,p_2,1/\kappa) = \frac{1}{r(p_1)r(p_2)}a(p_1,p_2,\kappa)\;,
    \end{aligned}
\end{equation}
from which we find for the $S$-matrices
\begin{equation}
    \begin{aligned}
        S^A(x,y,1/\kappa) = S^B(x,y,\kappa), \quad T^A(w,x,y,z,1/\kappa) = \frac{r(k_1)r(k_2)}{r(p_1)r(p_2)}T^B(w,x,y,z,\kappa).
    \end{aligned}    
\end{equation}
Using (\ref{abZ2}) one can check that the total action of $\mathbb{Z}_2$ on  say $\psi^{A}_{ee}$ is
\begin{equation}
    \psi^{A}_{ee} \rightarrow -r(p_1)^{-2}r(p_2)^{-2}r(k_1)^{-2}r(k_2)^{-2}\psi^{B}_{oo}.
\end{equation}
The other wavefunctions work similarly, with even and odd sites being exchanged by the $\Zset_2$ action. Therefore, if we want the general solution (\ref{eq:XZ4lincom}) to also be an eigenstate of $\Zset_2$ we need to impose $\alpha(1/\kappa)=\pm\beta(\kappa)$ and we find
\begin{equation}
    \mathbb{Z}_2\ket{\psi}_{\text{tot}} = \mp \big(r(p_1)r(p_2)r(k_1)r(k_2)\big)^{-2} \ket{\psi}^{\text{tot}}.
\end{equation}

It is also useful to study the symmetries related to permutations between the $p$ and $k$ momenta. Let us define the transformations: 
\be \label{MomentumSwaps}
 \sigma_p : \{p_1,p_2\} \leftrightarrow \{p_2,p_1\}\;, \sigma_k :\{k_1,k_2\} \leftrightarrow \{k_2,k_1\}\;,
 \sigma_{pk1} : \{p_1,p_2\} \leftrightarrow \{k_1,k_2\} \;.
\ee
(with the map  $\sigma_{pk2}: \{p_1,p_2\} \leftrightarrow \{k_2,k_1\}$ being a composition, e.g.  $\sigma_{pk2} =\sigma_{k} \circ \sigma_{pk_1} \circ \sigma_p$). We can now consider how these maps act on different parts of the wavefunctions, for example:
\begin{equation}
    \begin{aligned}
        &\sigma_{p}\psi_{oe}(\ell_1,\ell_2) = - \psi_{oe}(\ell_1,\ell_2)\\
        &\sigma_{k}\psi_{oe}(\ell_1,\ell_2) = - \psi_{oe}(\ell_1,\ell_2)\\
        &\sigma_{pk1}\psi_{oe}(\ell_1,\ell_2) = - \psi_{oe}(\ell_1,\ell_2)\\
        &\sigma_{pk2}\psi_{oe}(\ell_1,\ell_2) = - \psi_{oe}(\ell_1,\ell_2).
    \end{aligned}
\end{equation}
In Table \ref{Symmetries}, we summarise the action of the $\Zset_2$ maps, as well as the momentum permutation maps, on the $A$ and $B$ solutions.

\begin{table}[h]
\begin{center}
\begin{tabular}{ |p{2.8cm}|p{2.5cm}|p{2.5cm}|p{2.5cm}|}
\hline
{\textbf{Transformation}} &\multicolumn{3}{|c|}{\textbf{Wavefunction}} \\
\hline
 & $|\psi \rangle_{(r)}$ & $|\psi \rangle_{(A)}$ & $|\psi \rangle_{(B)}$ \\
\hline
$\mathbb{Z}_2$ & $-z_{(r)}|\psi \rangle_{(r)}$ & $-z_{(g)}|\psi \rangle_{(B)}$ & $-z_{(g)}|\psi \rangle_{(A)}$ \\
\hline
$\sigma_{p}$ & $-|\psi \rangle_{(r)}$ & $-|\psi \rangle_{(A)}$ & $-|\psi \rangle_{(B)}$ \\
\hline
$\sigma_{k}$ & $\cdot$ & $-|\psi \rangle_{(A)}$ & $-|\psi \rangle_{(B)}$ \\
\hline
$\sigma_{pk,1}$ & $\cdot$ & $-|\psi \rangle_{(A)}$ & $-|\psi \rangle_{(B)}$ \\
\hline
$\sigma_{pk,2}$ & $\cdot$ & $-|\psi \rangle_{(A)}$ & $-|\psi \rangle_{(B)}$ \\
\hline
$\sigma_{p} \circ \sigma_{k}$ & $\cdot$ & $+|\psi \rangle_{(A)}$ & $+|\psi \rangle_{(B)}$\\
\hline
$\sigma_{p} \circ \sigma_{pk,1}$ & $\cdot$ & $+|\psi \rangle_{(A)}$ & $+|\psi \rangle_{(B)}$\\
\hline
$\sigma_{p} \circ \sigma_{pk,2}$ & $\cdot$ & $+|\psi \rangle_{(A)}$ & $+|\psi \rangle_{(B)}$ \\
\hline
$\sigma^n$ &  & $(-1)^{n}|\psi \rangle_{(A)}$ & $(-1)^{n}|\psi \rangle_{(B)}$\\
\hline
\end{tabular}
\end{center}
\caption{\it A summary of the transformations of the wavefunctions considered in the text under the discrete symmetries of the problem. Here $\ket{\psi}_{(r)}$ is the restricted solution (\ref{restricted}) while the two general solutions are as given in (\ref{XY2magnonadditional}) with the coefficients in (\ref{FourtermA}),(\ref{FourtermB}) respectively. The respective $\Zset_2$ eigenvalues are
$z_{r} = (r(p_1)r(p_2))^{-2}(r(k_1)r(k_2))^{-1}$ and $z_{g} = (r(p_1)r(p_2))^{-2}(r(k_1)r(k_2))^{-2}$. The momentum maps are as defined in (\ref{MomentumSwaps}). ($\sigma^{n}$ means any $n$-composition of the sigma operators).} \label{Symmetries}
\end{table}

To conclude this section, we have exhibited the solution of the 2-magnon problem for the alternating Hamiltonian (\ref{HamiltonianXY}). The general solution includes scattering of the original momenta to another set of momenta which are not related by permutations. This feature, which is characteristic of staggered-type chains, can be thought of in terms of the \emph{discrete} diffractive framework discussed in \cite{Bibikov_2016} and, at least for the cases discussed there, is not by itself an impediment to solvability. Therefore, even though our system does exhibit diffractive scattering, and would appear to fail Sutherland's criterion of quantum integrability \cite{SutherlandBook}  (see also \cite{Lamacraft13}), we believe that it is important to understand how far one can extend the Bethe ansatz framework for the study of the three- and higher- magnon problem, and in particular whether a discrete-diffractive ansatz can play a role in its solution.\footnote{Sutherland's criterion, that a system is quantum integrable if all scattering processes are non-diffractive (meaning that the set of initial momenta is the same as the set of final momenta), applies to the three- and higher-magnon problem, while here we already see additional momenta at the 2-magnon level. We can expect, of course, that the additional momenta will be a feature of the higher-magnon problems as well.} We comment further on the applicability of Sutherland's criterion in the conclusions.

\subsubsection{Restricted  solution} \label{sec:RestrictedXY}

As mentioned in the previous section, the $k$-momenta are generally complex valued. This can lead to divergences coming from the exponents $\text{exp}(ik_1\ell_1+ik_2\ell_2)$ and $\text{exp}(ik_2\ell_1+ik_1\ell_2)$ in the wavefunctions. In fact, there are two ways this can happen: in the case of an infinite length spin chain, one of these terms will exponentially increase as the relative distance $|\ell_2 - \ell_1| \rightarrow \infty$; and, when considering the CoM case $K = 0$, the imaginary parts of $k_1,k_2$ diverge to complex infinity which again leads to exponentially increasing/decaying terms even for finite length chains. Thus, in these cases, the four term solution that we found in section \ref{sec:TwomagXY}  is not applicable. However, it is possible to take an appropriate linear combination of the two $\Zset_2$ conjugate solutions which removes the unwanted term. 

Let us write $k_1 = K/2 + \pi/2 - iv$, $k_2 = K/2 - \pi/2 + iv$, $v\geq 0$. Since in the wavefunction we have $\ell_2>\ell_1$, the term with swapped $k$ -momenta is the one leading to the divergence in the cases described above: $|\text{exp}(ik_2\ell_1+ik_1\ell_2)| \sim |e^{v(\ell_2-\ell_1)}| \rightarrow \infty$.

So let us try to solve the eigenvalue problem while imposing the boundary condition that there are no divergences at infinite relative distance. This requires that the swapped $k$ coefficients be set to zero. The ansatz will still be of the form (\ref{XY2magnonadditional}), however now each term will consists of just three Bethe wavefunctions:\footnote{To be clear, we could have chosen to remove any one of the four momentum pairs $(p_1,p_2),(p_2,p_1),(k_1,k_2),(k_2,k_1)$ from the ansatz. The specific choice of removing $(k_2,k_1)$ is the one that is relevant when we start with real $p_1,p_2$ and demand $p_1+p_2\ra 0$.} 
\begin{equation} \label{restrictedAnsatz}
    \begin{aligned}
\psi_{ee}^{(r)}(\ell_1, \ell_2) &= A_{ee}(p_1,p_2)e^{ip_1\ell_1+ip_2\ell_2} + A_{ee}(p_2,p_1)e^{ip_2\ell_1 + ip_1\ell_2}\\
        &+ A_{ee}(k_1,k_2)e^{ik_1\ell_1+ik_2\ell_2},\\
        \psi_{oe}^{(r)}(\ell_1, \ell_2) &= A_{oe}(p_1,p_2)e^{ip_1\ell_1+ip_2\ell_2} + A_{oe}(p_2,p_1)e^{ip_2\ell_1 + ip_1\ell_2}\\
        &+ A_{oe}(k_1,k_2)e^{ik_1\ell_1+ik_2\ell_2},\\
        \psi_{eo}^{(r)}(\ell_1, \ell_2) &= A_{eo}(p_1,p_2)e^{ip_1\ell_1+ip_2\ell_2} + A_{eo}(p_2,p_1)e^{ip_2\ell_1 + ip_1\ell_2}\\
        &+ A_{eo}(k_1,k_2)e^{ik_1\ell_1+ik_2\ell_2},\\
        \psi_{oo}^{(r)}(\ell_1, \ell_2) &= A_{oo}(p_1,p_2)e^{ip_1\ell_1+ip_2\ell_2} + A_{oo}(p_2,p_1)e^{ip_2\ell_1 + ip_1\ell_2}\\
        &+ A_{oo}(k_1,k_2)e^{ik_1\ell_1+ik_2\ell_2}.
    \end{aligned}
\end{equation}
We will call this the \emph{restricted} solution, and we will see that it is still possible to solve all the equations with this restriction.\footnote{Since we have established that, apart from the CoM case, there is no solution if we drop the $k$ momenta completely, this ansatz also contains the \emph{minimal} number of terms needed to solve the interacting equations.}. The solution is as follows:
\begin{equation} \label{restricted}
    \begin{aligned}
        &A_{ee}(p_1,p_2) = a(k_2,k_1)b(p_1,p_2) - b(k_2,k_1)a(p_1,p_2)\;,\\
        &A_{ee}(p_2,p_1) = -\big(a(k_2,k_1)b(p_2,p_1) - b(k_2,k_1)a(p_2,p_1) \big)\;,\\
        &A_{ee}(k_1,k_2) = -\big(a(p_2,p_1)b(p_1,p_2) - b(p_2,p_1)a(p_1,p_2) \big).
    \end{aligned}
\end{equation}
We observe that the $A$'s here are quadratic in the $a,b$-coefficients, unlike the general (\ref{FourtermA}),(\ref{FourtermB}) which are cubic.

Let us call $\ket{\psi}_{(r)}=\ket{p_1,p_2;k_1,k_2}_{(r)}$ the solution obtained by substituting (\ref{restricted}) in (\ref{XY2magnonadditional}). 
In the above we found this solution by directly solving the interacting equations with the restriction of no swapped $k$ momenta. But of course we should be able to recover this result from the four-term solution in equation (\ref{eq:XY4lincom}) by tuning the $\alpha,\ \beta$ coefficients in such a way that the swapped  $k$-momenta terms vanish. To show how this works, we set $\alpha = \alpha'/a(k_1,k_2)$ and $\beta = \beta'/b(k_1,k_2)$  in (\ref{eq:XY4lincom}). Then, for instance for $\psi_{ee}$ we can write
{\small
\begin{equation}
    \begin{aligned}
        &\alpha\  \psi^{A}_{ee}(\ell_1,\ell_2) + \beta\  \psi^{B}_{ee}(\ell_1,\ell_2)\\
        &= [a(k_2,k_1)b(k_1,k_2) - b(k_2,k_1)a(k_1,k_2)](\frac{\alpha'}{a(k_1,k_2)}a(p_1,p_2) + \frac{\beta'}{b(k_1,k_2)}b(p_1,p_2))e^{ip_1\ell_1+ip_2\ell_2}\\
        &-[a(k_2,k_1)b(k_1,k_2) - b(k_2,k_1)a(k_1,k_2)](\frac{\alpha'}{a(k_1,k_2)}a(p_2,p_1) + \frac{\beta'}{b(k_1,k_2)}b(p_2,p_1))e^{ip_2\ell_1+ip_1\ell_2}\\
        &-[a(p_2,p_1)b(p_1,p_2) - b(p_2,p_1)a(p_1,p_2)](\frac{\alpha'}{a(k_1,k_2)}a(k_1,k_2) + \frac{\beta'}{b(k_1,k_2)}b(k_1,k_2))e^{ik_1\ell_1+ik_2\ell_2}\\
        &+[a(p_2,p_1)b(p_1,p_2) - b(p_2,p_1)a(p_1,p_2)](\frac{\alpha'}{a(k_1,k_2)}a(k_2,k_1) + \frac{\beta'}{b(k_1,k_2)}b(k_2,k_1))e^{ik_2\ell_1+ik_1\ell_2}\\.
    \end{aligned}
\end{equation}
}
The swapped $k$-momenta term vanishes if we take 
\begin{equation}
    \frac{\alpha'}{\beta'} = -\frac{a(k_1,k_2)}{a(k_2,k_1)}\frac{b(k_2,k_1)}{b(k_1,k_2)}.
\end{equation}
The same is true for the $eo,oe$ and $oo$ wavefunctions. Substituting back and rearranging, we can express the restricted solution as
\begin{equation} \label{Restrictedlc}
    \begin{aligned}
       |\psi\rangle_{(r)} = \frac{1}{a(k_1,k_2)}\frac{c}{1+c}\ |\psi \rangle_{A} + \frac{1}{b(k_1,k_2)}\frac{1}{1+c}\ |\psi \rangle_{B},
    \end{aligned}
\end{equation}
where
\begin{equation}
    c= -\frac{b(k_2,k_1)}{b(k_1,k_2)}\frac{a(k_1,k_2)}{a(k_2,k_1)}.
\end{equation}

As mentioned, this is the solution that we expect to describe the 2-magnon problem for infinite chains. Also, it will play an important role in Appendix \ref{XYCoMlimit} as it will be the starting point for taking the centre-of-mass frame limit $K\ra 0$.

Let us take a closer look at the $S$-matrices appearing in the restricted solution, which we call  $S^{(r)}$ and $T^{(r)}$. They are
\begin{equation}
    \begin{aligned}
        S^{(r)}(p_1,p_2;k_1,k_2;\kappa) = \frac{A_{ee}(p_2,p_1)}{A_{ee}(p_1,p_2)}=-\frac{a(k_2,k_1)b(p_2,p_1) - b(k_2,k_1)a(p_2,p_1)}{a(k_2,k_1)b(p_1,p_2) - b(k_2,k_1)a(p_1,p_2)}
    \end{aligned}
\end{equation}
and
\begin{equation}
    \begin{aligned}
        T^{(r)}(p_1,p_2;k_1,k_2;\kappa) = \frac{A_{ee}(k_1,k_2)}{A_{ee}(p_1,p_2)}=-\frac{a(p_2,p_1)b(p_1,p_2) - b(p_2,p_1)a(p_1,p_2)}{a(k_2,k_1)b(p_1,p_2) - b(k_2,k_1)a(p_1,p_2)}
    \end{aligned}
\end{equation}
Note in particular that the $S$-matrix $S^{(r)}$ swapping $p_1 \leftrightarrow p_2$ now also depends on the $k$ momenta. It still satisfies the unitary condition, $S^{(3)}(p_1,p_2,k_1,k_2)S^{(3)}(p_2,p_1,k_1,k_2) = 1$, as well as the condition $S^{(3)}(p_1,p_1,k_1,k_2) = -1$. (Note that $p_2=p_1$ does not imply that $k_2=k_1$). It also smoothly reduces to the $XXX$ $S$-matrix as $\kappa\ra 1$. Importantly, it is $\Zset_2$ invariant:
\be
S^{(r)}(p_1,p_2;k_1,k_2;1/\kappa) = S^{(r)}(p_1,p_2;k_1,k_2;\kappa)
\ee
The importance of this is that, as we will see in Appendix \ref{XYCoMlimit}, in the limit $K\ra 0$  $S^{(r)}$ will become the centre-of-mass $S$-matrix (\ref{SCoMXY}), which determines the CoM spectrum of our system and thus needs to be $\Zset_2$ invariant.  

Similarly to the analysis carried out for the general solution, it can be shown that the full restricted solution $|\psi \rangle_{(r)}$ is a $\mathbb{Z}_2$ eigenstate
\begin{equation}
    \mathbb{Z}_2|\psi \rangle_{(r)} = -\big(r(p_1)r(p_2)\big)^{-2}\big(r(k_1)r(k_2)\big)^{-1} |\psi \rangle_{(r)}.
\end{equation}
Of course, unlike the general solution, the restricted solution only has the $\sigma_p$ momentum exchange symmetry, as the boundary condition has broken the symmetry between the four sets of momenta $(p_1,p_2),(p_2,p_1),(k_1,k_2),(k_2,k_1)$.

\subsection{Short closed chains} \label{sec:ClosedXY}

In this section we show how our 2-magnon solution above can be used to find the spectrum for closed chains, and check the answer against the explicit diagonalisation of the system for a specific example of a length-six chain. Note that in the $XY$ sector only even-length closed chains are allowed, as otherwise the gauge indices of the first and last sites cannot be matched.\footnote{Also $L$ needs to be larger than 2, as for $L=2$ double-trace contributions which are otherwise subleading in the planar limit become important \cite{Gadde:2010zi}.}  

As explained, we are not only interested in solutions satisfying the zero-momentum constraint, $K=0$, but also in solutions with nonzero $K$. This is because we wish to think of our solutions as part of an eventual multi-magnon solution, which does satisfy the momentum constraint.

Starting with the 1-magnon momenta, in the untwisted sector they are simply obtained by imposing periodicity, $e^{iLp}=1$. Even though they are of course equal to their corresponding values for the XXX limit at $\kappa=1$, the energies are different due to the dispersion relation (\ref{XYdispersion1}). Since the $S$-matrices of the general 2-magnon solution (\ref{SAmatrixXY}),(\ref{SBmatrixXY}) satisfy $S(p,0)=1$, we can always construct periodic 2-magnon wavefunctions by taking e.g. $p_2=0$ and $p_1$ to be a 1-magnon momentum. Thus the 1-magnon energies are 2-magnon energies as well. 

Proceeding to ``true'' 2-magnon energies, we saw above that the solution of the 2-magnon problem for our alternating chain requires the addition of a second set of momenta, $k_1,k_2$, on top of the original set of momenta $p_1,p_2$. These momenta are uniquely defined by the relations
\be \label{kdefinition}
p_1+p_2=k_1+k_2=K \;\; \text{and} \;\; E(p_1)+E(p_2)=E(k_1)+E(k_2)=E_2
\ee
with $E(p)$ the 1-magnon dispersion relation (which can belong to either the upper or lower branch).

We also saw that when going to the centre-of-mass frame where $K=0$, the $k$ momenta diverge and one needs to take a subtle limit to keep the wavefunction finite. However, once the limit is taken (the details are in Appendix \ref{XYCoMlimit}) one is left with the centre-of-mass $S$-matrix (\ref{SCoMXY}) for the $p$ momenta which allows us to write a standard 2-magnon Bethe ansatz.\footnote{We emphasise that the $p$ and $k$ momenta appear symmetrically in the wavefunction so which ones we call $p$ and $k$ is a matter of notation. We call $p$ the solutions of (\ref{kdefinition}) which stay finite in the CoM limit.}
For the untwisted sector, the periodicity equation coming from the Bethe ansatz is much like the usual one and reads:
\be 
e^{ipL}=1/S(p,-p)
\ee
with $S$ as in (\ref{SCoMXY}). Solving this equation numerically for fixed $L$, one can confirm that the momenta and corresponding energies agree with an explicit diagonalisation of the Hamiltonian. Since $S\ra e^{-ip}$ in the limit $\kappa\ra1$, the momenta reduce to those of the standard XXX model in this limit. In Table \ref{LengthSixXY} we tabulate the $K=0$ momenta for $L=6$, with the corresponding XXX momenta for comparison. 

Now recall that in the orbifold limit we also need to consider the twisted sector states \cite{Wang:2003cu,Ideguchi:2004wm,Beisert:2005he,Solovyov:2007pw}. These are spin chains which include the twist matrix $\gamma$ somewhere on the chain. Without loss of generality it can be taken to be the first or last site, and its effect is to modify the periodicity condition by a phase. In the $\Zset_2$ case this phase is just $-1$, which means that to get the full spectrum (still at the orbifold limit) one needs to consider also antiperiodic chains. Precisely the same is true in the interpolating theory. For 1-magnon states we simply impose $e^{ipL}=-1$, which gives the correct 1-magnon energies (and trivially some 2-magnon energies where one of the momenta is zero). For the 2-magnon twisted sector states we use the ansatz
\be
e^{ipL}=-1/S(p,-p)
\ee
with $S$ again as in (\ref{SCoMXY}), and find agreement with explicit diagonalisation of the Hamiltonian for an antiperiodic alternating chain. The $K=0$ 2-magnon momenta for an $L=6$ twisted chain are tabulated in Table \ref{LengthSixXYtwisted}. 

However, to get the full set of 2-magnon energies we also need to consider the case where $K\neq 0$. In this case our 2-magnon wavefunction is a linear combination of Bethe-like states, some of which depend on the $p$ and some on the $k$ momenta. Imposing periodicity on such a wavefunction requires additional steps which we outline below.

First of all, it is clear that one cannot impose periodicity on the $A$ or $B$ solutions (\ref{FourtermA},\ref{FourtermB}) separately. As in the CoM case, one needs to combine them. We will thus write
\be
\ket{\psi}_{\text{tot}}=A(p_1,p_2,k_1,k_2)+x B(p_1,p_2,k_1,k_2)
\ee
with $x$ a function which in principle depends on all the momenta. Now recall that in the standard XXX case, where we just have
\be
\ket{l_1,l_2}=A_{12}(p) e^{il_1 p_1+il_2 p_2}+A_{21}(p) e^{il_1 p_2+i l_2 p_1}
\ee
the periodicity requirement ($\ket{l_1+L,l_2}=\ket{l_1,l_2}$)  is imposed as
\be
A_{12}(p) e^{il_1p_1+i l_2 p_2}=A_{21}(p) e^{i l_2 p_2+i (l_1+L) p_1} \;\Rightarrow\;\; e^{iL p_1}=\frac{A_{12}(p)}{A_{21}(p)}
\ee
leading to the usual $S$-matrix $S_{12}=A_{21}/A_{12}$. It turns out that the only modification required in this case is the introduction of the ratio $x$. In particular, we will impose periodicity separately on the $p$ and $k$ momenta as
\be \label{periodiceeXY}
\begin{split} 
(A^{\text{ee},p}_{12}&+x B^{\text{ee},p}_{12})e^{i(l_1p_1+l_2p_2)}=(A^{\text{ee},p}_{21}+x B^{\text{ee},p}_{21}) e^{i (l_2p_2+(l_1+L) p_1)}\;,\quad\\ 
(A^{\text{ee},k}_{12}&+x B^{\text{ee},k}_{12})e^{i(l_1k_1+l_2k_2)}=(A^{\text{ee},k}_{21}+x B^{\text{ee},k}_{21}) e^{i (l_2k_2+(l_1+L) k_1)}\;,
\end{split}
  \ee
for the even-even parts. We also impose the same condition on the odd-odd parts, while for the even-odd and odd-even  ones we have
\be\begin{split}
(A^{\text{eo},p}_{12}&+x B^{\text{eo},p}_{12})e^{i(l_1p_1+l_2p_2)}=(A^{\text{oe},p}_{21}+x B^{\text{oe},p}_{21}) e^{i (l_2p_2+(l_1+L) p_1)}\;,\quad\\ 
(A^{\text{eo},k}_{12}&+x B^{\text{eo},k}_{12})e^{i(l_1k_1+l_2k_2)}=(A^{\text{oe},k}_{21}+x B^{\text{oe},k}_{21}) e^{i (l_2k_2+(l_1+L) k_1)}\;.
\end{split}
\ee
Periodicity of the wavefunction hinges on the $x$ factor in all of these relations being the same. Focusing on the even-even sector for now,we can simplify (\ref{periodiceeXY}) as 
\be
e^{iLp_1}=-\frac{a(p_1,p_2)+x~ b(p_1,p_2)}{a(p_2,p_1)+x~ b(p_2,p_1)} \;\quad\text{and}\quad
e^{iLk_1}=-\frac{a(k_1,k_2)+x~ b(k_1,k_2)}{a(k_2,k_1)+x~ b(k_2,k_1)}
\ee
where we have used the explicit expressions of $A_{12},A_{21},B_{12},B_{21}$ in terms of the $a$ and $b$ functions defined in (\ref{abcoeffsXY}). We can now solve for $x$ in terms of one set of momenta (the $k$'s, for instance), to write
\be
\bar{x}(k_1,k_2)=-\frac{a(k_1,k_2)+a(k_2,k_1)e^{iL k_1}}{b(k_1,k_2)+b(k_2,k_1)e^{iLk_1}}
\ee
Substituting this specific solution for $x$ (which we call $\bar{x}$) into the $p$ periodicity condition gives us the final form of the 2-magnon Bethe equation for the untwisted $XY$ sector:
\be \label{Bethefinal}
e^{iLp_1}=-\frac{a(p_1,p_2)+\bar{x}(k_1,k_2) b(p_1,p_2)}{a(p_2,p_1)+\bar{x}(k_1,k_2) b(p_2,p_1)}
\ee
The above was for the even-even part but similar considerations apply to the other parts of the wavefunction and lead to the same equation. Now recall that the $k$ momenta are algebraically defined in terms of the $p$ momenta by solving (\ref{kdefinition}), which is a quartic relation between $z(p_i)=e^{2ip_i}$ and $z(k_i)=e^{2ik_i}$ and can be straightforwardly solved. Equation (\ref{Bethefinal}) can thus be expressed purely in terms of the $p$ momenta, and its roots found numerically for given $L$ and $\kappa$. For short chains, these energies can be seen to match the eigenvalues found by explicit diagonalisation. As an example, in Table \ref{LengthSixXY} we show, for the untwisted sector, the momenta and corresponding energies for $L=6$ and $\kappa=0.9$, as compared to those of the XXX model of the same length. The value $\kappa=0.9$ is chosen since most of the $p$ momenta are still close enough to those for XXX to directly compare them. As $\kappa$ moves farther from 1, the general pattern is that more $k$ momenta which were real in the XXX limit will acquire imaginary parts.

\begin{table}[ht]
\begin{center} {\small \begin{tabular}{|ccc|cccc|} \hline \rowcolor{blue!15}
$K$ & $p_1^{XXX}$ &$ E^{XXX}$ & $p_1$ & $k_1$  & $\bar{x}$ & $E_{\kappa=0.9}$\\ \hline
$0$ &$1.25664$ & $2.76393$  & $1.28911$ & $\pi/2-i\infty$ &$1.69144$ &$2.83287$ \\
 $0$& $2.51327$ & $7.23607$ &$ 2.51285 $& $\pi/2-i\infty$ &$1.0834$ &$7.28474$ \\  
$\pi/3$&$2.38724$ & $5$ & $2.36719$ & $0.032045$&$0.761787$ &$ 4.95642$ \\
$2\pi/3$&$\scriptstyle\frac{\pi}3+0.732858i$ &$1.43845$ & $\scriptstyle\frac{\pi}3-0.7431i $&$ 0.22146$&$0.744253$ &$ 1.42822$\\
$2\pi/3$&$-2.76928$ & $5.56155$ &$\scriptstyle-2.76928+0.102878i$ &$\scriptstyle-2.76928-0.102878i $ &$9.34587$& $5.68203$ \\
$\pi$ &$\pi/2+i\infty$ & $2$&$\pi/2+i\infty$ &$1.0386$ &$0.545701$& $1.94906$ \\\hline
  \end{tabular}}
  \caption{{\bf Length-6, $\kappa\!=\!0.9$ Untwisted $XY$-Sector Chain:} \it The momenta and energies appearing in the untwisted $XY$ 2-magnon problem for the case $L=6,\kappa=0.9$. $K$ is the total momentum. On the left we list the $XXX$ ($\kappa=1$) momenta and energies for comparison. These can be found for instance in \cite{Karabachetal} (however, note that the XXX limit of our Hamiltonian has an extra factor of 2 compared to the usual XXX normalisation, so our energies are also rescaled by 2 relative to \cite{Karabachetal}). The $p_2,k_2$ momenta can be found as $p_2=K-p_1$, $k_2=K-k_1$. For the deformed theory we have chosen the $p$ momenta to be the ones closest to the corresponding momenta for $XXX$. $\bar{x}$ is the ratio between the $A$ and $B$ solutions, as discussed in the text. The energies included here are the ``true'' 2-magnon energies, i.e. those where both $p_1$ and $p_2$ in the XXX limit are nonzero.} \label{LengthSixXY}
 \end{center}   \end{table}

An intriguing feature of the Bethe solutions in Table \ref{LengthSixXY} is that the role of the $p$ and $k$ momenta appears to be exchanged at total momentum $K=0$ and $K=\pi$. In particular, at $K=0$ the $k$ momenta diverge and, as shown in Appendix \ref{XYCoMlimit}, they leave a relic in the wavefunction in the form of contact terms. At $K=\pi$ the $p$ momenta diverge while the $k$ momenta remain finite, and in this case there will be contact terms arising from the $p$ momenta (note that equations (\ref{eq:constraint1}) and (\ref{eq:constraint2}) are compatible also at $K=\pi$). It would be interesting to understand this duality better.

For the twisted sector everything works very similarly, apart from the fact that we need to impose antiperiodicity. This leads to the ratio between the two wavefunctions being:
\be
\bar{x}(k_1,k_2)=-\frac{a(k_1,k_2)-a(k_2,k_1)e^{iL k_1}}{b(k_1,k_2)-b(k_2,k_1)e^{iLk_1}}\;,
\ee
and the corresponding 2-magnon Bethe ansatz is now written in terms of this $\bar{x}$ as
\be
e^{iLp_1}=\frac{a(p_1,p_2)+\bar{x}(k_1,k_2)b(p_1,p_2)}{a(p_2,p_1)+\bar{x}(k_1,k_2)b(p_2,p_1)}\;.
\ee
This is again a function only of the $p$ momenta which can be solved numerically to obtain the required Bethe momenta. 
In Table \ref{LengthSixXYtwisted} we compare the twisted sector momenta and energies of the $L=6$ $\kappa=0.9$ chain to those of the antiperiodic XXX model. The energies given all agree perfectly with the explicit diagonalisation. In this case there are some degeneracies (e.g. for energies $E=2$ and $E=4$ in the twisted XXX limit) which we haven't attempted to resolve (as our interest is in the overall features of the Bethe ansatz) however it is clear that they can be straightforwardly reproduced with a more careful treatment.

\begin{table}[ht] 
\begin{center}  \begin{tabular}{|ccc|cccc|} \hline \rowcolor{blue!15}
$K$ & $p_1^{_{XXX}}$ & $E_{{}^{XXX}}$ &$p_1$ & $k_1$  &$\bar{x}$ &$E_{\kappa=0.9}$\\ \hline
    $0$&$1.88496$ & $5.23607$ &$ 1.83806 $& $\frac{\pi}2-i\infty$ &$5.97801$ &$5.15985$ \\
    $0$ &$0.628319$ & $0.763932$ &$ 0.627525$& $\frac{\pi}2-i\infty$ &$0.711862$ &$0.756877$ \\
    
$\pi/3$ &$3.02739$ & $6.78309$ & $3.02448$ & $2.0944$ &$1.20546$ & $6.83818$ \\
$\pi/3$ &$1.73909$ & $2.79493$ & $1.73092$ & $1.16646$ &$1.19773$ & $2.83988$ \\
$\pi/3$ &$\scriptstyle\frac{\pi}6+0.255761i$ & $0.421979$& $\scriptstyle\frac{\pi}6+0.256362i$ & $\scriptstyle\frac{2\pi}3-1.19117i$ &$0.82377$ & $0.419343$ \\
    $2\pi/3$ &$2.82661$ &$4.41421$ & $2.8254 $&$0.926182$&$0.930317$ &$ 4.43087$\\
    $2\pi/3$&$\scriptstyle\frac{\pi}3+0.632974i$ & $1.58579$ & $\scriptstyle\frac{\pi}3+0.654018i$&$1.81499$&$0.611008$ &$1.56061$ \\
\hline
  \end{tabular}
  \caption{{\bf Length-6, $\kappa\!=\!0.9$ Twisted $XY$-Sector Chain:} \it Some momenta appearing in the twisted $XY$ sector 2-magnon problem for the case $L=6,\kappa=0.9$. $K$ is the total momentum. For reference, on the left we list the twisted (antiperiodic) XXX momenta and energies. The $p_2,k_2$ momenta can be found as $p_2=K-p_1$, $k_2=K-k_1$. We have chosen the $p$ momenta for the interpolating theory to be the ones that are closest to the corresponding momenta for twisted $XXX$. $\bar{x}$ is the ratio between the $A$ and $B$ solutions, as discussed in the text.} \label{LengthSixXYtwisted}
 \end{center}   \end{table}

In the tables we also give the factor $\bar{x}$ defined in each case above. Note that $\bar{x}$ can be defined through either the $p$ or the $k$ momenta, so it has a specific value also when one of the two sets $(p_1,p_2)$ or $(k_1,k_2)$ diverges. It should perhaps be emphasised that, in taking the limit e.g. $k_{1,2}\ra \pi/2\mp i\infty$ (as discussed in Appendix \ref{XYCoMlimit}), one needs to use the functional dependence of the $k$ on the $p$ momenta, which involves the energy $E_2$. Therefore, the ratio $\bar{x}$ can take different limiting values as $k_{1,2}\ra \pi/2\mp i\infty$ for different energies, as seen in the tables.

It is worth noting that since the $\Zset_2$ operation $\kappa\ra 1/\kappa$ exchanges the $a$ and $b$ factors, it takes $\bar{x}(k_1,k_2)\ra 1/\bar{x}(k_1,k_2)$. This is what is required for the full wavefunction $\ket{\psi}_{\text{tot}}$ to be invariant under $\Zset_2$, which should be the case since the Hamiltonian is also $\Zset_2$ invariant in the periodic/antiperiodic case.\footnote{The $\kappa\ra1/\kappa$ operation can be undone by shifting the origin from an even to an odd site, which does not have an effect for a closed chain.}

\section{The ``dilute'' XZ sector} \label{sec:XZsector}

In this section we will study the ``$\SU(2)$-like'' sector composed of the $X$ and $Z$ fields. Unlike the $XY$ sector discussed previously, in the $XZ$ sector the two fields belong to different representations of the gauge groups ($X$ being bifundamental while $Z$ in the adjoint of each of the groups) so we expect the theory to be less symmetric than the $XY$ sector.

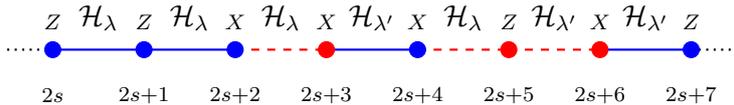
\begin{figure}[t]
  \begin{center}
  \begin{tikzpicture}[scale=0.6]
    \draw[-,thick,blue] (4,0)--(6,0);
    \draw[-,thick,blue] (6,0)--(8,0);
    \draw[-,thick,red,dashed] (8,0)--(10,0);
    \draw[-,thick,blue] (10,0)--(12,0);
    \draw[-,thick,red,dashed] (12,0)--(14,0);
    \draw[-,thick,red,dashed] (14,0)--(16,0);
    \draw[-,thick,blue] (16,0)--(18,0);
    
\draw[-,thick,dotted] (3,0)--(4,0);\draw[-,thick,dotted] (18,0)--(19,0);

\filldraw[blue] (4,0) circle (5pt);
    \filldraw[blue] (6,0) circle (5pt);
        \filldraw[blue] (8,0) circle (5pt);
        \filldraw[red] (10,0) circle (5pt);
            \filldraw[blue] (12,0) circle (5pt);
            \filldraw[red] (14,0) circle (5pt);
                \filldraw[red] (16,0) circle (5pt);
        \filldraw[blue] (18,0) circle (5pt);
        \node at (4,0.6){$\scriptstyle{Z}$};
        \node at (6,0.6){$\scriptstyle{Z}$};
        \node at (8,0.6){$\scriptstyle{X}$};
        \node at (10,0.6){$\scriptstyle{X}$};
        \node at (12,0.6){$\scriptstyle{X}$};
        \node at (14,0.6){$\scriptstyle{Z}$};
        \node at (16,0.6){$\scriptstyle{X}$};
        \node at (18,0.6){$\scriptstyle{Z}$};

        \node at (5,0.7) {$\Hcal_{\lambda}$}; 
        \node at (7,0.7) {$\Hcal_{\lambda}$};
        \node at (9,0.7) {$\Hcal_{\lambda}$}; 
        \node at (11,0.7) {$\Hcal_{\lambda'}$};
        \node at (13,0.7) {$\Hcal_{\lambda}$}; 
        \node at (15,0.7) {$\Hcal_{\lambda'}$};
        \node at (17,0.7) {$\Hcal_{\lambda'}$}; 

        \node at (4,-1) {$\scriptstyle 2s$};
        \node at (6,-1) {$\scriptstyle 2s+1$};

        \node at (8,-1) {$\scriptstyle 2s+2$};
        \node at (10,-1) {$\scriptstyle 2s+3$};
        \node at (12,-1) {$\scriptstyle 2s+4$};
        \node at (14,-1) {$\scriptstyle 2s+5$};
        \node at (16,-1) {$\scriptstyle 2s+6$};
        \node at (18,-1) {$\scriptstyle 2s+7$};

  \end{tikzpicture}
  \end{center}
      \caption{\it A section of the dilute spin chain for the XZ sector, for a specific configuration of states. We have chosen the gauge group to the left of site $2s$ to be the first one (depicted by a blue solid line). To make it easier to read off the relevant gauge group for each Hamiltonian, we have also coloured the first of each two sites on which the Hamiltonian acts the same as the region to the left of that site. Unlike the alternating chain depicted in (\ref{fig:alternating}), here the choice of which Hamiltonian acts on each pair of sites is not related to the even-odd nature of the site but rather to the number of $Z$ and $X$ fields that have been crossed until that point. \label{fig:dilute}}
  \end{figure}

Recall that in this sector the Hamiltonian is (\ref{eq:HXZdynamical}):
\be \label{HamiltonianXZ}
\Hcal(\lambda)=\left(\begin{array}{cccc} 0 &0&0&0\\ 0&\kappa &-1&0\\0&-1&1/\kappa&0\\ 0&0&0&0\end{array}\right) \;\;\text{,}\;\;
  \Hcal(\lambda')=\left(\begin{array}{cccc} 0 &0&0&0\\ 0&1/\kappa &-1&0\\0&-1&\kappa&0\\ 0&0&0&0\end{array}\right) \;\; \text{in the basis}
   \;\; \left(\begin{array}{c}XX\\ XZ\\ZX\\ZZ\end{array}\right),
  \ee
  where  $\lambda,\lambda'$  indicate whether the gauge group immediately to the left of the site where the Hamiltonian acts is the first or second gauge group respectively. Since $Z$ does not change the gauge group, the Hamiltonian also does not change when crossing a $Z$ field, however it will switch from $\Hcal(\lambda)$ to $\Hcal(\lambda')$ and vice versa when crossing an $X$ field (see Figure \ref{fig:dilute} for an illustration). 

  A consequence of the $XZ$ sector being less symmetric than the $XY$ sector is that there are two inequivalent vacua, one made up of $X$ fields and the other of $Z$ field. The $Z$ vacuum\footnote{Which is really two vacua, depending on $\lambda$. That is, either $\ket{\cdots \phi_1\phi_1\phi_1\cdots}$ or $\ket{\cdots \phi_2\phi_2\phi_2\cdots}$.} was studied in \cite{Gadde:2010zi}. The dispersion relation of $X$ excitations is trigonometric and the $S$-matrices arising in the 2-magnon problem are similar to those of the XXZ chain.  However, the dynamical nature of the chain becomes evident when considering the 3-magnon problem,  due to the fact that there are two 2-magnon $S$-matrices (mapped to each other by $\Zset_2$) and the two possible sequences of 2-body scatterings (particles 1+2 first, or particles 2+3 first) lead to incompatible equations. This leads to the standard YBE not being satisfied. We can thus think of excitations around the $Z$-vacuum as being described by a dynamical XXZ model. We do not have anything to add to the analysis of \cite{Gadde:2010zi}, so we will not consider the $Z$ vacuum in this work.  

  Instead, we wish to consider $Z$ excitations on a vacuum made up of $X$ fields. In $\Ncal=2$ language they will be of the type $\ket{\cdots Q_{12}Q_{21}\phi_1Q_{12}\phi_2 Q_{21}\cdots}$ Even though the $XZ$-sector Hamiltonian (\ref{HamiltonianXZ}) is rather different to the $XY$-sector Hamiltonian (\ref{HamiltonianXY}), it will turn out that the dispersion relation for the $X$ vacuum of the $XZ$ sector is identical to that of the $XY$ sector. So, even though the details of the wavefunction (such as the ratio functions) will be different, the overall approach to solving the 1- and 2-magnon problem and the final form of the solutions will be almost identical as well. In particular, for the 2-magnon problem, both the contact-term approach in the centre-of-mass frame and the more general additional-momenta approach will apply, as we will show in the following sections. 

\begin{figure}
\begin{center}
\includegraphics[width=12cm]{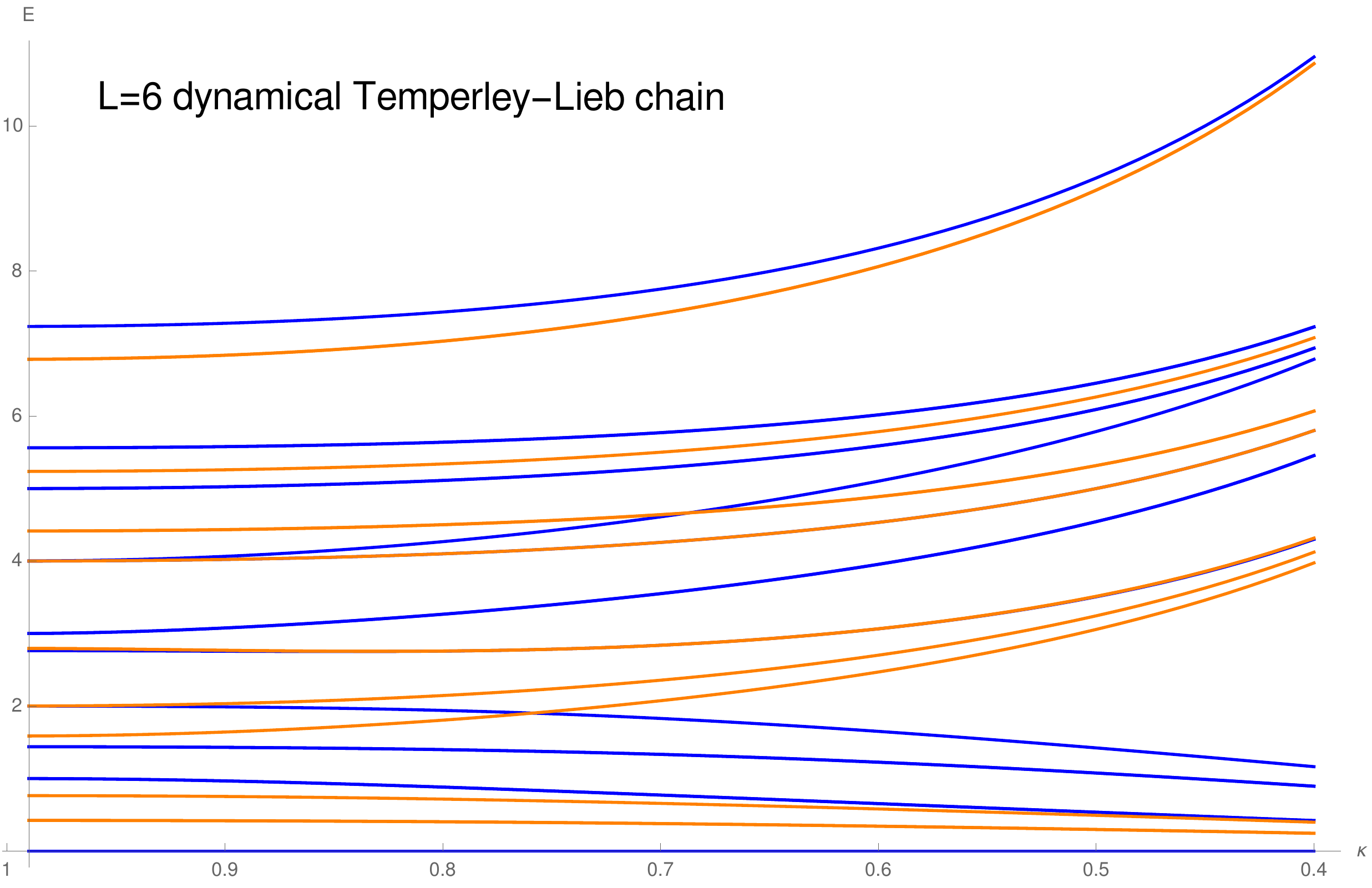}
\caption{\it The 2-magnon spectrum of the dynamical dilute Temperley-Lieb chain of the $XZ$ sector for $L=6$, as a function of the deformation parameter $\kappa$. As discussed in the text, we can only have even numbers of magnons for even-length chains in this sector. Blue denotes untwisted sector energies while orange twisted sector energies. Note the degeneracies at $E=2,4$ at the orbifold point $\kappa=1$, which get resolved as $\kappa$ decreases from 1. We have plotted down to $\kappa=0.4$ for clarity. As $\kappa$ decreases further, the eigenvalues bunch into groups with limiting values of $4/\kappa,2/\kappa$ and $0$.} \label{fig:L6XZ}
\end{center}
\end{figure}

\subsection{One magnon}

We start with the one magnon problem $H|p\rangle = E_1(p)|p\rangle$, where $\ket{p}$ indicates a $Z$ magnon. Due to the dynamical nature of the spin chain, there will be two equations to solve, since a $Z$ excitation on an arbitrary site $\ell$ can see either the parameter $\lambda$ or $\lp$ to its left. Specifically:

\be \label{OnemagXZeq}
\begin{split}
2/\kappa \ket{\ell}_\lambda-\ket{\ell-1}_{\lp}-\ket{\ell+1}_{\lp}=E_1\ket{\ell}_{\lambda}\\
2\kappa \ket{\ell}_{\lp}-\ket{\ell-1}_{\lambda}-\ket{\ell+1}_{\lambda}=E_1 \ket{\ell}_{\lp}
\end{split}
\ee
Here we use the fact that, if a $Z$ field at site $\ell$ sees $\lambda$ to its left, the states with a $Z$ field at site $\ell-1$ and $\ell+1$ necessarily see a $\lp$ to their left, as one needs to cross one less or more $X$ field to get to that site. (Similarly for a site with $\lp$ at site $\ell$). Of course, for a single $Z$ magnon, and given a reference point with fixed $\lambda$, the above equations can also be written in term of even and odd sites. However this is not general so we will avoid using this notation.

We therefore take for our ansatz a superposition of a single excitation on sites that see $\lambda$ or $\lp$:
\begin{equation} \label{OnemagXZ}
    |p\rangle = \displaystyle\sum_{\ell}\ \psi_{\lambda}(\ell)|\ell\rangle_\lambda + \displaystyle\sum_{\ell}\ \psi_{\lp}(\ell)|\ell\rangle_\lp \;,
\end{equation}
where
\begin{equation}
    \psi_\lambda(\ell) = A_\lambda(p)e^{ip\ell}, \quad \psi_{\lp}(\ell) = A_{\lp}(p)e^{ip\ell}.
\end{equation}
In (\ref{OnemagXZ}) the summations are over the sites corresponding to each value of $\lambda$. 

Substituting this ansatz in the equations (\ref{OnemagXZeq}) we find 
\begin{equation}
  2/\kappa A_\lambda e^{ips} - A_\lp e^{ip(s-1)} - A_\lp e^{ip(s+1)} = E_1(p)A_\lambda e^{ips}.
\end{equation}
and 
\begin{equation}
        2\kappa A_\lp e^{ipt} - A_\lambda e^{ip(t-1)} - A_\lambda e^{ip(t+1)} = E_1(p)A_\lp e^{ipt}.
\end{equation}
From these equations, we find that the ratio between $\lp$ and $\lambda$ sites is
\begin{equation}\label{eq:ratio}
    r(p;\kappa):=\frac{A_\lp(p)}{A_\lambda(p)} = \frac{1-\kappa ^2\mp\sqrt{\kappa ^4+2 \kappa ^2 \cos (2 p)+1}}{2 \kappa\cos(p) }, 
\end{equation}
and the eigenvalue is
\begin{equation}
    E_1(p;\kappa) = \kappa+\frac{1}{\kappa}\pm \frac{1}{\kappa}\sqrt{\kappa ^4+2 \kappa ^2 \cos (2 p)+1} \;.
\end{equation}

We see that the dispersion relation is precisely the same as for the $XY$ sector, however the ratio function $r(p)$ is different from the ratio between the odd and even sites defined there. 

 As in the $XY$ sector, the one magnon wavefunction has a $\mathbb{Z}_2$ symmetry. First, note that under $\kappa \leftrightarrow 1/\kappa$, 
\be 
E(1/\kappa) = E(\kappa) \quad \mbox{and} \quad     r(p;1/\kappa) = 1/r(p;\kappa) \, ,
\ee which again implies that the two coefficients $A_\lambda,\ A_\lp$ in our wavefunction are exchanged and that
\be
\mathbb{Z}_2      |p\rangle   =    \frac{1}{r(p;\kappa)}    |p\rangle
\ee

 \subsection{Two magnons}

 Moving on to two magnons, we will again treat separately the non-interacting equations (those where the $Z$ magnons are separated by at least one $X$ field) and the interacting equations (those where the $Z$ magnons are next to each other). There will again be four non-interacting equations as each magnon can see either $\lambda$ or $\lp$. 
{\small
 \be \label{NonInt2magXZ}
 \begin{split}
 &4\kappa^{-1} \ket{\ell_1,\ell_2}_{\lambda\lambda}\!-\!\ket{\ell_1\!-\!1,\ell_2}_{\lp\lambda}\!-\!\ket{\ell_1+1,\ell_2}_{\lp\lambda}\!-\!\ket{\ell_1,\ell_2\!-\!1}_{\lambda\lp}\!-\!\ket{\ell_1,\ell_2+1}_{\lambda\lp}=E_2\ket{\ell_1,\ell_2}_{\lambda\lambda}\;,\\
   &2(\kappa\!+\!\kappa^{-1})\ket{\ell_1,\ell_2}_{\lambda\lp}\!-\!\ket{\ell_1\!-\!1,\ell_2}_{\lp\lp}\!-\!\ket{\ell_1+1,\ell_2}_{\lp\lp}\!-\!\ket{\ell_1,\ell_2\!-\!1}_{\lambda\lambda}\!-\!\ket{\ell_1,\ell_2+1}_{\lambda\lambda}=E_2\ket{\ell_1,\ell_2}_{\lambda\lp}\;,\\
  &2(\kappa\!+\!\kappa^{-1})\ket{\ell_1,\ell_2}_{\lp\lambda}\!-\!\ket{\ell_1\!-\!1,\ell_2}_{\lambda\lambda}\!-\!\ket{\ell_1+1,\ell_2}_{\lambda\lambda}\!-\!\ket{\ell_1,\ell_2\!-\!1}_{\lp\lp}\!-\!\ket{\ell_1,\ell_2+1}_{\lp\lp}=E_2\ket{\ell_1,\ell_2}_{\lp\lambda} \;,\\
   &4\kappa \ket{\ell_1,\ell_2}_{\lp\lp}\!-\!\ket{\ell_1\!-\!1,\ell_2}_{\lambda\lp}\!-\!\ket{\ell_1+1,\ell_2}_{\lambda\lp}\!-\!\ket{\ell_1,\ell_2\!-\!1}_{\lp\lambda}\!-\!\ket{\ell_1,\ell_2+1}_{\lp\lambda}=E_2\ket{\ell_1,\ell_2}_{\lp\lp}  \;.\\
 \end{split}
 \ee
 }
 For the interacting equations, note that two consecutive $Z$ magnons necessarily have the same value of $\lambda$. So we have
 \be \label{Int2magXZ}
 \begin{split}
2 \kappa^{-1} \ket{\ell,\ell+1}_{\lambda\lambda}-\ket{\ell-1,\ell+1}_{\lp\lambda}-\ket{\ell,\ell+2}_{\lambda\lp}=E_2\ket{\ell,\ell+1}_{\lambda\lambda} \;,\\
2\kappa \ket{\ell,\ell+1}_{\lp\lp}-\ket{\ell-1,\ell+1}_{\lambda\lp}-\ket{\ell,\ell+2}_{\lp\lambda}=E_2\ket{\ell,\ell+1}_{\lp\lp}\;.
 \end{split}
 \ee
 
To solve these equations, we will proceed by direct analogy with Section \ref{sec:XYsector}. The main difference is that the eigenvalue equations are slightly different, and that we  do not use even-odd notation in the wavefunction, but $\lambda,\lp$ notation. 
We will start with the following ansatz
\begin{equation} \label{XZBethenaive}
    \begin{split}
\ket{p_1,p_2}=  &\displaystyle\sum_{\ell_1 < \ell_2} \psi_{\lambda\lambda}(\ell_1, \ell_2)|\ell_1\ell_2 \rangle + \displaystyle\sum_{\ell_1 < \ell_2} \psi_{\lambda\lp}(\ell_1, \ell_2)|\ell_1\ell_2 \rangle\\
        &+\displaystyle\sum_{\ell_1 < \ell_2} \psi_{\lp\lambda}(\ell_1, \ell_2)|\ell_1\ell_2 \rangle +\displaystyle\sum_{\ell_1 < \ell_2} \psi_{\lp\lp}(\ell_1, \ell_2)|\ell_1\ell_2 \rangle,
    \end{split}
\end{equation}
where
\begin{equation}
    \begin{aligned}
        & \psi_{\lambda\lambda}(\ell_1, \ell_2) = A_{\lambda\lambda}(p_1,p_2)e^{ip_1\ell_1+ip_2\ell_2},\\
        & \psi_{\lambda\lp}(\ell_1, \ell_2) = A_{\lambda\lp}(p_1,p_2)e^{ip_1\ell_1+ip_2\ell_2},\\
        & \psi_{\lp\lambda}(\ell_1, \ell_2) = A_{\lp\lambda}(p_1,p_2)e^{ip_1\ell_1+ip_2\ell_2},\\
        & \psi_{\lp\lp}(\ell_1, \ell_2) = A_{\lp\lp}(p_1,p_2)e^{ip_1\ell_1+ip_2\ell_2}.
    \end{aligned}
\end{equation}
Here the sites $\ell_1,\ell_2$ are assumed to be such that the $\lambda$ dependence is correct. For instance, if $\ell_2=\ell_1+3$ then the only options are $\psi_{\lambda\lambda}(\ell_1,\ell_1+3)$ or $\psi_{\lp\lp}(\ell_1,\ell_1+3)$, since there are two $X$ fields between the $Z$'s. Fixing $\lambda$ at a given site (e.g. $\ell=1$) will resolve the remaining ambiguity. 

Substituting in the non-interacting equations (\ref{NonInt2magXZ}), we find that the eigenvalue is additive,
\begin{equation}
    E_2(p_1,p_2) = E_1(p_1) + E_1(p_2),
\end{equation}
with the coefficients having the ratios
\begin{equation}\label{eq:straightr}
        \frac{A_{\lp\lp}(p_1,p_2)}{A_{\lambda\lambda}(p_1,p_2)} = r(p_1)r(p_2)\;\;,\quad \frac{A_{\lp\lambda}(p_1,p_2)}{A_{\lambda\lambda}(p_1,p_2)} = r(p_1)\;\;,\quad
        \frac{A_{\lambda\lp}(p_1,p_2)}{A_{\lambda\lambda}(p_1,p_2)} = r(p_2).
\end{equation}
As in the $XY$ sector, the usual Bethe approach of just adding the swapped momenta $(p_1,p_2)\leftrightarrow (p_2,p_1)$ does not lead to a solution of the interacting equations (\ref{Int2magXZ}). So in the next section we will improve the Bethe ansatz by adding contact terms, which as we will see gives a solution in the centre-of-mass frame $K=0$. To obtain more general solutions (for $K\neq 0$) we will need to add a second set of momenta, which we will do in Section \ref{sec:4termXZ}. Since the steps are very similar to that of Section \ref{sec:XYsector}, we will provide fewer details in this section.

 \subsubsection{Centre of mass solution} \label{CoMXZ}

We will start by improving the ansatz (\ref{XZBethenaive}) by adding the swapped momenta as well as contact terms: 
\begin{equation}
    \begin{aligned}
        \psi_{\lambda\lp}(\ell_1, \ell_2) &= A_{\lambda\lp}(p_1,p_2)e^{ip_1\ell_1+ip_2\ell_2} + A_{\lambda\lp}(p_2,p_1)e^{ip_2\ell_1 + ip_1\ell_2},\\
        \psi_{\lp\lambda}(\ell_1, \ell_2) &= A_{\lp\lambda}(p_1,p_2)e^{ip_1\ell_1+ip_2\ell_2} + A_{\lp\lambda}(p_2,p_1)e^{ip_2\ell_1 + ip_1\ell_2},\\
    \end{aligned}
\end{equation}
\begin{equation} \label{ContactAnsatzXZ}
    \begin{aligned}
        \psi_{\lambda\lambda}(\ell_1, \ell_2) &= A_{\lambda\lambda}(p_1,p_2)\big(1 + \delta_{\ell_2,\ell_1+1}\mathcal{A}(p_1,p_2)\big)e^{ip_1\ell_1+ip_2\ell_2} \\
        &+ A_{\lambda\lambda}(p_2,p_1)\big(1 + \delta_{\ell_2,\ell_1+1}\mathcal{A}(p_2,p_1)\big)e^{ip_2\ell_1 + ip_1\ell_2},\\
        \psi_{\lp\lp}(\ell_1, \ell_2) &= A_{\lp\lp}(p_1,p_2)\big(1 + \delta_{\ell_2,\ell_1+1}\mathcal{B}(p_1,p_2)\big)e^{ip_1\ell_1+ip_2\ell_2} \\
        &+ A_{\lp\lp}(p_2,p_1)\big(1 + \delta_{\ell_2,\ell_1+1}\mathcal{B}(p_2,p_1)\big)e^{ip_2\ell_1 + ip_1\ell_2}.
    \end{aligned}
\end{equation}
Note that in this case two neighbouring $Z$ fields have the same value of $\lambda$, so the contact terms enter in the $\psi_{\lambda\lambda}$ and $\psi_{\lp\lp}$ parts of the wavefunction. This is unlike the $XY$ sector, where two neighbouring $Y$ fields are necessarily at even-odd or odd-even sites, so in that case we modified the $\psi_{eo}$ and $\psi_{oe}$ terms. 

\mparagraph{Next-to-nearest neighbour magnons}

\mbox{}

The first step is to ensure that the contact terms do not spoil the non-interacting equations. There are two non-interacting equations where the contact terms enter, the $\lambda\lambda$ and $\lambda'\lambda'$ equations when $\ell_2 = \ell_1 + 2$. Cancelling the contact terms in these equations requires
\begin{equation}
    \begin{aligned}
        &\mathcal{A}(p_1,p_2) = -r(p_1)r(p_2)e^{i(p_1+p_2)}\mathcal{B}(p_1,p_2)\\
        &\mathcal{A}(p_1,p_2) = -r(p_1)r(p_2)e^{-i(p_1+p_2)}\mathcal{B}(p_1,p_2).
    \end{aligned}
\end{equation}
and the same relations for the swapped terms. These are only consistent when  $K = p_1+p_2 = n\pi$. We will focus on $K = 0$ since this is in the first Brillouin zone $(-\pi/2,\pi/2]$. So in the rest of this section we will be in the centre-of-mass frame. Our constraints are:
\begin{equation} \label{ConstraintXZfinal}
    \begin{aligned}
        &\mathcal{A}(p_1,p_2) = -r(p_1)r(p_2)\mathcal{B}(p_1,p_2)\\
        &\mathcal{A}(p_2,p_1) = -r(p_1)r(p_2)\mathcal{B}(p_2,p_1).
    \end{aligned}
\end{equation}

\mparagraph{Interacting equations}

\mbox{}

Substituting (\ref{ConstraintXZfinal}) in the $(\lambda,\lambda)$ and $(\lambda',\lambda')$ interacting equations (\ref{Int2magXZ}), we obtain two $S$-matrices, defined as $S^{\lambda\lambda}=A_{\lambda\lambda}(p_2,p_1)/A_{\lambda\lambda}(p_1,p_2)$ and $S^{\lambda'\lambda'}=A_{\lambda'\lambda'}(p_2,p_1)/A_{\lambda'\lambda'}(p_1,p_2)$. Equating these two $S$-matrices we obtain a relation between  $\mathcal{B}(p_2,p_1)$ and $\mathcal{B}(p_1,p_2)$. It is:
\begin{equation}\label{eq:contTerm}
    \begin{aligned}
        \mathcal{B}(p_2,p_1) = F(p_1,p_2)\mathcal{B}(p_1,p_2) + \mathcal{G}(p_1,p_2),
    \end{aligned}
\end{equation}
where 
\begin{equation}\label{eq:gterm}
    \begin{aligned}
        F(p_1,p_2) = e^{i(p_1-p_2)}\frac{f(p_2,p_1)}{f(p_1,p_2)}, \quad \mathcal{G}(p_1,p_2) = -\frac{1}{r(p_1)r(p_2)e^{ip_2}}\frac{g(p_1,p_2)}{f(p_1,p_2)}
    \end{aligned}
\end{equation}
with
\begin{equation}
    \begin{aligned}
        f(p_1,p_2) &= \kappa  \left(-e^{i \left(p_1+p_2\right)}\right) \left(\text{E}_2-2 \kappa \right) r\left(p_2\right)-\left(\text{E}_2 \kappa -2\right) \left(r\left(p_2\right)+e^{i p_1} \left(\text{E}_2-2 \kappa
        \right)\right)\\
        &+r\left(p_1\right) \left(e^{i \left(p_1+p_2\right)} \left(2-\text{E}_2 \kappa \right)-\left(\text{E}_2-2 \kappa \right) \left(\kappa +e^{i p_1} \left(\text{E}_2 \kappa -2\right)
        r\left(p_2\right)\right)\right)
    \end{aligned}
\end{equation}
and 
\begin{equation}
    \begin{aligned}
        g(p_1,p_2) &= -\kappa  r\left(p_1\right){}^2 \left(-e^{i p_2} \left(-1+e^{2 i p_1}\right) \left(\text{E}_2-2 \kappa \right) r\left(p_2\right)-e^{2 i \left(p_1+p_2\right)}+1\right)\\
        &-e^{i p_1} \left(-1+e^{2 i p_2}\right) r\left(p_1\right)
        \left(-\text{E}_2 \kappa +\kappa  \left(\text{E}_2-2 \kappa \right) r\left(p_2\right){}^2+2\right)\\
        &-r\left(p_2\right) \left(e^{i p_2} \left(-1+e^{2 i p_1}\right) \left(\text{E}_2 \kappa -2\right)+\kappa  \left(-1+e^{2 i
        \left(p_1+p_2\right)}\right) r\left(p_2\right)\right)\;.
    \end{aligned}
\end{equation}
Substituting these expressions back into one of the equations for the $S$-matrix we obtain our final expression:
\begin{equation}\label{eq:SMatCoMXZ}
    \begin{aligned}
        S_{\text{CoM}}(p) &= -\frac{4 \left(\kappa  \left(\text{E}_2-\kappa \right)-1\right) r(p)+e^{i p} \left(\text{E}_2-2 \kappa \right) \left(\text{E}_2 \kappa -2\right) \left(r(p)^2+1\right)}{e^{3 i p} \left(\text{E}_2-2 \kappa \right)
        \left(\text{E}_2 \kappa -2\right) \left(r(p)^2+1\right)+4 e^{4 i p} \left(\kappa  \left(\text{E}_2-\kappa \right)-1\right) r(p)}.
    \end{aligned}
\end{equation}
This S-matrix obeys unitarity $S_{\text{CoM}}(-p)S_{\text{CoM}}(p) = 1$ and in the orbifold limit $\kappa \rightarrow 1$, $S_{\text{CoM}}(p) = e^{-ip}$ which matches the $\mathfrak{su}(2)$-sector S-matrix for $\mathcal{N}=4$ super Yang-Mills in the CoM case.

\mparagraph{Final form of the contact terms}

\mbox{}

The last step in finalising our wavefunction is to substitute the solution for $\mathcal{B}(p_2,p_1)$ (\ref{eq:contTerm}) into the ansatz (\ref{ContactAnsatzXZ}).

The equation~(\ref{eq:contTerm}) for $\mathcal{B}(p_2,p_1)$ has an interesting effect on our ansatz. First, one can show that
\begin{equation}
    \begin{aligned}
        F(p) = - e^{-2ip}S_{\text{CoM}}(p,-p)^{-1}
    \end{aligned}
\end{equation}
and that 
\begin{equation}
        \mathcal{G}(p) = -F(p)\mathcal{G}(p)= e^{-2ip}S_{\text{CoM}}(p,-p)^{-1} \mathcal{G}(-p)   
\end{equation}
Therefore, we can write
\begin{equation}\label{eq:contTerm}
    \begin{aligned}
        \mathcal{B}(-p,p) = - e^{-2ip}S_{\text{CoM}}(p,-p)^{-1}\mathcal{B}(p,-p) + e^{-2ip}S_{\text{CoM}}(p,-p)^{-1} \mathcal{G}(-p).
    \end{aligned}
\end{equation}
We see that the $\mathcal{B}(-p,p)$ term in the swapped part of the wavefunctions contains a $\mathcal{B}(p,-p)$ term which cancels the one from the direct part, leaving only $\mathcal{G}(p)$ as a contact term. As we saw in the $XY$ sector, we have the option of keeping $\mathcal{G}(p)$ in the swapped part, bringing it to the direct part using (\ref{eq:contTerm}), or distributing it in any other way. Choosing to move the contact term to the direct part, and factoring out an overall factor of $A_{\lambda'\lambda'}$, we can write our final centre-of-mass wavefunction as
\begin{equation}\label{eq:CoMwav1XZ}
    \begin{aligned}
        \psi_{\lambda\lambda'}(\ell_1, \ell_2) &= r(p)e^{ip\ell_1+i(-p)\ell_2} + r(p) S_{\text{CoM}}(p)e^{i(-p)\ell_1 + ip\ell_2},\\
        \psi_{\lambda'\lambda}(\ell_1, \ell_2) &= r(p)e^{ip_1\ell_1+i(-p)\ell_2} + r(p) S_{\text{CoM}}(p)e^{i(-p)\ell_1 + ip\ell_2},\\
    \end{aligned}
\end{equation}
and
\begin{equation}\label{eq:CoMwav2XZ}
  \begin{aligned}
    \psi_{\lambda\lambda}(\ell_1, \ell_2) &= \big(1 - r(p)^{2} \delta_{\ell_2,\ell_1+1}\mathcal{G}(-p)\big)e^{ip\ell_1+i(-p)\ell_2} \\
        &+S_{\text{CoM}}(p)e^{i(-p)\ell_1 + ip\ell_2}\;,\\
      \psi_{\lp\lp}(\ell_1, \ell_2) &=  r(p)^{2}\big(1 + \delta_{\ell_2,\ell_1+1}\mathcal{G}(-p)\big)e^{ip\ell_1+i(-p)\ell_2} \\
        &+ r(p)^{2}S_{\text{CoM}}(p)e^{i(-p)\ell_1 + ip\ell_2},
    \end{aligned}
\end{equation}
where $S_{\text{CoM}}(p)$ is given by equation~(\ref{eq:SMatCoMXZ}) and $\mathcal{G}(-p)$ is given through equation~(\ref{eq:gterm}).

\subsubsection{General solution} \label{sec:4termXZ}

Having found the centre-of-mass solution for two $Z$ excitations in the $X$ vacuum, we now proceed to study the case $K\neq0$. As before, and still following the approach of \cite{Bell_1989, Medvedetal91}, we will do this by introducing a second set of momenta $(k_1,k_2)$
\begin{equation} \label{GeneralSolutionXZ}
    \begin{aligned}
        \psi_{\lambda\lambda}(\ell_1, \ell_2) &= A_{\lambda\lambda}(p_1,p_2)e^{ip_1\ell_1+ip_2\ell_2} + A_{\lambda\lambda}(p_2,p_1)e^{ip_2\ell_1 + ip_1\ell_2}\\
        &+ A_{\lambda\lambda}(k_1,k_2)e^{ik_1\ell_1+ik_2\ell_2} + A_{\lambda\lambda}(k_2,k_1)e^{ik_2\ell_1+ik_1\ell_2},\\
        \psi_{\lambda\lp}(\ell_1, \ell_2) &= A_{\lambda\lp}(p_1,p_2)e^{ip_1\ell_1+ip_2\ell_2} + A_{\lambda\lp}(p_2,p_1)e^{ip_2\ell_1 + ip_1\ell_2}\\
        &+ A_{\lambda\lp}(k_1,k_2)e^{ik_1\ell_1+ik_2\ell_2} + A_{\lambda\lp}(k_2,k_1)e^{ik_2\ell_1+ik_1\ell_2},\\
        \psi_{\lp\lambda}(\ell_1, \ell_2) &= A_{\lp\lambda}(p_1,p_2)e^{ip_1\ell_1+ip_2\ell_2} + A_{\lp\lambda}(p_2,p_1)e^{ip_2\ell_1 + ip_1\ell_2}\\
        &+ A_{\lp\lambda}(k_1,k_2)e^{ik_1\ell_1+ik_2\ell_2} + A_{\lp\lambda}(k_2,k_1)e^{ik_2\ell_1+ik_1\ell_2},\\
        \psi_{\lp\lp}(\ell_1, \ell_2) &= A_{\lp\lp}(p_1,p_2)e^{ip_1\ell_1+ip_2\ell_2} + A_{\lp\lp}(p_2,p_1)e^{ip_2\ell_1 + ip_1\ell_2}\\
        &+ A_{\lp\lp}(k_1,k_2)e^{ik_1\ell_1+ik_2\ell_2} + A_{\lp\lp}(k_2,k_1)e^{ik_2\ell_1+ik_1\ell_2}.
    \end{aligned}
\end{equation}
The $k$ momenta are such that $k_1+k_2=p_1+p_2$ and $E(k_1)+E(k_2)=E(p_1)+E(p_2)$. Since the dispersion relation in the $XZ$ sector is the same as that of the $XY$ sector, they are also given in terms of the energy and the $p$ momenta by (\ref{ksols}). 

Each of the four terms (i.e. the direct $p$, swapped $p$, direct $k$ and swapped $k$ terms) are solutions of the non-interacting equations if the appropriate version of (\ref{eq:straightr}) holds. Substituting those relations, there will be four remaining coefficients $A_{\lambda\lambda}(p_1,p_2)$, $A_{\lambda\lambda}(p_2,p_1)$, $A_{\lambda\lambda}(k_1,k_2)$ and $A_{\lambda\lambda}(k_2,k_1)$. Through the interacting equations, we will fix these coefficients in terms of just $A_{\lambda\lambda}(p_1,p_2)$ and thus obtain a solution.

\mparagraph{Interacting equations}

\mbox{}

As before, it is convenient to simplify the interacting equations by combining them with the non-interacting equations which, although unphysical, are still satisfied for $\ell_2=\ell_1+1$. In this way obtain the equations
\begin{equation}
    2/\kappa\ \psi_{\lambda\lambda}(s,s+1) - \psi_{\lp\lambda}(s+1,s+1)-\psi_{\lambda\lp}(s,s) = 0,
\end{equation}
and
\begin{equation}
    2\kappa\ \psi_{\lp\lp}(t,t+1) - \psi_{\lambda\lp}(t+1,t+1)-\psi_{\lp\lambda}(t,t) = 0.
\end{equation}
Solving these one obtains two $\mathbb{Z}_2$-conjugate solutions for the remaining coefficients. The first one is 
\begin{equation}
    \begin{aligned}
        &A_{\lambda\lambda}(p_1,p_2) = \big(a(k_2,k_1)b(k_1,k_2) - b(k_2,k_1)a(k_1,k_2)\big)a(p_1,p_2)\\
        &A_{\lambda\lambda}(p_2,p_1) = -\big(a(k_2,k_1)b(k_1,k_2) - b(k_2,k_1)a(k_1,k_2)\big)a(p_2,p_1)\\
        &A_{\lambda\lambda}(k_1,k_2) = -\big(a(p_2,p_1)b(p_1,p_2) - b(p_2,p_1)a(p_1,p_2)\big)a(k_1,k_2)\\
        &A_{\lambda\lambda}(k_2,k_1) = -\big(a(p_1,p_2)b(p_2,p_1) - b(p_1,p_2)a(p_2,p_1)\big)a(k_2,k_1)
    \end{aligned}
\end{equation}
and the second solution is given by
\begin{equation}
    \begin{aligned}
        &B_{\lambda\lambda}(p_1,p_2) = \big(a(k_2,k_1)b(k_1,k_2) - b(k_2,k_1)a(k_1,k_2)\big)b(p_1,p_2)\\
        &B_{\lambda\lambda}(p_2,p_1) = -\big(a(k_2,k_1)b(k_1,k_2) - b(k_2,k_1)a(k_1,k_2)\big)b(p_2,p_1)\\
        &B_{\lambda\lambda}(k_1,k_2) = -\big(a(p_2,p_1)b(p_1,p_2) - b(p_2,p_1)a(p_1,p_2)\big)b(k_1,k_2)\\
        &B_{\lambda\lambda}(k_2,k_1) = -\big(a(p_1,p_2)b(p_2,p_1) - b(p_1,p_2)a(p_2,p_1)\big)b(k_2,k_1)
    \end{aligned}
\end{equation}
where $a,b$ are the coefficients

\begin{equation} \label{abXZ}
\boxed{\begin{aligned}
  a(p_1,p_2, \kappa) &= e^{i(p_1+p_2)}r(p_1) - 2\kappa\ e^{ip_1} r(p_1)r(p_2)+r(p_2)\\
      b(p_1,p_2, \kappa) &= \kappa\ e^{i(p_1+p_2)}r(p_2) - 2 e^{ip_1} + \kappa\ r(p_1).
     \end{aligned}}
\end{equation}
These coefficients clearly reduce to the standard XXX form $1-2e^{ip_1}+e^{i(p_1+p_2)}$ as $\kappa\ra1$. Note, however, that
they are different from the equivalent ones for the $XY$ sector in (\ref{abcoeffsXY}). To avoid cluttering the notation we use the same labels as in Section \ref{sec:XYsector}, it being understood that all instances of these coefficients in this section refer to (\ref{abXZ}). We will also drop the explicit dependence on $\kappa$ unless required for clarity.

The most general solution will be a linear combination of these two solutions:
\begin{equation}
    \psi_{\lambda\lambda}(\ell_1,\ell_2) = \alpha\ \psi^{A}_{\lambda\lambda}(\ell_1,\ell_2) + \beta\ \psi^{B}_{\lambda\lambda}(\ell_1,\ell_2),
\end{equation}
where $\psi^A_{\lambda\lambda}$ and $\psi^B_{\lambda\lambda}$ are of the form in (\ref{GeneralSolutionXZ}) with the $A$ and $B$ coefficients respectively. The same holds true for the other wavefunctions but with appropriate placements of $r(p)$.

In summary, the total eigenstate that solves the two magnon problem is given by
\begin{equation}\label{eq:XZ4lincom}
    |\psi \rangle_{\text{tot}} = \alpha\ |p_1,p_2,k_1,k_2 \rangle_{A} +\beta\ |p_1,p_2,k_1,k_2 \rangle_{B},
\end{equation}
where
\begin{equation}
    \begin{split}
      |p_1,p_2,k_1,k_2 \rangle_{i} = \displaystyle\sum_{\ell_1<\ell_2}\ &\big(
      \psi_{\lambda\lambda}^i(\ell_1,\ell_2)|\ell_1\ell_2 \rangle + \psi_{\lambda\lp}^i(\ell_1,\ell_2)|\ell_1\ell_2 \rangle\\
        &+  \psi_{\lp\lambda}^i(\ell_1,\ell_2)|\ell_1\ell_2 \rangle + \psi_{\lp\lp}^i(\ell_1,\ell_2)|\ell_1\ell_2 \rangle\big)\;,\quad i = A,B.
    \end{split}
\end{equation}

\mparagraph{Properties of the $S$-matrices}

Since each of the two general solutions contains four Bethe-like terms, we can define three $S$-matrices. Two of them refer to scattering of the $p$ or $k$ momenta among themselves:
\begin{equation} \label{SAmatXZ}
    S^A(p_1,p_2,\kappa) = \frac{A_{\lambda\lambda}(p_2,p_1)}{A_{\lambda\lambda}(p_1,p_2)}=-\frac{a(p_2,p_1)}{a(p_1,p_2)},\quad S^A(k_1,k_2,\kappa) = \frac{A_{\lambda\lambda}(k_2,k_1)}{A_{\lambda\lambda}(k_1,k_2)}=-\frac{a(k_2,k_1)}{a(k_1,k_2)}.
\end{equation}
There is also an $S$-matrix scattering the $p$  momenta to the $k$ momenta, which we call $T$:
\begin{equation}
        T^{A}(p_1,p_2,k_1,k_2,\kappa) = \frac{A_{\lambda\lambda}(k_1,k_2)}{A_{\lambda\lambda}(p_1,p_2)}
        = -\frac{a(p_2,p_1)b(p_1,p_2) \!-\!b(p_2,p_1)a(p_1,p_2)}{a(k_2,k_1)b(k_1,k_2) \!-\! b(k_2,k_1)a(k_1,k_2)}\frac{a(k_1,k_2)}{a(p_1,p_2)}.
\end{equation}
Similarly, for the solution with B coefficients,
\begin{equation} \label{SBmatXZ}
    S^B(p_1,p_2,\kappa) = \frac{B_{\lambda\lambda}(p_2,p_1)}{B_{\lambda\lambda}(p_1,p_2)}=-\frac{b(p_2,p_1)}{b(p_1,p_2)},\quad S^B(k_1,k_2,\kappa) = \frac{B_{\lambda\lambda}(k_2,k_1)}{B_{\lambda\lambda}(k_1,k_2)}=-\frac{b(k_2,k_1)}{b(k_1,k_2)},
\end{equation}
and 
\begin{equation}
        T^{B}(p_1,p_2,k_1,k_2,\kappa)= \frac{B_{\lambda\lambda}(k_1,k_2)}{B_{\lambda\lambda}(p_1,p_2)} 
        = -\frac{a(p_2,p_1)b(p_1,p_2) \!-\! b(p_2,p_1)a(p_1,p_2)}{a(k_2,k_1)b(k_1,k_2) \!-\! b(k_2,k_1)a(k_1,k_2)}\frac{b(k_1,k_2)}{b(p_1,p_2)}.
\end{equation}
These two solutions are related by $\mathbb{Z}_2$ symmetry. More precisely, recalling that $r(p,1/\kappa) = 1/r(p,\kappa)$, one finds
\begin{equation}
    \begin{aligned}
        a(p_1,p_2,1/\kappa) = \frac{1}{\kappa}\frac{1}{r(p_1)r(p_2)}b(p_1,p_2,\kappa), \quad b(p_1,p_2,1/\kappa) = \frac{1}{\kappa}\frac{1}{r(p_1)r(p_2)}a(p_1,p_2,\kappa).
    \end{aligned}
\end{equation}
From this, we find that
\begin{equation}
    \begin{aligned}
        S^A(x,y,1/\kappa) = S^B(x,y,\kappa), \quad T^A(w,x,y,z,1/\kappa) = \frac{r(k_1)r(k_2)}{r(p_1)r(p_2)}T^B(w,x,y,z,\kappa).
    \end{aligned}    
\end{equation}
The $S$-matrices satisfy unitarity and reduce to $-1$ for equal momenta:
\be
S^{A}(y,x)S^{A}(x,y) = 1\;,\;S^{B}(y,x)S^{B}(x,y) = 1\;,\;S^A(p,p)=-1\;,S^B(p,p)=-1
\ee
They also smoothly reduce to the $XXX$ $S$-matrix as $\kappa\ra 1$. However, unlike the $XXX$ $S$-matrix and the corresponding $XY$-sector $S$-matrices, $S^{A,B}(0,p)\neq 1$. Note that the $T$-matrix is again not a phase. 

\mparagraph{Symmetries}

Since $\mathbb{Z}_2$ maps the $A$ coefficient solutions to the $B$ coefficient solutions and vice versa,
to obtain a $\Zset_2$-invariant solution we need to combine them as 
\begin{equation}
    |\psi \rangle_{(gen)} = \alpha\ |p_1,p_2,k_1,k_2 \rangle_{A}+\beta\ |p_1,p_2,k_1,k_2 \rangle_{B}\;,
\end{equation}
where we must have $\alpha(1/\kappa) =\pm \beta(\kappa)$. Then we have
\begin{equation}
    \mathbb{Z}_2|\psi \rangle_{(gen)} = \mp\frac{1}{\kappa^{3}}\big(r(p_1)r(p_2)r(k_1)r(k_2)\big)^{-2} |\psi \rangle_{(gen)}.
\end{equation}
Note that the eigenvalue is slightly different from the one of the $XY$ sector (see e.g. Table \ref{Symmetries}).

One can also apply the momentum permutation maps (\ref{MomentumSwaps}) to the $A$ and $B$ solutions. The eigenvalues are exactly the same as those in Table \ref{Symmetries}.

\subsubsection{Restricted Solution} \label{sec:RestrictedXZ}

Also for the $XZ$ case, the presence of additional ``$k$'' momenta which can take complex values (for real values of the original $p$ momenta) is problematic when considering large distances $\ell_2-\ell_1\gg 1$ on the spin chain, as one of the two $k$ Bethe wavefunctions will diverge. Also, even for short chains the $k$ momenta diverge as we take the centre-of-mass limit $K=0$. For these reasons we would like to restrict the general solution to remove one of the two $k$ wavefunctions. Expressing $k_1 = K/2 + \pi/2 - iv$, $k_2 = K/2 - \pi/2 + iv$, with $v\geq 0$ the term we need to remove is the one with the swapped $k$-momenta. We make an ansatz simply consisting of three terms:
\begin{equation} \label{RestrictedwavXZ}
    \begin{aligned}
        \psi_{\lambda\lambda}^{(r)}(\ell_1, \ell_2) &= A_{\lambda\lambda}(p_1,p_2)e^{ip_1\ell_1+ip_2\ell_2} + A_{\lambda\lambda}(p_2,p_1)e^{ip_2\ell_1 + ip_1\ell_2}\\
        &+ A_{\lambda\lambda}(k_1,k_2)e^{ik_1\ell_1+ik_2\ell_2},\\
        \psi_{\lambda\lp}^{(r)}(\ell_1, \ell_2) &= A_{\lambda\lp}(p_1,p_2)e^{ip_1\ell_1+ip_2\ell_2} + A_{\lambda\lp}(p_2,p_1)e^{ip_2\ell_1 + ip_1\ell_2}\\
        &+ A_{\lambda\lp}(k_1,k_2)e^{ik_1\ell_1+ik_2\ell_2},\\
        \psi_{\lp\lambda}^{(r)}(\ell_1, \ell_2) &= A_{\lp\lambda}(p_1,p_2)e^{ip_1\ell_1+ip_2\ell_2} + A_{\lp\lambda}(p_2,p_1)e^{ip_2\ell_1 + ip_1\ell_2}\\
        &+ A_{\lp\lambda}(k_1,k_2)e^{ik_1\ell_1+ik_2\ell_2},\\
        \psi_{\lp\lp}^{(r)}(\ell_1, \ell_2) &= A_{\lp\lp}(p_1,p_2)e^{ip_1\ell_1+ip_2\ell_2} + A_{\lp\lp}(p_2,p_1)e^{ip_2\ell_1 + ip_1\ell_2}\\
        &+ A_{\lp\lp}(k_1,k_2)e^{ik_1\ell_1+ik_2\ell_2}.
    \end{aligned}
\end{equation}
This is the minimal number of terms needed to solve the interacting equations. We find the following solution
\begin{equation} \label{RestrictedXZ}
    \begin{aligned}
        &A_{\lambda\lambda}(p_1,p_2) = a(k_2,k_1)b(p_1,p_2) - b(k_2,k_1)a(p_1,p_2)\\
        &A_{\lambda\lambda}(p_2,p_1) = -\big(a(k_2,k_1)b(p_2,p_1) - b(k_2,k_1)a(p_2,p_1) \big)\\
        &A_{\lambda\lambda}(k_1,k_2) = -\big(a(p_2,p_1)b(p_1,p_2) - b(p_2,p_1)a(p_1,p_2) \big).
    \end{aligned}
\end{equation}
with the other coefficients related to these three via (\ref{eq:straightr}). 
We observe that the $A$'s here are quadratic in the $a,b$-coefficients unlike the general solution which is cubic. On the other hand, the $a$ and $b$ terms with $p$ and $k$ dependence are mixed in the restricted solution, while they appear in a more factorised form in the general one. 

One can of course also recover this result by suitably combining the $A$ and $B$ solutions. The details are precisely the same as for the $XY$ case (Section \ref{sec:RestrictedXY}) so we do not repeat them here. The correct linear combination is the same as the one given (\ref{Restrictedlc}), with of course the $a$ and $b$ coefficients given by (\ref{abXZ}) and not (\ref{abcoeffsXY}). 

The restricted solution (\ref{RestrictedXZ}) will be our starting point in Appendix \ref{XZCoMlimit}, where we will show how carefully taking the $p_2\ra -p_1$ limit leads to the centre-of-mass solution (\ref{eq:CoMwav2XZ}).

\mparagraph{Symmetries}

Factoring out the $A_{\lambda\lambda}$ coefficient, we can define two $S$-matrices:
\begin{equation} \label{SmatXZ}
    \begin{aligned} 
        S^{(r)}(p_1,p_2,k_1,k_2) = \frac{A_{\lambda\lambda}(p_2,p_1)}{A_{\lambda\lambda}(p_1,p_2)}=-\frac{a(k_2,k_1)b(p_2,p_1) - b(k_2,k_1)a(p_2,p_1)}{a(k_2,k_1)b(p_1,p_2) - b(k_2,k_1)a(p_1,p_2)}\;,
    \end{aligned}
\end{equation}
and
\begin{equation} \label{TmatXZ}
    \begin{aligned}
        T^{(r)}(p_1,p_2,k_1,k_2) = \frac{A_{\lambda\lambda}(k_1,k_2)}{A_{\lambda\lambda}(p_1,p_2)}=-\frac{a(p_2,p_1)b(p_1,p_2) - b(p_2,p_1)a(p_1,p_2)}{a(k_2,k_1)b(p_1,p_2) - b(k_2,k_1)a(p_1,p_2)}\;.
    \end{aligned}
\end{equation}
The $S^{(r)}$ matrix satisfies the unitarity and fermionic property
\be
S^{(r)}(p_1,p_2,k_1,k_2)S^{(r)}(p_2,p_1,k_1,k_2) = 1 \;\;,\;\; S^{(r)}(p_1,p_1,k_1,k_2) = -1\;,
\ee
and smoothly reduces to the XXX $S$-matrix in the $\kappa\ra 1$ limit. It is also $\Zset_2$ invariant:
\be
S^{(r)}(p_1,p_2,k_1,k_2,1/\kappa) = S^{(r)}(p_1,p_2,k_1,k_2,\kappa)\;.
\ee
As before, the $T^{(r)}$ matrix is not a phase.

It can also be shown that $|\psi \rangle_{(r)}$ is a $\mathbb{Z}_2$ eigenstate, with a slightly different eigenvalue to the $XY$ sector restricted solution. 
\begin{equation}
    \mathbb{Z}_2|\psi \rangle_{(r)} = -\frac{1}{\kappa^{2}}\big(r(p_1)r(p_2)\big)^{-2}\big(r(k_1)r(k_2)\big)^{-1} |\psi \rangle_{(r)}.
\end{equation}
Apart from this difference, all the transformations tabulated in Table \ref{Symmetries} apply to the $XZ$ restricted solution.

\subsection{Short closed chains}

In this section we will consider closed chains in the $XZ$ sector and compare their explicit diagonalisation to the two-$Z$ magnon solution around the $X$ vacuum constructed above. It should be noted that due to the dilute nature of these chains, not all configurations of magnons are possible. For instance, odd-$Z$-magnon states on an even-length closed chain are not allowed, as there is no way to match the gauge indices, or alternatively guarantee that the value of $\lambda$ to the left of the first site is the same as that to the right of the last site. Let us e.g. consider the state $\ket{XXZX}$. If the value of the dynamical parameter is $\lambda$ to the left of the first site, making that $X$ a $Q_{12}$, then its value is $\lambda'$ to the right of the last site, making that $X$ a $Q_{12}$ again. The gauge indices of these fields cannot be contracted to create a closed chain. However if we consider a state with two $Z$ magnons, such as $\ket{XZZX}$, the last field is a $Q_{21}$ and we now can contract the gauge indices. Therefore, the two-$Z$ magnon problem is restricted to closed chains of even length. 

Another difference to the $XZ$ sector is that the length-$L$ Hamiltonian is not just a direct product of the (alternately) $\Hcal_{eo}$ and $\Hcal_{oe}$ Hamiltonians acting on each pair of sites, and it is not possible to choose a reference site (e.g. the first one) with a fixed $\lambda$, as the action of the Hamiltonian on sites $(L,1)$ can change that value. Therefore, to reproduce the length-$L$ Hamiltonian we work on the full basis of two-$Z$ excitations, with both possibilities for $\lambda$ on the first site. (E.g. for the length-6 examples we exhibit, this basis is 30-dimensional).

Despite these differences, it is clear that the solution of the 2-magnon problem in the $X$ vacuum of the $XZ$ sector is very similar to that of the alternating $XY$ sector. It also requires two sets of momenta, $p_i$ and $k_i$, giving the same total momentum $K$ and energy $E_2$. The main diffferences stem from the fact that the $a$ and $b$ coefficients (\ref{abXZ}) entering the $S$ matrices are different from (\ref{abcoeffsXY}). Also, as we will focus on even-length chains there is no 1-magnon problem, and thus also no ``trivial'' 2-magnon energies just obtained by periodicity.\footnote{Accordingly, the $S$-matrices of the $XZ$ sector general solution (\ref{SAmatXZ}),(\ref{SBmatXZ}) do not reduce to 1 as one of the momenta becomes zero.}

In the CoM frame, we will again have, for the untwisted and twisted sectors respectively:
\be 
e^{ipL}=1/S(p,-p) \quad \text{and}\quad e^{ipL}=-1/S(p,-p) \;,
\ee
with $S$ now as in (\ref{eq:SMatCoMXZ}). Solving these for fixed $L$, one can confirm that the momenta and corresponding energies agree with an explicit diagonalisation of the Hamiltonian. Since $S\ra e^{-ip}$ in the limit $\kappa\ra1$, the momenta reduce to those of the standard XXX model as they should. As an example, the momenta and energies for $L=6$ and $\kappa=0.9$ are tabulated in Table \ref{LengthSixTableXZ} for the untwisted, and Table \ref{LengthSixTableXZtwisted} for the twisted case. 

Periodicity for the general (non-CoM) solution is also easy to impose by suitably combining the two wavefunctions related by $\Zset_2$:
\be \label{xlcXZ}
\ket{\psi}_{\text{tot}}=A(p_1,p_2,k_1,k_2)+x B(p_1,p_2,k_1,k_2)
\ee
where now
\be \label{periodicee}
\begin{split}
(A^{{\lambda\lambda,p}}_{12}&+x B^{{\lambda\lambda,p}}_{12})e^{i(l_1p_1+l_2p_2)}=(A^{{\lambda\lambda,p}}_{21}+x B^{{\lambda\lambda,p}}_{21}) e^{i (l_2p_2+(l_1+L) p_1)}\;,\quad\\ 
(A^{{\lambda\lambda,k}}_{12}&+x B^{{\lambda\lambda,k}}_{12})e^{i(l_1k_1+l_2k_2)}=(A^{{\lambda\lambda,k}}_{21}+x B^{{\lambda\lambda,k}}_{21}) e^{i (l_2k_2+(l_1+L) k_1)}
\end{split}
  \ee
and similarly for the $\lp\lp$ parts, while for the $\lambda\lambda'$  ones we have
\be\begin{split}
(A^{{\lambda\lp,p}}_{12}&+x B^{{\lambda\lp,p}}_{12})e^{i(l_1p_1+l_2p_2)}=(A^{{\lp\lambda,p}}_{21}+x B^{{\lp\lambda,p}}_{21}) e^{i (l_2p_2+(l_1+L) p_1)}\;,\quad\\ 
(A^{{\lambda\lp,k}}_{12}&+x B^{{\lambda\lp,k}}_{12})e^{i(l_1k_1+l_2k_2)}=(A^{{\lp\lambda,k}}_{21}+x B^{{\lp\lambda,k}}_{21}) e^{i (l_2k_2+(l_1+L) k_1)}
\end{split}
\ee
which all lead to 
\be
\bar{x}(k_1,k_2)=-\frac{a(k_1,k_2)+a(k_2,k_1)e^{iL k_1}}{b(k_1,k_2)+b(k_2,k_1)e^{iLk_1}}\;.
\ee
in terms of which the Bethe ansatz is
\be \label{BethefinalXZ}
e^{iLp_1}=-\frac{a(p_1,p_2)+\bar{x}(k_1,k_2) b(p_1,p_2)}{a(p_2,p_1)+\bar{x}(k_1,k_2) b(p_2,p_1)}\;.
\ee
This is the same expression as for the $XY$ sector, however now the $a,b$ functions are those in (\ref{abXZ}). 

As an example, in Table \ref{LengthSixTableXZ} we show the momenta and corresponding energies for $L=6$ and $\kappa=0.9$. All the energies coming from the Bethe ansatz solution have been verified to match with the explicit diagonalisation of the Hamiltonian. We note that their values are different from the corresponding ones for the $XY$ sector at $\kappa=0.9$, illustrating that these models are different, even though their dispersion relation is the same. 
\begin{table}[ht] 
  \begin{center} {\small \begin{tabular}{|ccc|cccc|} \hline  \rowcolor{blue!15}
$K$ & $p_1^{XXX}$ &$ E^{XXX}_2$& $p_1$ & $k_1$  &$\bar{x}$& $E_2$\\ \hline
$0$ &$1.25664$ & $2.76393$&$ 1.26769$& $\pi/2-i\infty$ &$-1.40161$ &$2.75581$ \\
$0$ & $2.51327$ & $7.23607$&$2.51088  $& $\pi/2-i\infty$ &$-0.877401$ &$7.28012$ \\  
$\pi/3$&$2.38724$ & $5$  & $2.39814$  & $3.1335$&$1.37095$ &$5.02498$ \\
$2\pi/3$ &$\scriptstyle\frac{\pi}3+0.732858i$ &$1.43845$& $\scriptstyle{\frac{\pi}3+0.742053i}$&$1.87214$&$0.983762$ &$1.43$\\
    $2\pi/3$&$-2.76928$ & $5.56155$ &$-2.6909$ & $-2.65293$ & $2.39$& $5.57835$ \\
$\pi$ &$\pi/2+i\infty$ & $2$ &$\pi/2+i\infty$ & $2.07826$ & $1.36864$& $2.03304$ \\
    \hline 
  \end{tabular}}
  \caption{ {\bf Length-6, $\kappa\!=\!0.9$ Untwisted XZ Chain:} \it The momenta appearing in the untwisted $X$-vacuum $XZ$-sector 2-magnon wavefunction for the case $L=6,\kappa=0.9$. The corresponding momenta and energies for the XXX model at $\kappa=1$ are given for comparison. The $p_2$ momenta can be found as $p_2=K-p_1$ and similarly for $k_2$. We have chosen the $p$ momenta to be the ones that are closest to the corresponding momenta for $XXX$. As explained in the main text, $\bar{x}$ is the proportionality constant between the $A$ and $B$ solutions. } \label{LengthSixTableXZ}
 \end{center}   \end{table}

Analogously we can study the twisted sector by imposing antiperiodicity on (\ref{xlcXZ}). The Bethe ansatz is now
\be \label{BethefinalXZtwisted}
e^{iLp_1}=\frac{a(p_1,p_2)+\bar{x}(k_1,k_2) b(p_1,p_2)}{a(p_2,p_1)+\bar{x}(k_1,k_2) b(p_2,p_1)}\;,
\ee
with $\bar{x}$ defined by
\be
\bar{x}(k_1,k_2)=-\frac{a(k_1,k_2)-a(k_2,k_1)e^{iL k_1}}{b(k_1,k_2)-b(k_2,k_1)e^{iLk_1}}\;.
\ee
Table \ref{LengthSixTableXZtwisted} contains the data for the twisted sector of the $L=6$, $\kappa=0.9$ chain, where again the energies have been matched to the explicit diagonalisation of the Hamiltonian. In this case there are degeneracies (e.g. for the energies $E=2$ and $E=4$ in the twisted XXX limit) which we have not attempted to fully resolve as our interest is in checking the overall features of the Bethe ansatz. However it is clear that those energies can also be reproduced with a slightly more careful treatment.
\begin{table}[ht] 
\begin{center}  \begin{tabular}{|ccc|cccc|} \hline \rowcolor{blue!15}
$K$ & $p_1^{XXX}$ &$ E^{XXX}_2$& $p_1$ & $k_1$  & $\bar{x}$ &$E_2$\\ \hline
$0$ &$0.628319$ & $0.763932$ &$0.62559 $& $\frac{\pi}2-i\infty$ &$0.79193$ &$0.752397$ \\  
$0$ &$1.88496$ & $5.23607$ &$ 1.86567$& $\frac{\pi}2-i\infty$ &$1.2656$ &$5.25901$ \\
$\pi/3$ &$3.02739$ & $6.78309$& $3.02508$ & $\scriptstyle\frac{2\pi}3-0.86918i$ & $1.09069$&$ 6.83937$ \\
 $\pi/3$ &$1.73909$ & $2.79493$ & $2.97503$ & $2.48291$ & $3.17803$& $2.76882$ \\
 $\pi/3$ &$\scriptstyle\frac{\pi}6+0.255761i$ & $0.421979$ & $\scriptstyle\frac{\pi}6+0.258222i$ & $\scriptstyle\frac{2\pi}3-1.19173i$ & $0.837968$& $0.417685$ \\
    $2\pi/3$ &$2.82661$ &$4.41421$ & $2.82675 $&$0.925401 $ &$1.01882$&$ 4.43349$\\
    $2\pi/3$&$\scriptstyle\frac{\pi}3+0.632974i$ & $1.58579$  & $\scriptstyle\frac{\pi}3+0.591745i $&$1.78061$ &$1.8021$&$1.6412 $\\
\hline
  \end{tabular}
    \caption{ {\bf Length-6, $\kappa\!=\!0.9$ Twisted XZ Chain:} \it Some of the momenta appearing in the twisted $XZ$-sector 2-magnon wavefunction for the case $L=6,\kappa=0.9$, with the corresponding ones for the twisted XXX model for comparison. The $p_2$ momenta can be found as $p_2=K-p_1$ and similarly for $k_2$. As explained in the main text, $\bar{x}$ is the proportionality constant between the first and second solutions. } \label{LengthSixTableXZtwisted}
 \end{center}   \end{table}  
It would of course be interesting to compare the above energies (and the 2-magnon energies we can compute for any length $L$) to one-loop anomalous dimensions of specific operators in the $\Ncal=2$ gauge theory. We leave such checks for future work.

\section{Elliptic parametrisation} \label{sec:Elliptic}

The motivation of this section is to understand the most appropriate elliptic parametrisation for each of our sectors. Having this understanding would allow us to, following \cite{Thacker:1998sh},  boost the quantum plane (non-spectral parameter dependent) $R$-matrix of Section \ref{sec:QPlanes}, in order to introduce the spectral parameter dependence.

As emphasised, the dispersion relation appearing in the $XY$ sector and the $X$-vacuum of the $XZ$ sector is naturally uniformised by elliptic functions. In this section we provide more details of this parametrisation and show how it allows us to uncover an interesting relation between these two sectors.

\mparagraph{$XY$ sector}

Let us start with the dispersion relation (\ref{XYdispersion2}) in the upper branch:
\be \label{Epm}
E(p) = \kappa + \frac{1}{\kappa}\pm\frac{1}{\kappa}\sqrt{1+\kappa^{2}e^{-2ip}}\sqrt{1+\kappa^{2}e^{2ip}}\;.
\ee
Recall that the rapidity variable which uniformises the dispersion relation should be such that 
\be
\frac{\p p(v)}{\p v}=E(p) \;\Rightarrow \;\;v(p)=\int \frac{\diff p}{E(p)}\;.
\ee
The solutions of this equation are not very easy to work with, so we shift the energy temporarily to $E'=E-\frac{1}{\kappa}-\kappa$. It should be pointed out that, although this shift is by a constant, it is not completely harmless since it brings the dispersion relation to $XY$-model/free-fermion form, effectively dropping the contribution from the $\sigma^z\otimes\sigma^z$ terms in the Hamiltonian which are responsible for the $1/\kappa+\kappa$ term.\footnote{See \cite{deLeeuw:2020bgo} for discussion of the free-fermionic condition in the context of AdS/CFT integrability.} We can now follow the treatment in  \cite{Thacker:1998sh}, which is in the context of the XY model, to express this relation using Jacobi elliptic functions with modular parameter $m=\kappa^4$. We find
\be
e^{ip}=i\kappa\sn (v/\kappa|m_{XY})\;,
\ee
in terms of which the energy is expressed as
\be
E'(v)=\frac{\diff p}{\diff v}=\frac{i}{\kappa}\frac{\cn(v/\kappa|m) \dn(v/\kappa|m)}{\sn(v/\kappa|m)}\;.
\ee
We note that in the above parametrisation $p=0$ corresponds to $v_0=-i\kappa K(1-m)/2$, with $K(m)$ the elliptic integral of the first kind with modular parameter $m$. One could thus have shifted the rapidity as $v\ra v+v_0$ so that $v=0$ corresponds to $p=0$  \cite{Thacker:1998sh}.
Another important quantity entering our wavefunctions in Section \ref{sec:XYsector} is the ratio function, which for this sector is
\be
r_{XY}(p)=\frac{e^{ip}\sqrt{1+\kappa^2e^{-2ip}}}{\sqrt{1+\kappa^2 e^{2ip}}} \rightarrow
r_{XY}(v)=\frac{\kappa \cn(v/\kappa,m)}{\dn(v/\kappa,m)}\;.
  \ee

\mparagraph{$XZ$ sector}

The dispersion relation of the $XZ$ sector is the same as the $XY$ sector. However, for reasons to be clear momentarily, for this sector we will choose not to split the square root and write the dispersion relation as (\ref{XYdispersion1})
\be \label{Epmt}
E(p)=\frac{1}{\kappa}+\kappa\pm\frac{1}{\kappa}\sqrt{(1+\kappa^2)^2-4\kappa^2\sin^2p}\;.
\ee
Then the natural modular parameter that presents itself is $\tilde{m}=\frac{4\kappa^2}{(1+\kappa^2)^2}$, and we will parametrise (\ref{Epmt}) using elliptic functions depending on this parameter. 
\footnote{This approach is closer to the standard elliptic function parametrisation of the higher-loop $\Ncal=4$ dispersion relation, where the elliptic parameter is $m=-4g^2$ with $g$ the Yang-Mills coupling (see e.g. \cite{Beisert:2006qh,Arutyunov:2007tc})} Subtracting the constant we find
\be
\sin(p)=\sn((\kappa+\kappa^{-1}) v|\tilde{m})\;,
\ee
and the energy is now
\be
E'(v)=(\kappa+\kappa^{-1}) \dn((\kappa+\kappa^{-1})v|\tilde{m})\;.
\ee
The ratio function for the $XZ$ sector is expressed as
\be
r_{XZ}(v)=\frac{(\kappa-\kappa^{-1})-(\kappa+\kappa^{-1})\dn(-(\kappa+\kappa^{-1})v|\tilde{m})}{2\cn(-(\kappa+\kappa^{-1})v|\tilde{m})}\;.
\ee

An interesting relation between the ratio functions arises if we use the Gauss transformation (or descending Landen transformation) (e.g. \cite{AbramowitzStegun})\footnote{In the $\Ncal=4$ higher-loop context, Landen transformations were used in \cite{Kostov:2007kx}.}
\be\begin{split}
&\cn\left((1+\sqrt{m})v\Big|\frac{4\sqrt{m}}{(1+\sqrt{m})^2}\right)=\frac{\cn(v|m)\dn(v|m)}{1+\sqrt{m}\sn(v|m)^2}\;,\\
&\dn\left((1+\sqrt{m})v\Big|\frac{4\sqrt{m}}{(1+\sqrt{m})^2}\right)=\frac{1-\sqrt{m}\sn(v|m)^2}{1+\sqrt{m}\sn(v|m)^2}\;.
\end{split}
\ee
It can then be straightforwardly shown that
\be
r_{XZ}((\kappa+\kappa^{-1})v|\tilde{m})=\frac{1}{r_{XY}(v/\kappa|m)}\;.
\ee
So, despite being quite different functions of the momentum, the ratio functions of the two sectors can be mapped to each other by a modular identity, for the same value of the rapidity.\footnote{One could of course have redefined the ratio in one of the sectors to be the opposite ratio, so as to get precise matching.}. This mapping hints that there exists
a unifying form of the Bethe ansatz equations in the two sectors, which should become evident with the correct elliptic uniformisation.

\mparagraph{Theta function parametrisation}

It is also instructive to switch to a theta function parametrisation. We will only show this explicitly for the choice of
modular parameter $m=\kappa^4$ as in the $XY$ sector. Using standard relations (e.g. \cite{Gomez96})  we can write
\be \label{Momentumtheta}
  e^{ip}=i\sqrt{k} \sn(v/\kappa)=i\sqrt{k}\frac{1}{\sqrt{k}}\frac{\theta_1(u)}{\theta_4(u)}=i\frac{\theta_1(u)}{\theta_4(u)}\;,
  \ee
where the arguments of the Jacobi and theta functions are related as $v/\kappa=2 K(m) u$, . For the $XY$ ratio function we have:
  \be \label{Ratiotheta}
  r(u)=\frac{\sqrt{k}\cn (v/\kappa)}{\dn (v/\kappa)}=\sqrt{k}\left(\sqrt{\frac{k'}{k}}\frac{\theta_2(u)}{\theta_4(u)}\right)\left(\frac{1}{\sqrt{k'}}\frac{\theta_4(u)}{\theta_3(u)}\right)
  =\frac{\theta_2(u)}{\theta_3(u)}\;.
\ee
Here we use the nome
\be
q=e^{i\pi\tau} \;,\;\text{where} \;\;\tau=i\frac{K'(m)}{K(m)}\;,
\ee
with $m=\kappa^4$, $K'(m)=K(1-m)$ and our theta function conventions as in \cite{Gomez96,Thetavocabulary}. In particular we have
\be
\theta_1(u+1)=-\theta_1(u)\;,\;\;\theta_1(u+\tau)=-e^{\pi i (2u+\tau)}\theta_1(u)\;.
\ee
and the zeroes of $\theta_1(u)$ are at $u=m+n\tau$ with $m,n$ integer. As above, it is usually convenient to shift the rapidity $u$ by $\tau/4$ in order to get that $u=0$ gives $p=0$, and that real rapidities correspond to real momenta. 

It is intriguing that $e^{ip}$ and the ratio function $r(p)$,  (which both appear in the $a(p_1,p_2)$ and $b(p_1,p_2)$ terms entering the $S$-matrices) can be exchanged by the simple shift $u\ra u+1/2$, given that $\theta_{1,3,4}(u+1/2)=\theta_{2,4,3}(u)$, $\theta_{2}(u+\half)=-\theta_1(u)$.
\be
e^{ip} \overset{{}_{}^{ u\ra u+1/2}}\longrightarrow  ir(p;\kappa) \overset{{}_{}^{ u\ra u+1/2}}\longrightarrow -e^{ip} \overset{{}_{}^{ u\ra u+1/2}}\longrightarrow -ir(p;\kappa) \overset{{}_{}^{ u\ra u+1/2}}\longrightarrow e^{ip}. 
\ee
Also, under shifts $u\ra u+\tau/2$, given the tranformations $\theta_{1,4}(u+\tau/2)=ie^{-\pi i (u+\tau/4)}\theta_{4,1}(u)$ and
$\theta_{2,3}(u+\tau/2)=e^{-\pi i (u+\tau/4)}\theta_{3,2}(u)$, we have
\be
e^{ip} \overset{{}_{}^{ u\ra u+\tau/2}}\longrightarrow -e^{-ip} =e^{i(\pi-p)}\;\quad\text{and}\;\quad r(p;\kappa) \overset{{}_{}^{ u\ra u+\tau/2}}\longrightarrow \frac{1}{r(p;\kappa)}\;.
\ee
We can put the above components together to write the 1-magnon wavefunction  as
\begin{equation}
    |p\rangle = \displaystyle\sum_{\ell \in 2\mathbb{Z}}\ e^{ip\ell}|\ell\rangle \!+\! \displaystyle\sum_{\ell \in 2\mathbb{Z}+1}\ r(p;\kappa)e^{ip\ell}|\ell\rangle=i\displaystyle\sum_{\ell \in 2\mathbb{Z}}\left(\frac{\theta_1(u)}{\theta_4(u)}\right)^\ell|\ell\rangle+i\displaystyle\sum_{\ell \in 2\mathbb{Z}+1}\frac{\theta_2(u)}{\theta_3(u)}\left(\frac{\theta_1(u)}{\theta_4(u)}\right)^\ell|\ell\rangle
    \, .
\ee
The reduced energy is also rather simple to express in terms of theta functions:
\be \label{Energytheta}
E'(u)=-i\frac{\theta_4(0)^2}{\theta_2(0)\theta_3(0)}\frac{\theta_2(u)\theta_3(u)}{\theta_1(u)\theta_4(u)}\;,
\ee
and given also the relation
\be
\kappa^2=k=\sqrt{m}=\left(\frac{\theta_2(0)}{\theta_3(0)}\right)^2\;,
\ee
the energy shift $E(0)-E'(0)=\kappa+1/\kappa$ of the total energy $E(u)$ can also be expressed in terms of theta functions. Finally, the $\Zset_2$ eigenvalue of the 1-magnon solution is \eqref{eq:Z2EigenvalueXY}
\be
\Zset_2\ket{p} =\frac{\theta_3(u)}{\theta_2(u)} \ket{p}.
\ee
Similarly, for the 2-magnon eigenproblem we can express the wavefunctions (for the CoM, general and restricted solutions)
and their energy and $\Zset_2$ eigenvalues (summarised in Table \ref{Symmetries}) in terms of theta functions. Even though we derived these solutions by explicitly solving the coordinate Bethe ansatz, we could have attempted to guess the wavefunctions simply by knowing their eigenvalues and modular properties. We believe that point of view will be helpful in the solution of the multi-magnon problem and also in the study of more general ADE orbifold eigenproblems. 

In Appendix \ref{appendix:Alternating} we provide a simple example of a dynamical alternating chain starting from Felder's elliptic $R$-matrix and using the algebraic Bethe ansatz formalism. The expressions for the momenta, ratio function and one-magnon energy above are exactly what one obtains from the ABA approach, apart from the $\kappa+1/\kappa$ prefactor (i.e. our $E'(u)$ is the total energy there). This is natural, as the model we consider in \ref{appendix:Alternating} is at a free-fermion point. Although this is a minor difference for one magnon, it leads to the two-magnon problem we treated in Section \ref{sec:XYsector} being very different (and considerably more involved) than the two-magnon problem of the model in the appendix, whose $S$-matrix is just $S=-1$. This hints that we need to look for a different uniformisation which is better adapted to the full energy $E(u)$ rather than the reduced one $E'(u)$.

\section{Conclusions and outlook}
\label{sec:conclusions}

In this work we took a fresh look at the spin chains related to the one-loop spectral problem of the $\Zset_2$ marginally deformed orbifold of $\Ncal=4$ SYM. We focused on the holomorphic $\SU(3)$ scalar sector and in particular two different $\SU(2)$ subsectors. One of these, the $XY$-sector, can be described by a relatively standard alternating Heisenberg chain, while the other, the $XZ$ sector, can be understood as  a slightly more exotic dynamical Temperley-Lieb model. Both cases are examples of dynamical spin chains, where the Hamiltonian depends on an additional parameter whose value is dynamically determined at each site of the chain.

Following \cite{Bundzik:2005zg,Mansson:2008xv,Dlamini:2019zuk} and pursuing the quantum plane limit approach we  discovered that novel quantum groups govern the dynamics of our spin chains. An in-depth study of  the RTT relations as well as the higher order relations is needed in order to fully characterise these quantum groups.
However, the $R$-matrix in each $\SU(2)$ sector is triangular (or is related to a triangular $R$-matrix with the same RTT relations), which implies that it can be factorised by a Drinfeld twist (\ref{Ftwist}). In this way the quantum plane can be seen to arise by twisting the trivial quantum plane of the original $\Ncal=4$ theory. The presence of a twist also guarantees that the quantum deformed generators can be made to act in a consistent way on the quantum plane coordinates, as long as one works in a quasi-Hopf setting \cite{Dlamini:2019zuk}. 
We will leave the study of the $SU(2)_\kappa$ quantum group as well as its generalisation to the full scalar sector for future work. One important aspect will be to clarify whether the Drinfeld twist of the $XZ$ sectors is generic or whether it obeys any cocycle conditions. If in particular it satisfies a shifted cocycle condition \cite{Babelon:1995rz,Enriquez_1998,Jimbo:1999zz} the corresponding $R$-matrix would satisfy the dynamical YBE (without spectral parameter) and that would provide additional support for our claim that the dynamical formalism is relevant for our type of spin chains.

With the quantum plane limit approach we were able to obtain the $R$-matrix at one specific value of the rapidity. With Baxterization \cite{Jones1990}\footnote{Baxterisation refers to starting from one point in the $R$-matrix parameter space (one Yang-Baxter integrable $R$-matrix) and building from it the algebraic variety where the spectral
parameter lives.} 
we should be able to write down the  $R$-matrix at any value of the rapidity.
This is equivalent to boosting the rapidity to a generic value after using the correct elliptic parametrisation \cite{Thacker:1998sh}. More traditionally, starting from the finite limit of the quantum plane we can affinise and obtain the $R$-matrix as a function of the rapidity via the adjoint action (as in  \cite{Gomez96} and  \cite{Jimbo:1999zz})
\be
R(u ; \kappa) = Adj(u^d \otimes 1) R(\kappa)  \, ,
\ee
where $d$ is the derivation of the affine algebra. Obtaining a rapidity-dependent $R(u;\kappa)$ is work in progress.

Generalising the quantum plane approach to include fermions will help us upgrade the quantum group of the full scalar sector to ultimately an $PSU(2,2|4)_\kappa$ which we conjecture exists and is the correct quantum group which $(i)$ leaves the complete Lagrangian invariant\footnote{The $SU(3)_\kappa$ symmetry leaves the superpotential invariant. To obtain the quantum supercharges we need to consider the Lagrangian in components and check the supersymmetry variations for the broken $Q$ generators looking for a quantum deformation of the usual supersymmetry algebra.} and $(ii)$ will uniquely fix the scattering of this model to all loops. This can be done by combining our approach with Beisert's \cite{Beisert:2005tm}  to construct the all loop Scattering matrix of  the quantum  $SU(2|2)_\kappa  \subset PSU(2,2|4)_\kappa$.
 
In fact we conjecture that there should be a very general theorem at work.
When breaking the $\mathcal{N}=4$ SCA down to some  $\mathcal{N}\geq 1$ SCA, the naively broken generators are not lost, but they get upgraded to quantum generators! All theories obtained from the   $\mathcal{N}=4$  SYM mother theory by explicitly breaking the $SU(4)_R$ symmetry, like orbifolds, orientifolds and marginal deformations,
 enjoy both at the level of the Lagrangian but also at the level of the spin chains a quantum version of $PSU(2,2|4)$, the precise form of it depends on the R-symmetry breaking pattern and can be reconstructed looking at the quantum planes as we did in Section \ref{sec:QuasiHopf} following \cite{Mansson:2008xv,Dlamini:2019zuk}, and then their super extensions.
On the string theory (AdS) side, all these daughter theories are obtained by breaking the $SU(4)_R\sim SO(6)$ of the transverse to the D3 branes directions.
 This quantum symmetry can always be understood as the symmetry of the non-commutative $\mathbb{R}^6$ plane 
 (seen by open strings) due to the B-field \cite{Seiberg:1999vs}.
 This noncommutativity has been related to marginal deformations in \cite{Kulaxizi:2006pp,Kulaxizi:2006zc,Dlamini:2016aaa}.

Another important realization in this paper is that on the one hand, the  $XZ$ sector is a dilute model but also that the full  holomorphic $\SU(3)$ sector is a  dynamical 15-vertex model. An urgent next step for our program, which is work in progress,  is  to work out  the RSOS description of this 15-vertex model,  which is expected to be similar to the elliptic (off-critical) models described in \cite{Morin-Duchesne:2018apx,Morin-Duchesne:2020pbt}.\footnote{Typically, elliptic RSOS models are off-critical and one expects to recover criticality by a suitable limit of the elliptic nome.} The symmetries and the periodicity  of the problem allowed us to determine the adjacency graph of the dilute RSOS model associated to the $\Zset_2$ quiver theory depicted in Figure \ref{fig:z2adjacency}.
The adjacency graph of the corresponding RSOS model is dual to the brane-tiling diagram of the quiver theory.

Having the adjacency graph together with Figure \ref{fig:15vertex}  should be enough to fix the Boltzmann
face  weights $\BW{a}{b}{c}{d}{u}$ of our model and finally check whether they obey the star triangle relation. A positive answer would  immediately mean integrability \`a la Felder. A worst case scenario, as we know that our system is of quasi-Hopf type in the quantum plane limit, would be to obtain a dressed star triangle relation leading to a more complicated and exotic quasi-Hopf dYBE.
For this part of our project the known literature is not ready to accommodate our needs. As far as we know,
for RSOS models, the dYBE is not known or derived unless they are dense like the ABF model. In this dense case, starting from the star-triangle relation and applying the vertex-face map \eqref{eq:ew} Felder's  dYBE  can be obtained. As explained, however, we are interested in dilute models for which (as far as we are aware) the equivalent of Felder's  dYBE has never been derived. We remind the reader that our focus on Felder's approach is motivated by the fact that it clarifies the connection with  elliptic quantum groups.

 An important feature of our theories is the appearance of elliptic-type dispersion relations. This fits with known examples of (integrable) elliptic chains, which can be usefully described by a dynamical $R$-matrix \cite{Felder:1994be}.
 In Sections \ref{sec:XYsector} and   \ref{sec:XZsector} we showed how the coordinate Bethe ansatz approach can be adapted to these types of spin chains to obtain a solution of the 1- and 2-magnon problems. 
 These explicit solutions are essential as we do not know the $R$-matrix description (as a function of the rapidity) and we are trying to discover it. Any guess for such an $R$-matrix (based on the quantum symmetries that we have outlined or the 15-vertex RSOS model) must, through the Algebraic Bethe ansatz, lead to the same eigenvectors and eigenvalues that we derived here using the coordinate Bethe ansatz.

In addition, it is interesting that for the general solution of the 2-magnon problem, we need to introduce additional $k$ momenta on top of the $p$ momenta. This is a known property of  staggered-type models   \cite{BaxterBook}.
These $k$ momenta can be understood as resonances, 
which means that already at the 2-magnon level, the system seems to fail Sutherland's criterion of quantum integrability \cite{SutherlandBook}, i.e. the absence of diffractive scattering. For a discussion of  different known criteria of integrability and their validity see \cite{Caux:2010by}.
For Sutherland's criterion to be applicable, diffractive scattering  should be  detectable as asymptotic states at infinity. In the center of mass solution, for which the introduction of nearest-neighbour contact terms is enough, the $k$ resonances decouple (see Appendices \ref{XYCoMlimit} and  \ref{XZCoMlimit} for the detailed computation), however, for generic momenta the $k$ momenta resonances are unavoidably present.
It is important to mention that through our study of  the periodic chains we discovered examples of eigenvalues where both the $p$ and the  $k$ momenta are real (see Tables \ref{LengthSixXY}, \ref{LengthSixXYtwisted}, \ref{LengthSixTableXZ} and \ref{LengthSixTableXZtwisted}).
We hope that the correct elliptic uniformisation of the two magnon eigenvector will combine both the $p$ and the  $k$ momenta in one rapidity, however at this stage it is not clear how and if it is at all possible to achieve this.
What is more, it is still possible that the three-magnon problem can be tackled using a variant of the discrete-diffractive scattering approach of \cite{Bibikov_2016}. Work on establishing (or disproving) this possibility is in progress.  
 Yet another important question we leave open here is whether the three- and higher- magnon problems can be approached in a similar way. It certainly appears to require new elements, which a better understanding of the underlying quantum symmetries might help to eventually uncover. 

We wish to emphasise, that, regardless of any integrable features, several important features of the types of spin chains considered here can be studied. For instance, in \cite{Southern98} the multi-magnon excitations have been studied using the recursion method \cite{RecursionMethod}. Translating the  multi-magnon excitations of our chain to the language of gauge theory single trace local operators, we can immediately acquire their anomalous dimensions, and then compare them to explicit one-loop computations. 

Yet an other interesting direction is to study this model from the string theory side.
The AdS string sigma-model was never considered nor constructed for marginally deformed orbifold theories.
We would need to start with type IIB string theory on $AdS_5 \times S^5/\mathbb{Z}_2$,  adding the appropriate constant NS-NS B-fields in either Green-Schwarz, hybrid or  pure-spinor formalism.
This AdS sigma model, and its $AdS_5 \times S^5/\Gamma$ generalization (with $\Gamma \in ADE$), would also be relevant, independently of integrability, for many other currently popular directions of research \cite{Beccaria:2021ksw,Galvagno:2021bbj,Galvagno:2020cgq,Zarembo:2020tpf,Niarchos:2020nxk,Niarchos:2019onf,Mitev:2015oty,Mitev:2014yba}. Intriguingly (at least for AdS/CFT practitioners),   RSOS lattice models 
are associated to (often non-unitary) minimal model CFTs in the continuum limit \cite{Bazhanov:1987zu,Saleur:1993aw,Read:2001pz,Morin-Duchesne:2018apx}. Their elliptic uplifts which are relevant for us are off-critical, however, and one could then speculate about whether those
off-critical models could capture a specific closed subsector of the full string theory sigma model.

We finally wish to conclude with some remarks on orbifolds of higher rank.
 \begin{figure}[h]
 \begin{center}
   \begin{tikzpicture}[scale=0.5]
     
  \draw[->,violet,thick] (-0.1,0.1)--(2.9,5.3);  \draw[->,violet,thick] (3.1,5.1)-- (0.1,-0.1);
  \draw[->,yellow,thick] (3.1,5.3)--(6.1,0.1);   \draw[->,yellow,thick] (5.9,-0.1)--(2.9,5.1);
  \draw[->,cyan,thick] (6,-0.12)--(0,-0.12);   \draw[->,cyan,thick] (0,0.12)--(6,0.12);
 \draw[->,violet,thick] (-0.1,0.1)--(1.4,2.7); \draw[->,violet,thick] (3.1,5.1)-- (1.6,2.5);
 \draw[->,yellow,thick] (3.1,5.29)--(4.6,2.7); \draw[->,yellow,thick] (5.9,-0.1)--(4.4,2.5);
 \draw[->,cyan,thick] (6,-0.12)--(3,-0.12);  \draw[->,cyan,thick] (0,0.12)--(3,0.12);
 
  \draw[fill=blue] (0,0) circle (2ex);
\draw[fill=red] (3,5.19) circle (2ex);
\draw[fill=green] (6,0) circle (2ex);

\draw[->,blue,thick] (-0.1,-0.7) arc (-20:-310:1);
\draw[->,red,thick] (2.4,5.3) arc (-130:-410:1);
\draw[->,green,thick] (6.4,0.5) arc (100:-170:1);

  \node at (-0.5,-0.5) {$1$};\node at (3,5.9) {$2$};\node at (6.5,-0.5) {$3$};
\node at (0.6,3) {$Q_{12}$};\node at (2.3,2) {$\Qt_{21}$};
\node at (5.3,3) {$Q_{23}$};\node at (4,2) {$\Qt_{32}$};
\node at (3,0.8) {$\Qt_{13}$};\node at (3,-0.7) {$Q_{31}$};

\node at (-1.8,-1.8) {$\phi_1$};\node at (3,7.6) {$\phi_2$};\node at (7.8,-1.8) {$\phi_3$};

\end{tikzpicture}\vspace{-0.5cm}
 \end{center}
 \caption{\it The $Z_3$ quiver. The $Q$ bifundamental fields (components of $X$ in $\Ncal=4$ language) connect the nodes in a clockwise direction while the $\tQ$ bifundamental fields (components of $Y$ in $\Ncal=4$ language) connect the nodes in an anticlockwise direction. The $\phi$ adjoint fields (components of $Z$ in $\Ncal=4$ language) return to the same node.} \label{Z3}
\end{figure}
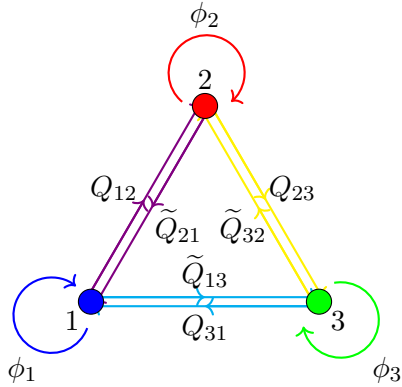
The elliptic nature of the spin chains appears to be a somewhat special feature of the $\Zset_2$ orbifold. Let us consider, for instance, a $\Zset_3$ orbifold of $\Ncal=4$ SYM (see Figure \ref{Z3}), marginally deformed so that all three gauge couplings $g_1,g_2,g_3$ are all different. By a straightforward generalisation of the discussion in Section \ref{sec:XZsector},
the  dispersion relation of one  $Z$ excitation in the $X$ vacuum is obtained as a solution of the cubic equation:
\be
 {E}^3-2 {E}^2 \left(g_1^2+g_2^2+g_3^2\right)+3 {E} \left(g_1^2g_2^2+g_1^2 g_3^2+g_2^2 g_3^2\right)+2 g_1^2 g_2^2 g_3^2 \left(\cos (3 p)-1\right)=0
 \, .
 \ee
 The $\SU(3)$ sector of the $\Zset_3$ quiver will also be described by a dynamical 15-vertex model, but in this case with three dynamical parameters $\lambda_1,\lambda_2,\lambda_3$. The corresponding adjacency graph is depicted in Figure \ref{fig:z3adjacency}. Note that as discussed in Section \ref{sec:Dynamical} for the $\Zset_2$ case, the adjacency graph in Figure \ref{fig:z3adjacency} is the dual graph of the bipartite graph of the $\Zset_3$ quiver theory \cite{Hanany:2005ve,Franco:2005rj,Franco:2005sm,Yamazaki:2008bt}.
 What is more, the 15 vertices
 of the $\Zset_3$  vertex model,
  will be the same as the ones in Figure \ref{fig:15vertex} for the   $\Zset_2$ case with the only essential difference that
 the periodicities will change as $\lambda \sim \lambda+6\eta$.

 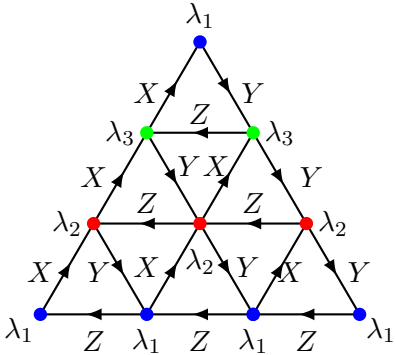
\begin{figure}[ht]
\begin{center}
\begin{tikzpicture}[scale=0.7]
  \draw[thick] (0,0)--(1,2*0.86) node [sloped,pos=0.5,allow upside down,scale=1.5]{\arrowIn};
  \draw[thick] (1,2*0.86)--(2,0) node [sloped,pos=0.5,allow upside down,scale=1.5]{\arrowIn};
  \draw[thick] (2,0)--(0,0) node [sloped,pos=0.5,allow upside down,scale=1.5]{\arrowIn};

  \draw[thick] (1,2*0.86)--(2,4*0.86) node [sloped,pos=0.5,allow upside down,scale=1.5]{\arrowIn};
  \draw[thick] (2,4*0.86)--(3,2*0.86) node [sloped,pos=0.5,allow upside down,scale=1.5]{\arrowIn};
    \draw[thick] (3,2*0.86)--(1,2*0.86) node [sloped,pos=0.5,allow upside down,scale=1.5]{\arrowIn};

    \draw[thick] (2,0)--(3,2*0.86) node [sloped,pos=0.5,allow upside down,scale=1.5]{\arrowIn};
  \draw[thick] (3,2*0.86)--(4,0) node [sloped,pos=0.5,allow upside down,scale=1.5]{\arrowIn};
  \draw[thick] (4,0)--(2,0) node [sloped,pos=0.5,allow upside down,scale=1.5]{\arrowIn};

  \draw[thick] (2,4*0.86)--(3,6*0.86) node [sloped,pos=0.5,allow upside down,scale=1.5]{\arrowIn};
  \draw[thick] (3,6*0.86)--(4,4*0.86) node [sloped,pos=0.5,allow upside down,scale=1.5]{\arrowIn};
  \draw[thick] (4,4*0.86)--(2,4*0.86) node [sloped,pos=0.5,allow upside down,scale=1.5]{\arrowIn};

    \draw[thick] (3,2*0.86)--(4,4*0.86) node [sloped,pos=0.5,allow upside down,scale=1.5]{\arrowIn};
  \draw[thick] (4,4*0.86)--(5,2*0.86) node [sloped,pos=0.5,allow upside down,scale=1.5]{\arrowIn};
    \draw[thick] (5,2*0.86)--(3,2*0.86) node [sloped,pos=0.5,allow upside down,scale=1.5]{\arrowIn};

    \draw[thick] (4,0*0.86)--(5,2*0.86) node [sloped,pos=0.5,allow upside down,scale=1.5]{\arrowIn};
    \draw[thick] (5,2*0.86)--(6,0*0.86) node [sloped,pos=0.5,allow upside down,scale=1.5]{\arrowIn};
    \draw[thick] (6,0*0.86)--(4,0*0.86) node [sloped,pos=0.5,allow upside down,scale=1.5]{\arrowIn};

    \filldraw[blue,thick] (0,0) circle (3pt);  \filldraw[blue,thick] (2,0) circle (3pt);
    \filldraw[blue,thick] (4,0) circle (3pt);  \filldraw[blue,thick] (6,0) circle (3pt);
  \filldraw[blue,thick] (4,0) circle (3pt);  \filldraw[red,thick] (1,2*0.86) circle (3pt);
  \filldraw[red,thick] (3,2*0.86) circle (3pt);  \filldraw[red,thick] (5,2*0.86) circle (3pt);
  \filldraw[green,thick] (2,4*0.86) circle (3pt);
   \filldraw[blue,thick] (3,6*0.86) circle (3pt);  \filldraw[green,thick] (4,4*0.86) circle (3pt);

   \node at (0,0.8){$X$};\node at (1,2.6){$X$};\node at (2,0.9){$X$};\node at (2,4.2){$X$};\node at (3.3,2.8){$X$};\node at (4.7,0.8){$X$};
   \node at (1.1,0.8){$Y$};\node at (3.9,0.9){$Y$};\node at (2.8,2.8){$Y$};\node at (5.1,2.6){$Y$};\node at (4,4.2){$Y$};
   \node at (6,0.8){$Y$};
   \node at (1,-0.5){$Z$}; \node at (3,-0.5){$Z$};\node at (2,2.1){$Z$};\node at (4,2.1){$Z$};\node at (3,3.8){$Z$};
   \node at (5,-0.5){$Z$};
   \node at (-0.4,-0.3){$\lambda_1$};\node at (0.5,2*0.86){$\lambda_2$};\node at (1.5,4*0.86){$\lambda_3$};
   \node at (3,6*0.86+0.5){$\lambda_1$}; \node at (4.5,4*0.86){$\lambda_3$};
   \node at (2,-0.5){$\lambda_1$};\node at (4,-0.5){$\lambda_1$};\node at (6.4,-0.3){$\lambda_1$};\node at (5.5,2*0.86){$\lambda_2$};\node at (3,1){$\lambda_2$};
   \end{tikzpicture}
\end{center}
\caption{\it The adjacency diagram of the dilute RSOS model associated to the $\Zset_3$ quiver theory, in the dynamical $\Ncal=4$-like picture. Vertices of the same colour/height $\lambda$ are identified. This is the dual graph to the brane-tiling diagram of the quiver theory.} \label{fig:z3adjacency}  
\end{figure}

 More generally, for any $\Ncal=2$ ADE quiver the corresponding one magnon dispersion relation is obtained as a solution of a higher order equation (enjoying the symmetry of the quiver). These dispersion relations appear to be more naturally described by hyperelliptic functions.  This bring to mind the Chiral Potts model \cite{BaxterBook,McCoyBook,AuYang:1987zc,McCoy:1987pt,Baxter_1998}. The study of these ADE $\mathcal{N}=2$ spin chains is work in progress and we believe that
due to their dilute nature, they will have similar features to the off-critical ADE models of \cite{Warnaar:1993zn}. From the point of view of the dynamical 15-vertex model, the most general ADE quiver has adjacency graph the dual graph of its brane tiling. We finally wish to conjecture that a large class of $\Ncal=1$ superconformal quiver theories can also be described by a dynamical vertex model coming from an RSOS model with adjacency graph the dual graph of its brane tiling.

To conclude, we hope that we have convinced the reader that the spin chains arising in the planar limit of $\mathcal{N}=2$ 4d SCFTs deserve further study. 
The interplay between different structures in mathematics like Elliptic quantum groups and quasi-Hopf algebras, statistical mechanical models like dilute elliptic vertex and RSOS models, as well as alternating bond, staggered models of interest to the condensed matter  community (arising in known materials),
with 4d gauge theory physics and string theory renders this research direction very exciting.
We plan to report on further progress in uncovering their underlying structures in upcoming work.

\acknowledgments

We are grateful to Gleb Arutyunov, Alejandra Castro, Paul Pearce,  Volker Schomerus and Ole Warnaar for useful correspondence and comments on the manuscript.  The work of EP was partially supported by the DFG via the Emmy
Noether program ``Exact results in Gauge theories'' PO 1953/1-1 and the GIF Research Grant I-1515-303./2019.
RR acknowledges financial support by the National Institute for Theoretical Physics (NITheP) during the initial stages of this project, as well as financial support by the University of Pretoria during the final stages. Part of this work was performed while KZ was visiting the University of Athens (supported by the Hellenic Foundation for Research and Innovation (HFRI) and the General Secretariat for Research and Technology (GSRT), under grant agreement No 2344) as well as the Niels Bohr Institute, and he would like to thank these institutes for their hospitality.

\appendix

\section{Taking the centre-of-mass limit in the XY sector} \label{XYCoMlimit}

In this appendix we show how the generalised Bethe ansatz (\ref{GeneralisedBethe}) for the XY sector, which involves a second set of momenta, reduces to the more standard Bethe ansatz (\ref{eq:CoMwav2}) in the centre-of-mass frame $K=0$. Understanding this limit will also clarify the origin of the contact terms that appear in the CoM frame.

To start, note (as can be seen from inspection of (\ref{ksols}) that as $K\ra 0$, the $k$ momenta scale as $k_1\ra\pi/2-i\infty$ and $k_2\ra \pi/2+i\infty$. So the centre-of-mass limit needs to be taken very carefully.

As discussed, the relevant wavefunction on which to take the limit is the restricted solution (\ref{restricted}), which contains only the direct $k$ momenta $k_1,k_2$ (and not the swapped ones $k_2,k_1$). Let us write the $S$ and $T$ matrices arising in that solution as found in section \ref{sec:RestrictedXY} (for general $p_1,p_2$):
\begin{equation}
    \begin{aligned}
        S^{(r)}(p_1,p_2,k_1,k_2) &= -\frac{a(k_2,k_1)b(p_2,p_1) - b(k_2,k_1)a(p_2,p_1)}{a(k_2,k_1)b(p_1,p_2) - b(k_2,k_1)a(p_1,p_2)}\\
        &= -\frac{f~ b(p_2,p_1) - a(p_2,p_1)}{f~b(p_1,p_2) -  a(p_1,p_2)},
    \end{aligned}
\end{equation}
and
\begin{equation}
    \begin{aligned}
        T^{(r)}(p_1,p_2,k_1,k_2) &= -\frac{a(p_2,p_1)b(p_1,p_2) - b(p_2,p_1)a(p_1,p_2)}{a(k_2,k_1)b(p_1,p_2) - b(k_2,k_1)a(p_1,p_2)}\\
        &= -\frac{1}{b(k_2,k_1)}\frac{a(p_2,p_1)b(p_1,p_2) - b(p_2,p_1)a(p_1,p_2)}{f ~b(p_1,p_2) -  a(p_1,p_2)},
    \end{aligned}
\end{equation}
where we have defined 
\begin{equation} \label{fratio}
    f = \frac{a(k_2,k_1)}{b(k_2,k_1)} \;.
\end{equation}
Now, we choose $k_1, k_2$ to be of the form $k_1 = K/2 +\pi/2 - iv,\ k_2 = K/2-\pi/2 + iv$, where $v \geq 0$, since this leads to a decaying exponential in the wavefunctions. Noting the $1/sin^2(K)$ term in (\ref{ksols}), it is clear that for $K$ sufficiently small the argument of the $\arccos$ will eventually exceed 1 and the $\arccos$ term will become (positive) imaginary. As $K$ is further decreased towards zero, $v\ra \infty$.

What we will now show is that, in the limit where $p_2 \rightarrow -p_1$,
\begin{equation}
    \begin{aligned}
        &1)\quad S^{(r)}(p_1,p_2,k_1,k_2)\ \longrightarrow\ S_{CoM}(p,-p),\\
        &2)\quad r(k_1)T^{(r)}(p_1,p_2,k_1,k_2)e^{-ik_1}\ \longrightarrow\ r(p)e^{-ip}\mathcal{G}(-p).
    \end{aligned}
\end{equation}
We will thus completely recover the centre-of-mass wavefunctions that were originally found using contact term method in section \ref{sec:CoMXY}. For point 2, which we show later in this section, note that this comes from the fact that only nearest-neighbour terms for the $(k_1,k_2)$-term survive the limit. Consequently, looking at the wavefunctions in equation~(\ref{restrictedAnsatz}), the $(k_1,k_2)$-terms for even-even and odd-odd completely vanish (as they do not have nearest neighbour terms); for the odd-even and and even-odd terms, only $(2\ell-1,2\ell)$ and $(2\ell,2\ell+1)$ survives, thus generating the $\delta_{\ell_1+1,\ell_2}$ which was entered by hand in the CoM wavefunctions.\footnote{We note that this behaviour resembles that of bound states for the spin-$1/2$ XXX chain whose imaginary part can diverge  and lead to localised interactions (see e.g. the introduction to the Bethe ansatz \cite{Karabachetal}).}

To take the limits of the $S$  matrix, it is clearly important to understand the limit of the function $f$ in (\ref{fratio}) as $p_2\ra-p_1$, as the other terms remain finite and non-zero in the limit. An important result concerns the limits of the ratio functions appearing in the $a$ and $b$ coefficients. They are:
\be
    \begin{aligned}
        r(k_1) = -\frac{1}{\kappa}, \quad r(k_2) = \kappa \;,\quad \text{as} \;\;p_2\ra -p_1\;.
    \end{aligned}
    \ee
We note the minus sign appearing in $r(k_1)$, which is important to get the right limit. 

Using this result, and also recalling that $r(-p)=1/r(p)$, we find in the limit
\begin{equation}\label{eq:fform1}
    f = -\frac{-\kappa +2 e^{i k_2}+\kappa  e^{i (p_1+p_2)}}{\kappa  (2 \kappa  e^{i k_2}+e^{i (p_1+p_2)}-1)}\;.
\end{equation}
Here we have taken the $K\ra0$ limit only of the ratio functions, and still need to take it in the exponents. The naive limit of
this expression gives $0/0$, so to obtain a definite limit, we first express $k_2$ in terms of $p_1,p_2$ using (\ref{ksols}) and then set $p_2 = -p_1 + \epsilon$, so that the limit becomes $\epsilon \rightarrow 0$. Since the equations for $k_1,k_2$ contain an arccos, we are able to remove the exponents using the  formula
\begin{equation}
    \arccos(z)  = -i\ln\big(z+\sqrt{z^{2}-1}\big)\;.
\end{equation}
After some analysis, we can extract the divergent terms $\text{csc}(\epsilon)$ from both the numerator and denominator of equation~(\ref{eq:fform1}). To take the limit $\epsilon \rightarrow 0$, we apply l'Hopital's rule to finally obtain the complete limit for $f$:
\begin{equation}
    \begin{aligned}
        f(p,-p) = \frac{-1+\sqrt{1+\kappa ^2 e^{-2 i p_1}} \sqrt{1+\kappa ^2 e^{2 i p_1}}}{\kappa ^2-\sqrt{1+\kappa ^2 e^{-2 i p_1}} \sqrt{1+\kappa ^2 e^{2 i p_1}}} .
    \end{aligned}
\end{equation}
Substituting this back into $S(p_1,p_2,k_1,k_2)$ and setting $p_2 \rightarrow -p_1$ for the other terms (which are harmless), we find that 
\begin{equation}
    S(p_1,p_2,k_1,k_2) \longrightarrow S_{\text{CoM}}(p,-p).
\end{equation}
where $S_{\text{CoM}}(p,-p)$ is given in (\ref{SCoMXY}). We have thus confirmed that the CoM $S$-matrix obtained using the contact-term approach is a limiting case of the $S$-matrix for the restricted solution.

The final step to confirm that the full centre-of-mass solution arises as a limit of the restricted solution is to check that the contact term$\mathcal{G}$  arises from one of the terms with additional momenta. Let us focus on the following term for the odd-even part of the wavefunction (\ref{restricted}):
\begin{equation}
    r(k_1)T(p_1,p_2,k_1,k_2)e^{ik_1\ell_1 + ik_2\ell_2}.
\end{equation}
We set $\ell_2 = 2\ell$ and $\ell_1 = 2\ell -(2m+1)$ where $m\geq0$. Therefore, $m=0$ corresponds to the nearest-neighbour case. After using that $k_1+k_2\ra 0$ in the limit, we are left with
\begin{equation}
    \begin{aligned}
        r(k_1)T(p_1,p_2,k_1,k_2)e^{-ik_1(2m+1)} = -r(k_1)\frac{e^{-ik_1(2m+1)}}{b(k_2,k_1)}\frac{a(p_2,p_1)b(p_1,p_2) - b(p_2,p_1)a(p_1,p_2)}{f~b(p_1,p_2) - \ a(p_1,p_2)} \;.
    \end{aligned}
\end{equation}
The factor $f$ was determined above, and proceeding similarly we find that the term 
\begin{equation}
    \frac{e^{-ik_1(2m+1)}}{b(k_2,k_1)} \overset{\epsilon\ra 0}\longrightarrow
\frac{e^{-i k_1 (2 m+1)}}{-2 \kappa  e^{i k_2}-e^{i \left(p_1+p_2\right)}+1}   \;.
\end{equation}
 This factor vanishes in the $\epsilon\ra 0$ limit when  $m>0$. However, $m = 0$ gives a non-zero result, which is the nearest-neighbour term and produces the $\delta_{\ell_1,\ell_1+1}$ contact term that we originally inserted by hand in the CoM method. More precisely, upon taking $\epsilon \rightarrow 0$ the $m=0$ term reduces to:
 \be
\frac{\kappa }{-2 \kappa ^2+2 \sqrt{1+\kappa ^2 e^{-2 i p}} \sqrt{1+\kappa ^2 e^{2 i p}}}\;.
 \ee
Putting everything together we conclude that out of all the $(k_1,k_2)$-dependent terms in the odd-even part of (\ref{restricted}) the only one that survives the centre-of-mass limit is
\begin{equation}
    \begin{aligned}
      r(k_1)T(p_1,p_2,k_1,k_2)e^{i\ell_1 k_1+i(\ell_1+1) k_2} \overset{\epsilon\ra 0}\longrightarrow   r(p)e^{-ip}\mathcal{G}(-p).
    \end{aligned}
\end{equation}
which is precisely the centre-of-mass contact term appearing in the odd-even part of (\ref{restricted}). A similar procedure applies to the even-odd part. We have thus shown how the general method of \cite{Bell_1989, Medvedetal91,Bibikov_2016} of adding additional momenta also applies in the centre-of-mass frame, as long as the limit is taken carefully. All terms containing $k$ momenta in the wavefunction vanish, apart from the nearest-neighbour term which produces the centre-of-mass contact term. This explains the physical origin of the contact terms. We end by noting that in the standard XXX Bethe ansatz at $\kappa=1$, the momenta diverge in the limit $K\ra \pi$. Moving slightly away from $\kappa=1$, we see that the $k$ momenta are finite in the CoM limit at $K=\pi$. So in a sense the role of the $p$ and $k$ momenta is exchanged at $K=\pi$ and this might allow us to think of the $\kappa\neq 1$ theory and the additional-momenta Bethe ansatz as a regularisation of the XXX Bethe ansatz at $K=\pi$.\footnote{Recall that one always obtains a better understanding of the Bethe roots by regularising the Bethe ansatz, can be done e.g. by introducing a phase that changes the periodicity properties (see e.g. \cite{Staudacher:2010jz} for a discussion).}

\section{Taking the centre-of-mass limit in the XZ sector} \label{XZCoMlimit}

In our study of the $X$ vacuum of the $XZ$ sector we also applied two, superficially very different, methods to solve the 2-magnon problem: A contact-term method which required the centre-of-mass condition $K=0$ (section \ref{CoMXZ}) and the approach with additional momenta (sections \ref{sec:4termXZ} and \ref{sec:RestrictedXZ}). In this section we show how the two methods are related, and in particular that the CoM $S$-matrix (\ref{eq:SMatCoMXZ}) arises as a limit of the restricted $XZ$ $S$-matrix (\ref{SmatXZ}). As most features are exactly parallel to the $XY$ case of the previous appendix, we will just show the main points.

\mparagraph{Recovering the centre-of-mass $S$-matrix:}

Recall the restricted solution $S$-matrix in this sector is
\begin{equation} 
    \begin{aligned}
        S^{(res)}(p_1,p_2,k_1,k_2) &=-\frac{a(k_2,k_1)b(p_2,p_1) - b(k_2,k_1)a(p_2,p_1)}{a(k_2,k_1)b(p_1,p_2) - b(k_2,k_1)a(p_1,p_2)}\\
        &=-\frac{f\cdot b(p_2,p_1) - a(p_2,p_1)}{f\cdot b(p_1,p_2) - a(p_1,p_2)},
    \end{aligned}
\end{equation}
where 
\begin{equation}
    f = \frac{a(k_2,k_1)}{b(k_2,k_1)}= \frac{e^{i \left(p_1+p_2\right)} r\left(k_2\right)+r\left(k_1\right) \left(1-2 \kappa  e^{i k_2} r\left(k_2\right)\right)}{\kappa  e^{i \left(p_1+p_2\right)} r\left(k_1\right)+\kappa  r\left(k_2\right)-2 e^{i k_2}}.
\end{equation}
Due to the divergence of the $k$ momenta as $p_2\ra -p_1$, the limit needs to be taken carefully. 
We start by simplifying the ratio functions $r(k_1)$, $r(k_2)$ by taking the limit in some of their terms:
\begin{equation}
\begin{split}
  r(k_1)& \rightarrow \frac{e^{-i k_1} \left(-\kappa ^2+\sqrt{\kappa ^4+\kappa ^2 e^{2 i k_1}+1}+1\right)}{\kappa }\;,\\
    r(k_2)& \rightarrow \frac{e^{i k_2} \left(-\kappa ^2+\sqrt{\kappa ^4+\kappa ^2 e^{-2 i k_2}+1}+1\right)}{\kappa }.
\end{split}
\end{equation}
We then substitute these expressions into $f$ and set $k_1 = p_1+p_2 - k_2$. Next, to remove the divergent parts from $k_2$,  we set $p_2 = -p_1 + \epsilon$. After some manipulation, we find that we can write
\begin{equation}\label{eq:XZlimk2}
    e^{2ik_2} = \frac{\sin(\epsilon)^{2}}{x}\;,\text{where}\;\;\lim_{\epsilon\ra 0} x= -\frac{4 \left(\kappa ^4+2 \kappa ^2 \cos (2 p)+1\right)}{\kappa ^2}.
\end{equation}
Substituting back into $f$ and taking the limit $\epsilon \rightarrow 0$,  and then substituting $f$ into (\ref{SmatXZ})  we find precisely $S_{CoM}$ as in (\ref{eq:SMatCoMXZ}).\\

\mparagraph{Recovering the contact term}

To complete the matching of the solutions we need to also recover the contact term $\mathcal{G}(p)$ in (\ref{eq:gterm}). We will focus on the $(\lambda,\lambda)$ and $(\lp,\lp)$ terms in the restricted wavefunction, as there are no neighbouring $Z$ magnons in the  $\psi_{\lambda\lp}^{(r)},\psi_{\lp\lambda}^{(r)}$ wavefunctions. As in the previous appendix, we set $\ell_2 = \ell_1 + 2m + 1$, where $m\geq0$ (in $\psi_{\lambda\lambda},\psi_{\lp,\lp}$, for a $Z$ excitation at site $\ell_1$, the excitation at $\ell_2$ is an odd number of sites away). We will show that for $m=0$ (in other words, the nearest-neighbour case), we have a finite non-zero result in the $K=0$ limit, while the terms with  $m > 0$ vanish in the limit. We will thus recover the required contact terms. 

Recalling the form of $T$ from (\ref{TmatXZ}),
a generic term in  $\psi_{\lambda\lambda}^{(r)}$ will be
\begin{equation} \label{Twavlimit}
    \begin{aligned}
        &-r(k_1)r(k_2) T^{(3)}(p_1,p_2,k_1,k_2)e^{ik_1\ell_1 + ik_2(\ell_1 +2m+1) }\\
        &=-r(k_1)r(k_2)\frac{e^{ik_2(2m+1)}}{b(k_2,k_1)}\frac{a(p_2,p_1)b(p_1,p_2) - b(p_2,p_1)a(p_1,p_2)}{f\cdot b(p_1,p_2) - a(p_1,p_2)}e^{i(k_1+k_2)\ell_1 }\;.
    \end{aligned}
\end{equation}
where we used that $k_1+k_2=0$ in the limit.
We have already determined the limit for $f$, and $r(k_1)r(k_2)\ra -1$ in the limit, so it just remains to study the behaviour of the prefactor
\begin{equation}
   \frac{e^{ik_2(2m+1)}}{b(k_2,k_1)}.
\end{equation}
Setting $p_2 = -p_1 + \epsilon$ and using equation~(\ref{eq:XZlimk2}), it is straightforward to confirm that for $m > 0$, the above expression vanishes as we send $\epsilon\ra 0$. However, if $m = 0$ we find
\begin{equation}
       \frac{e^{ik_2}}{b(k_2,k_1)} \rightarrow \frac{1}{2 \sqrt{\kappa ^4+2 \kappa ^2 \cos (2 p)+1}-2 \kappa ^2}\;.
\end{equation}
 Substituting this (and $f$) back into (\ref{Twavlimit}), we find a precise match with the term
\begin{equation}
    -r(p)^{2}e^{-ip}\mathcal{G}(-p),
\end{equation}
which arises in the solution (\ref{eq:CoMwav2XZ}). The case of $\psi^{(r)}_{\lambda'\lambda'}$ works similarly. We conclude that, also for the $XZ$ sector, the contact term wavefunction (\ref{eq:CoMwav2XZ}) is a limiting case of the restricted wavefunction (\ref{RestrictedwavXZ}),(\ref{RestrictedXZ}) in the centre-of-mass limit.

\section{Example of a dynamical alternating chain} \label{appendix:Alternating}

In section \ref{sec:Dynamical} we argued that the dynamical parameter provides a useful way to describe Hamiltonians which are not fixed at each site but are dynamically determined based on the spin chain configuration, and in particular on the number and type of states which have been crossed before arriving at the sites on which the Hamiltonian acts. To make this statement more concrete, in this appendix we show how one can obtain an alternating Hamiltonian from Felder's dynamical $\SU(2)$ $R$-matrix. The Hamiltonian will be that of an alternating XX model, so not directly relevant to either the alternating XXX Hamiltonian of the $XY$ sector or the dynamical Temperley-Lieb Hamiltonian of the $XZ$ sector (as we do not know how to obtain those models from an $R$-matrix).  However we hope that this simple example will illustrate how the dynamical parameter might be used to construct alternating-type models more generally.

The derivation below is essentially the same as that of \cite{Jones74}, who used Baxter's original site-dependent transformation starting from the XYZ model. Here we just put it into a more modern context to emphasise the link to elliptic quantum groups, which might eventually also play a role in our case.  One can also equivalently perform the same computation in loop model language, by suitably adapting the treatment of the ABF model in \cite{Bianchini:2014bfa}.

Let us start from Felder's dynamical $R$-matrix \cite{Felder:1994be} but in symmetrised form, where (as in (\ref{Feldersym})) one makes use of the available gauge freedom to rescale the off-diagonal components so that they are equal (see e.g. the appendix of \cite{Deguchi_2002}). Let us actually write the permuted $\Rhat=P R$:
\be \label{Feldersym}
\Rhat=\left(\begin{array}{cccc}
  \gamma& 0 & 0& 0\\
  0& \beta_+ & \alpha &0\\
  0&\alpha & \beta_-&0\\
  0&0&0&\gamma\end{array}\right)\;,
\ee
where
\be \label{Feldercoeff}
  \gamma(u)=\frac{\ttheta(2\eta-u)}{\ttheta(2\eta)}\;,\;
  \alpha(u,\lambda)=\frac{\sqrt{\ttheta(\lambda-2\eta)\ttheta(\lambda+2\eta)}}{\ttheta(-\lambda)}\frac{\ttheta(u)}{\ttheta_1(2\eta)}\;\;\text{and}\;\;
  \beta_\pm(u,\lambda)=\frac{\ttheta(\lambda\pm u)}{\ttheta(\lambda)}.
\ee
Our conventions for $\theta$ functions are as in \cite{Gomez96, Thetavocabulary}. In particular the elliptic nome will be $q=e^{i\pi\tau}$.  The elliptic parameter is related to the deformation parameter as $m=k^2=\kappa^4$ (see section \ref{sec:Elliptic}), and the imaginary period of the torus is $\tau=i K(1-m)/K(m)$, with $K$ the complete elliptic integral of the first kind.

To obtain the alternating behaviour we need, will have to set the step $2\eta$ to be imaginary, in particular we will take it to be proportional to the imaginary period $\tau$. The usual theta function $\theta_1$ is not periodic under such a shift, so, following \cite{Baxter:1972wg,Deguchi_2002,Levy_Bencheton_2013}, we have redefined our theta functions as:
\be \label{Ttheta}
\ttheta(u)=e^{i\pi u^2/(4\eta)}\theta_1(u)\;.
\ee
This modified theta function is antiperiodic under a shift by $\tau$, and periodic under a shift by $2\tau$:
\be\label{thetaperiodicity}
\ttheta(u+\tau)=-\ttheta(u) \;,\;\; \ttheta(u+2\tau)=\ttheta(u)\;.
\ee
We will choose $2\eta=\tau/2$, and also set $\lambda=\half+\frac{\tau}2$. So the dynamical shift is
\be
\lambda\ra \lambda'=\lambda+2\eta=\lambda+\tau/2\;.
\ee
From the above periodicity we see that repeating the shift, $\lambda\ra 4\eta$, takes us back to the original $R$-matrix as shifts by $\tau$ just change an overall sign which cancels in the ratios. 

A standard theta-function identity relates the elliptic modulus to the ratio of two of the theta functions evaluated at the origin. Specifically, given our choice $m=\kappa^4$, we will have
\be
\kappa^2=k=\sqrt{m}=\left(\frac{\theta_2(0)}{\theta_3(0)}\right)^2\;.
\ee
This relation will be important later on, when checking how the Hamiltonian is dynamically modified between even and odd sites. 

The $\Rhat$-matrix (\ref{Feldersym}) satisfies the dynamical Yang-Baxter relation in Hecke form (which is equivalent to (\ref{dYBE1}) for $R$):
 \be \label{dYBE}
  \begin{split}
    \Rhat_{23}(u,\lambda+2\eta h^{(1)}) ~&\Rhat_{12}(u+v,\lambda)~\Rhat_{23}(v,\lambda+2\eta h^{(1)}) \\=
  &\Rhat_{12}(v,\lambda)~\Rhat_{23}(u+v,\lambda+2\eta h^{(1)})~ \Rhat_{12}(u,\lambda)\;,
\end{split}
  \ee  
where we recall that the notation is that $\lambda$ is shifted by $2\eta$ times the $\SU(2)$ weight of the line that is crossed. Note that the $\SU(2)$ weights are $\pm 1$ for $X$ and $Y$ respectively, however the effect of crossing an $X$ of a $Y$ is the same, as $\lambda+2\eta \sim \lambda-2\eta$ by the above periodicity (\ref{thetaperiodicity}). So, with our choices, $\lambda$ and $\lambda'$ are the only two values of the dynamical parameter that will appear.

The normalisation of the $R$-matrix implies that
\be
\Rhat(u)\Rhat(-u)=\gamma(u)\gamma(-u)\;.
\ee
At $u=0$, we find $\Rhat(0)=I$, i.e. $R=P$. So $u=0$ is a good point to expand around and give a Hamiltonian.\footnote{Recall that in Section \ref{sec:QuasiHopf} we saw that choosing $u=1/2$ with the same parameters $\lambda$ and $\eta$ gives us the $R$-matrix in the quantum plane limit, however not for the $XY$ sector (whose quantum plane is trivial) but for the $XZ$ sector.}  However, to find this Hamiltonian it is best to work with the transfer matrix following the Algebraic Bethe Ansatz approach (which for the dynamical YBE has been detailed in \cite{Felder:1996xym,Deguchi_2002}). We will assume that the reader has familiarity with the main features of the ABA construction (see e.g. \cite{Faddeev:1994nk,Gomez96, Nepomechie:1998jf, Staudacher:2010jz, Levkovich-Maslyuk:2016kfv} for introductions) and focus mostly on the differences arising due to the dynamical nature of the $R$-matrix. Let us start by considering the transfer matrix for the $X$-vacuum with a single $Y$ impurity. We will focus on a small part of the chain around the impurity. The overall transfer matrix will be a trace in the auxiliary space, which we draw by a horizontal line:
 \be
\begin{tikzpicture}[scale=0.5,baseline={(-0.5cm,0)}]
\node at(-2,2) {${T=}$};
  \draw[->,thick,black!10!green] (1,1)--(1,3);
  \draw[->,thick,black!10!green] (2,1)--(2,3);
  
  \draw[->,thick,black!10!green] (4,1)--(4,3);
  \draw[->,thick,black!10!green] (5,1)--(5,3);
  \draw[->,thick,black!10!green] (6,1)--(6,3);

  \draw[->,thick,black!10!green] (8,1)--(8,3);
  \draw[->,thick,black!10!green] (9,1)--(9,3);

  \draw[->,thick,black!20!red] (0,1.5)--(10,2.5);
  
  \node at (1,0) {$X$};
  \node at (2,0) {$X$};
  \node at (3,0) {$\cdots$};
  \node at (4,0) {$X$};
  \node at (5,0) {$Y$};
  \node at (6,0) {$X$};
  \node at (7,0) {$\cdots$};
  \node at (8,0) {$X$};
  \node at (9,0) {$X$};

  \node at(-0.6,1.5) {$X$};

\node at (10.6,2.5) {$X$};

\end{tikzpicture}
\begin{tikzpicture}[scale=0.5,baseline={(-0.5cm,0)}]
\node at(-2,2) {$+$};
  \draw[->,thick,black!10!green] (1,1)--(1,3);
  \draw[->,thick,black!10!green] (2,1)--(2,3);
  
  \draw[->,thick,black!10!green] (4,1)--(4,3);
  \draw[->,thick,black!10!green] (5,1)--(5,3);
  \draw[->,thick,black!10!green] (6,1)--(6,3);

  \draw[->,thick,black!10!green] (8,1)--(8,3);
  \draw[->,thick,black!10!green] (9,1)--(9,3);

  \draw[->,thick,black!20!red] (0,1.5)--(10,2.5);
  
  \node at (1,0) {$X$};
  \node at (2,0) {$X$};
  \node at (3,0) {$\cdots$};
  \node at (4,0) {$X$};
  \node at (5,0) {$Y$};
  \node at (6,0) {$X$};
  \node at (7,0) {$\cdots$};
  \node at (8,0) {$X$};
  \node at (9,0) {$X$};

  \node at(-0.6,1.5) {$Y$};

\node at (10.6,2.5) {$Y$};

\end{tikzpicture}
\ee
The Hamiltonian is the first-order in $u$ term in $T$, so we can write $T=\Omega(1-u\Hcal+\cdots)$. We note that most of the terms in the left diagram will be $R_{XX}^{XX}=\gamma(u)$, which expands as $\gamma(u)=1-u() +\cdots$. On the other hand, the diagram on the right will necessarily contain lots of $R_{YX}^{YX}=\alpha(u)$, which expands as $\alpha(u)=O(u)$ so will go beyond linear order in $u$. So the term on the right will only be relevant for higher charges (or very short chains), so we will only focus only the term on the left for the purpose of deriving the Hamiltonian.

Now note that at  $u=0$ the $\Rhat$-matrix (\ref{Feldersym}) becomes the unit matrix, and the corresponding $R$-matrix (\ref{Rsym}) becomes the permutation matrix $P$. In other words,  $R_{XX}^{XX}=R_{XY}^{YX}=R_{YX}^{XY}=R_{YY}^{YY}=1$ while $R_{XY}^{XY}=R_{YX}^{YX}=0$. It is easy to check that this leaves only one nonzero term in the transfer matrix acting on the single-$Y$ state. Focusing on a small region around $Y$, with all $R$-matrices not shown to the left and right being $R_{XX}^{XX}$, we find that the transfer matrix at $u=0$ will treat a $Y$ impurity as follows:
\be
\begin{tikzpicture}[scale=0.5,baseline={(0,1cm)}]
  
  \draw[->,thick,black!10!green] (2,1)--(2,3.5);
  \draw[->,thick,black!10!green] (4,1)--(4,3.5);
  \draw[->,thick,black!10!green] (6,1)--(6,3.5);
  \draw[->,thick,black!10!green] (8,1)--(8,3.5);
  
  \draw[->,thick,black!20!red] (1,1.5)--(9,2.5);
  
  \node at (1,0) {$\cdots$};
  \node at (2,0) {$X$};
  \node at (4,0) {$Y$};
  \node at (6,0) {$X$};
    \node at (8,0) {$X$};
  \node at (9,0) {$\cdots$};

  \node at (1,4) {$\cdots$};
  \node at (2,4) {$X$};
  \node at (4,4) {$X$};
  \node at (6,4) {$Y$};
  \node at (8,4) {$X$};
  
  \node at (9,4) {$\cdots$};

  \node at (3,2.2) {$X$};
  \node at (5,2.45) {$Y$};
  \node at (7,2.7) {$X$};
  
  \node at (0,1.5) {$X$};
  \node at (9.7,2.8) {$X$};
  
  \node at (16.5,2) {$=R_{XX}^{XX}(\lambda')~R_{YX}^{XY}(\lambda)~R_{XY}^{YX}(\lambda')~R_{XX}^{XX}(\lambda).$};

  \node at (1.3,2.3){$\lambda$};
\end{tikzpicture}
\ee
In writing out this expression we start from the left in the diagram, following the arrow of the auxiliary space line, but, as we think of the transfer matrix as an operator acting to the right, the order of $R$-matrices as they are written out is reverse of the order of the corresponding vertices in the diagram. We have also indicated the dynamical parameter to the right of the first $R$-matrix, which we have arbitrarily taken to be $\lambda$ (i.e. not $\lambda'$). Each time one crosses a line (including of course the auxiliary line) the parameter flips from $\lambda$ to $\lambda'$ and back. As in this case we are interested in $R(u=0)$ where the $\lambda$ dependence disappears, $\lambda$ will not play a role, but note that the shifted $Y$ magnon sees the same value ($\lambda$) as the initial one before the action of the transfer matrix, and thus is a $\Qt_{12}$ field.

We see that $Y$ has been pushed one site to the right. The same will be true for any number of $Y$'s at $u=0$ and we conclude that $T(0)= \Omega$, the shift operator, as required. So to obtain the Hamiltonian we will now look for the first-order in $u$ terms in $T(u)$, which we will need to multiply  by $\Omega^{-1}$. There will be three such terms, which we can represent graphically as below:
\be
\begin{tikzpicture}[scale=0.5,baseline={(0,1cm)}]
  \node at (-2,2){$\Scal:$};
  
  \draw[->,thick,black!10!green] (2,1)--(2,3.5);
  \draw[->,thick,black!10!green] (4,1)--(4,3.5);
  \draw[->,thick,black!10!green] (6,1)--(6,3.5);
  \draw[->,thick,black!10!green] (8,1)--(8,3.5);
  
  \draw[->,thick,black!20!red] (1,1.5)--(9,2.5);
  
  \node at (1,0) {$\cdots$};
  \node at (2,0) {$X$};
  \node at (4,0) {$Y$};
  \node at (6,0) {$X$};
    \node at (8,0) {$X$};
  \node at (9,0) {$\cdots$};

  \node at (1,4) {$\cdots$};
  \node at (2,4) {$X$};
  \node at (4,4) {$X$};
  \node at (6,4) {$Y$};
  \node at (8,4) {$X$};
  
  \node at (9,4) {$\cdots$};

  \node at (3,2.2) {$X$};
  \node at (5,2.4) {$Y$};
  \node at (7,2.7) {$X$};
  
  \node at (0,1.5) {$X$};
  \node at (9.7,2.8) {$X$};
  
  \node at (16.5,3) {$=R_{XX}^{XX}(\lambda')~R_{YX}^{YX}(\lambda)~R_{XY}^{YX}(\lambda')~R_{XX}^{XX}(\lambda)$};
  \node at (15,1) {$=\gamma(u) \beta_-(u,\lambda)\beta_+(u,\lambda')\gamma(u)\;.$};

    \node at (1.3,2.3){$\lambda$};
\end{tikzpicture}
\ee

\be
\begin{tikzpicture}[scale=0.5,baseline={(0,1cm)}]
 \node at (-2,2){$\Lcal:$};
  
  \draw[->,thick,black!10!green] (2,1)--(2,3.5);
  \draw[->,thick,black!10!green] (4,1)--(4,3.5);
  \draw[->,thick,black!10!green] (6,1)--(6,3.5);
  \draw[->,thick,black!10!green] (8,1)--(8,3.5);
  
  \draw[->,thick,black!20!red] (1,1.5)--(9,2.5);
  
  \node at (1,0) {$\cdots$};
  \node at (2,0) {$X$};
  \node at (4,0) {$Y$};
  \node at (6,0) {$X$};
    \node at (8,0) {$X$};
  \node at (9,0) {$\cdots$};

  \node at (1,4) {$\cdots$};
  \node at (2,4) {$X$};
  \node at (4,4) {$Y$};
  \node at (6,4) {$X$};
  \node at (8,4) {$X$};
  
  \node at (9,4) {$\cdots$};

  \node at (3,2.2) {$X$};
  \node at (5,2.4) {$X$};
  \node at (7,2.7) {$X$};
  
  \node at (0,1.5) {$X$};
  \node at (9.7,2.8) {$X$};
  
  \node at (16.5,3) {$=R_{XX}^{XX}(\lambda')~R_{XX}^{XX}(\lambda)~R_{XY}^{XY}(\lambda')~R_{XX}^{XX}(\lambda)$};
  \node at (14.2,1) {$=\gamma(u) \gamma(u)\alpha(u,\lambda')\gamma(u)\;.$};
   \node at (1.3,2.3){$\lambda$}; 
\end{tikzpicture}
\ee

\be
\begin{tikzpicture}[scale=0.5,baseline={(0,1cm)}]
  \node at (-2,2){$\Rcal:$};
 
  \draw[->,thick,black!10!green] (2,1)--(2,3.5);
  \draw[->,thick,black!10!green] (4,1)--(4,3.5);
  \draw[->,thick,black!10!green] (6,1)--(6,3.5);
  \draw[->,thick,black!10!green] (8,1)--(8,3.5);
  
  \draw[->,thick,black!20!red] (1,1.5)--(9,2.5);
  
  \node at (1,0) {$\cdots$};
  \node at (2,0) {$X$};
  \node at (4,0) {$Y$};
  \node at (6,0) {$X$};
    \node at (8,0) {$X$};
  \node at (9,0) {$\cdots$};

  \node at (1,4) {$\cdots$};
  \node at (2,4) {$X$};
  \node at (4,4) {$X$};
  \node at (6,4) {$X$};
  \node at (8,4) {$Y$};
  
  \node at (9,4) {$\cdots$};

  \node at (3,2.2) {$X$};
  \node at (5,2.4) {$Y$};
  \node at (7,2.7) {$Y$};
  
  \node at (0,1.5) {$X$};
  \node at (9.7,2.8) {$X$};
  
  \node at (16.5,3) {$=R_{YX}^{XY}(\lambda')~R_{YX}^{YX}(\lambda)~R_{XY}^{YX}(\lambda')~R_{XX}^{XX}(\lambda)$};
   \node at (15.6,1) {$=\beta_-(u,\lambda') \alpha(u,\lambda)\beta_+(u,\lambda')\gamma(u)$};
    \node at (1.3,2.3){$\lambda$};
\end{tikzpicture}
\ee
The coefficients refer to (\ref{Feldercoeff}) where we note that $\gamma(u)$ does not depend on the dynamical parameter. We have named these contributions $\mathcal{S},\mathcal{L}$ and $\mathcal{R}$ for ``stay'', ``left'' and ``right'' as, after the action of $\Omega^{-1}$, they correspond to the $Y$ field staying on the same site or moving one site to the left or right. There are no terms with $Y$ shifted more to the left, while terms with $Y$ shifted more to the right contain higher powers of $\alpha(u)$ and thus $u$, and can be ignored to linear order in $u$. The Hamiltonian will thus be nearest-neighbour, just like in the non-dynamical case (and unlike the situation when one adds alternating inhomogeneities $u_i$ at each site (e.g. \cite{deVega:1991rc}), which shift e.g. $\alpha(u)$ to $\alpha(u-u_i)$ and modify the above argument). 

Considering now the full chain of length $L$, the vacuum configuration will just contribute a factor of $\gamma(u)^L$  and we are interested in the difference due to presence of the magnon. We thus divide our 4-site block by $\gamma(u)^4$ to obtain the final terms that will contribute to the Hamiltonian.\footnote{We could have avoided this step by starting from a rescaled (\ref{Feldersym}) where the $\gamma(u)$ terms are 1.} We conclude that for a $Y$ which initially has $\lambda$ to its left, the contributions to the Hamiltonian are
\be
  \Scal=\frac{\beta_-(u,\lambda)\beta_+(u,\lambda')}{\gamma(u)^2}\;,\;\;
  \Lcal=\frac{\alpha_s(u,\lambda')}{\gamma(u)}\;,\;\;
  \Rcal=\frac{\beta_-(u,\lambda')\alpha_s(u,\lambda)\beta_+(u,\lambda')}{\gamma(u)^3}\;.
\ee
  Repeating the above derivation but with a $\lambda'$ to the left of the first $R$-matrix (which also implies a $\lambda'$ to the left of the initial $Y$ magnon, we find:
\be
  \Scal'=\frac{\beta_-(u,\lambda')\beta_+(u,\lambda)}{\gamma(u)^2}\;,\;\;
  \Lcal'=\frac{\alpha_s(u,\lambda)}{\gamma(u)}\;,\;\;
  \Rcal'=\frac{\beta_-(u,\lambda)\alpha_s(u,\lambda')\beta_+(u,\lambda)}{\gamma(u)^3}\;.
  \ee
Substituting the values of the coefficients from (\ref{Feldercoeff}), we obtain:
\be
\Scal=\frac{\ttheta(2\eta)^2\ttheta(2\eta+\lambda-u)\ttheta(u+\lambda)}{\ttheta(u-2\eta)^2\ttheta(\lambda)\ttheta(2\eta+\lambda)}
=\frac{\theta_4(0)^2\theta_2(u)\theta_3(u)}{\theta_4(u)^2\theta_2(0)\theta_3(0)}\;,
\ee
\be
\Lcal=\frac{\sqrt{\ttheta(\lambda)\ttheta(\lambda+4\eta)}}{\ttheta(\lambda+2\eta)}\frac{\ttheta(u)}{\ttheta(u-2\eta)}
=-i\frac{\ttheta(\lambda)}{\ttheta(\lambda+2\eta)}\frac{\ttheta(u)}{\ttheta(u-2\eta)}
=-i\frac{\theta_3(0)}{\theta_2(0)}\frac{\theta_1(u)}{\theta_4(u)}\;,
\ee
and
\be
\begin{split}
\Rcal&=\frac{\sqrt{\ttheta(\lambda+2\eta)\ttheta(\lambda-2\eta)}}{\ttheta(\lambda)}\frac{\ttheta(-u+\lambda+2\eta)\ttheta(u+\lambda+2\eta)}{\ttheta(\lambda+2\eta)^2}\frac{\ttheta(2\eta)^2\ttheta(u)}{\ttheta(u-2\eta)^3}\\
&=-i\frac{\theta_2(0)}{\theta_3(0)}\frac{\theta_2(0)^2}{\theta_2(u)^2}\frac{\theta_4(0)^2\theta_1(u)}{\theta_4(u)^3}\;,
\end{split}
\ee
where to obtain the final result we have first gone from the $\ttheta$ notation back to standard $\theta_1$ functions by simplifying the exponent in (\ref{Ttheta}), and then used standard theta function identities (e.g. \cite{Gomez96,Thetavocabulary}) to express e.g. $\theta_1(u+\tau/2)$ in terms of $\theta_4(u)$ etc. 

For the Hamiltonian we are interested in the $O(u)$ terms in the series expansion of the above coefficients. Given that $\theta_1(u)$ is odd while the other three functions are even, we immediately find:
\be
\Scal=1+O(u^2)\;\;,\quad\; \Lcal'=-i\frac{1}{\kappa} \frac{\theta_1'(0)}{\theta_4(0)}u+O(u^2)\;\;,\quad\; \Rcal=-i\kappa \frac{\theta_1'(0)}{\theta_4(0)}u+O(u^2)\;.
\ee
where we have substituted $\kappa=\theta_2(0)/\theta_3(0)$. 

The computation for the primed operators is completely analogous, with the only essential difference being the shifts of the dynamical parameters in $\Lcal$ and $\Rcal$. We find
\be
\Scal'=1+O(u^2)\;\;,\quad\; \Lcal'=-i\kappa \frac{\theta_1'(0)}{\theta_4(0)}u+O(u^2)\;\;,\quad\; \Rcal'=-i\frac{1}{\kappa} \frac{\theta_1'(0)}{\theta_4(0)}u+O(u^2)\;.
\ee
To compare with the field theory Hamiltonian, let us focus on the $O(u)$ terms and divide by the prefactor $i\theta_1'(0)/\theta_4(0)$.
Let us use the same convention as in Section \ref{sec:XYsector}, that a magnon on an even site sees a $\lambda'$ dynamical parameter, while a magnon on an odd site sees a $\lambda$. So to act on a $Y$ magnon on an even site $2r$ of the chain we need to use $\Lcal'$ and $\Rcal'$:
\be
\Lcal' \ket{XXXYXX} =-\frac{1}{\kappa}\ket{XXYXXX}\;,\quad
\Rcal' \ket{XXXYXX} =-\kappa\ket{XXXXYX}\;
\ee
The action of the $\Lcal'$ operator can be identified with the action of $\Hcal_{2r-1,2r}$ and thus with $(\Hcal_{oe}){XY}^{YX}=-\frac{1}{\kappa}$, while the $\Rcal'$ operator with the action of $\Hcal_{2r,2r+1}$, and so with $(\Hcal_{eo})_{YX}^{XY}=-\kappa$. Considering now a $Y$ magnon on an odd site $2r+1$, the transfer matrix acts as:
\be
\Lcal \ket{XXYXXX} =-\kappa\ket{XYXXXX}\;\quad
\Rcal \ket{XXYXXX} =-\frac{1}{\kappa}\ket{XXXYXX}\;.
\ee
This is interpreted as the action of $\Hcal_{2r,2r+1}$ with $(\Hcal_{eo})_{XY}^{YX}=-\kappa$, and of $\Hcal_{2r+1,2r+2}$ with $(\Hcal_{oe}){YX}^{XY}=-\frac{1}{\kappa}$. We thus conclude that the Hamiltonian acts differently on (even,odd) sites and (odd,even) sites, just as expected from the field theory. Using the same basis as in Section \ref{sec:XYsector} we conclude that our alternating Hamiltonian is:
\be \label{dynamicalXX}
\Hcal_{\text{oe}}=\Hcal(\lambda)=\left(\begin{array}{cccc}0 &0&0&0\\0& 0& -1/\kappa &0 \\ 0 &-1/\kappa &0 &0\\0 & 0& 0&0\end{array}\right)\;,\;\;
\Hcal_{\text{eo}}=\Hcal(\lambda+\gamma)=\left(\begin{array}{cccc}0 &0&0&0\\0& 0& -\kappa &0 \\ 0 &-\kappa &0 &0\\0 & 0& 0&0\end{array}\right)\;.
  \ee
Each of these Hamiltonians is that of the XX model, $\Hcal_{\ell,\ell+1}=-(\sigma_x\otimes \sigma_x+\sigma_y\otimes \sigma_y)$. However the coefficient alternates between $\kappa$ and $1/\kappa$ depending on the site. We have thus found an alternating XX model (and not the alternating XXX model studied in Section \ref{sec:XYsector}).

The 2-magnon problem proceeds analogously, defining the necessary terms such as $(\Scal,\Lcal)$, $(\Lcal,\Rcal)$ etc. and taking special care of the case where the two magnons are adjacent. The Hamiltonian still agrees with that of a dynamical XX model.

This model is integrable by construction and can be studied by ABA using the techniques of \cite{Felder:1996xym}. However, due to the elliptic-root-of-unity condition $4\eta=\tau$ the naive ABA misses certain eigenvectors which need to be constructed via a limiting process, described in \cite{Deguchi_2002}. Note also that it is not surprising that we have obtained an XX model as the condition $\eta=\tau/4$ is similar to the usual free-fermionic conditions on the XYZ model (often expressed via the deformation parameter $\gamma=2\eta$ as e.g. $\gamma=1/2$ \cite{BaxterBook}) which lead to an XY model. Also in our case, this condition has led to the vanishing of the $\sigma_z\otimes \sigma_z$ terms which would need to be added to obtain the alternating XXX model which describes the $XY$ sector. 

We will not reproduce the ABA solution here (it is given in \cite{Deguchi_2002}, specialised to our choice $\lambda=\tau/2+1/2,\eta=\tau/4$), as our goal was just to show how to obtain an alternating Hamiltonian. However, it is straightforward to check that the momentum of a single magnon, the ratio functions between even and odd sites, and the 1-magnon energy coming from the ABA fully agree with (\ref{Momentumtheta}), (\ref{Ratiotheta}) and (\ref{Energytheta}), respectively (with the total energy being just $E'(u)$, i.e. $E(u)$ minus the constant). This is natural, as for a single magnon the difference between the alternating XX of this appendix and the $XY$-sector alternating XXX model is just an additive constant. However, for two magnons the interacting equations differ between the two cases, and this leads to the alternating XX model just having an $S$-matrix $S=-1$, while the solution of the alternating XXX model has all the new features discussed in Section \ref{sec:XYsector}, with nontrivial $S$-matrices such as (\ref{SAmatrixXY}) and (\ref{TAmatrixXY}).


\bibliography{interpolating}
\bibliographystyle{utphys}

\end{document}